\documentclass[a4paper]{cas-sc}

\usepackage[dvipsnames]{xcolor}

\usepackage[utf8]{inputenc} 
\usepackage[english]{babel}
\DeclareUnicodeCharacter{2041}{\^y}
\usepackage{makecell}

\usepackage{float}
\usepackage{comment}
\usepackage{epstopdf}
\usepackage{adjustbox}
\usepackage{subfigure}
\usepackage{wrapfig}
\usepackage{multicol}
\usepackage{framed}

\usepackage{tikz}
\usetikzlibrary{arrows.meta,
                chains,
                positioning,
                shapes.geometric,
                shapes,
                arrows,
                }
\usetikzlibrary{calc}

\usepackage[ruled,vlined]{algorithm2e}

\usepackage{cellspace}
\usepackage{mathtools}
\let\oldref\ref
\renewcommand{\ref}[1]{(\oldref{#1})}
\usepackage{amsmath}
\usepackage{tikz}

\usepackage{listings}
\newlength\tindent
\newcommand{\abs}[1]{|#1|}

\usepackage{natbib}

\def\tsc#1{\csdef{#1}{\textsc{\lowercase{#1}}\xspace}}
\tsc{WGM}
\tsc{QE}
\tsc{EP}
\tsc{PMS}
\tsc{BEC}
\tsc{DE}

\begin{document}
\let\WriteBookmarks\relax
\def\floatpagepagefraction{1}
\def\textpagefraction{.001}
\shorttitle{Parallel domain decomposition with multiple layers of ghost cells}
\shortauthors{V. Delmas and  A. Soulaïmani}

\title [mode = title]{Parallel high-order resolution of the Shallow-water equations on real large-scale meshes with complex bathymetries}

\author[1]{Vincent Delmas}
\cormark[1]
\cortext[cor1]{Corresponding author}

\credit{Conceptualization of this study, Methodology, Software}

\address[1]{Department of Mechanical Engineering, École de Technologie Supérieure (ÉTS), 1100 Notre-Dame Ouest, Montréal, QC H3C 1K3, CANADA}

\author[1]{Azzeddine Soulaïmani}

\begin{abstract}
The resolution of the Shallow-water equations is of practical interest in the study of inundations and often requires very large and dense meshes to accurately simulate river flows. Those large meshes are often decomposed into multiple sub-domains to allow for parallel processing. When such a decomposition process is used in the context of distributed parallel computing, each sub-domain requires an exchange of one or more layers of ghost cells at each time step of the simulation due to the spatial dependency of numerical methods. In the first part of this paper, we show how the domain decomposition and ghost-layer generation process can be performed in a parallel manner for large meshes, and show a new way of storing the resulting sub-domains with all their send/receive information within a single CGNS mesh file. The performance of the ghost-layer generation process is studied both in terms of time and memory on 2D and 3D meshes containing up to 70 million cells. In the second part of the paper, the program developed in the first part is used to generate the domain decomposition of large meshes of practical interest in the study of natural free surface flows. We use our in-house multi-CPU multi-GPU (MPI+CUDA) solver to show the impact of multiple layers of ghost cells on the execution times of the first-order HLLC, and second-order WAF and MUSCL methods. Finally, the parallel solver is used to perform the second-order resolution on real large-scale meshes of rivers near Montréal, using up to 32 GPUs on meshes of 13 million cells.
\end{abstract}

\begin{highlights}
\item The domain decomposition with the per-zone send/receive cell information is stored in a single CGNS file.
\item The Parallel-CGNS library is used for parallel IO to improve performance.
\item A second-order multi-GPU solver is developed to solve the Shallow-water equations on a 13 million cell mesh.
\item The differences between the first-order and second-order solutions are shown on real large-scale meshes.
\end{highlights}


\maketitle

\section{Introduction}
In many scientific and engineering computational applications, the need for very large simulation meshes arises either from the need for a very dense mesh or for a mesh that covers a very large area. In either case, distributed programming is often the only way to tackle simulations using such large meshes. One common way of dealing with such large-size meshes is to partition them into sub-domains that will be handled by one process. However, the spatial dependency of numerical methods means that information must be exchanged at each time step of the simulations. This memory exchange is often performed using an implementation of the Message Passing Interface (MPI), and it must be as fast as possible to ensure good scaling across multiple compute units. One way to facilitate the memory exchange is to add to each sub-domain new \textit{ghost cells}, also called \textit{overlapping cells}, which correspond to cells that will be received from adjacent sub-domains at each step of the computation. In that context, generating a high-quality domain decomposition with the appropriate, well-numbered, layers of ghost cells is key to fast simulations scaling well across many compute units.

\bigbreak
This paper follows from a previous work \citep{delmasvsoulaima}, where a multi-CPU multi-GPU solver was developed using CUDA and a CUDA-Aware version of OpenMPI to solve the Shallow-water equations. That earlier work presented a sequential pre-processing step to perform the domain decomposition and add a single layer of ghost cells to each sub-domain. In this paper, the goal is to improve this pre-processing step by using parallel processing, adding multiple layers of ghost cells to each sub-domain, and then storing the entire domain decomposition inside a single mesh file. The goal is to use the multi-GPU solver to perform the parallel high-order resolution of the Shallow-water equations on large-scale domains with complex bathymetries, such as a full simulation of the Montréal archipelago.

\bigbreak The first part of this paper presents the process of generating the domain decomposition with multiple layers of ghost cells in a parallel manner. The ParMETIS \citep{parmetis} library is used to perform the graph partitioning in parallel, and the multiple layers of ghost layers are added with a parallel algorithm derived from \cite{pat}. The performance of this generation process is studied in detail for 2-dimensional and 3-dimensional meshes both in terms of execution time and memory. To minimize the number of files generated by the domain decomposition, the CFD General Notation System (CGNS) file format is used to store the entire domain decomposition, including the send/receive information necessary for the memory exchange, inside a single mesh file. The time taken by the Input/Output (IO) operations is studied when using the simple sequential CGNS library and the Parallel CGNS (PCGNS) library.

\bigbreak
In the second part of this paper, the domain decomposition process presented in the first part is used to perform the parallel high-order resolution of the Shallow-water equations. To this end, the first-order HLLC, and second-order WAF and MUSCL methods are used in our in-house multi-GPU solver \citep{delmasvsoulaima}. The quality of the resolution is first displayed in a simple test case, and then  the necessity of having multiple layers of ghost cells when using a second-order method is revealed. The scaling of the solver across multiple compute units is studied in detail for the first- and second-order methods using multiple layers of ghost cells. Finally, the multi-GPU solver is used to perform a second-order resolution on real meshes of rivers near Montréal, using up to 32 GPUs on meshes of 13 million cells.

\section{Related work}
The problem of decomposing a mesh in multiple sub-domains is known as the graph partitioning problem and has spawned massive research in the domain of parallel computing. The goal of the domain decomposition is to partition the problem into multiple chunks that are to be handled by a compute unit such as a CPU core or a GPU in the following simulations. Many very good graph partitioning libraries exist, the most commonly used are \textit{SCOTCH} \citep{scotch} and \textit{METIS} \citep{metis}. Both of these libraries have parallel version that make use of the Message Passing Interface (MPI) in the name of \textit{PT-SCOTCH} \citep{ptscotch} and \textit{ParMETIS} \citep{parmetis}. We use \textit{ParMETIS} in this paper but the same work could be accomplished using \textit{PT-SCOTCH} or any MPI parallel implementation of a graph partitioning library.\\

If the decomposed mesh is to be used in a shared memory system, the pre-processing step can end after the domain decomposition as each process will be able to access any data. However, if the mesh is to be used on distributed memory systems, the spatial dependency of the numerical methods will require each process to access data owned by other processes. The most common way of dealing with this issue is to add layers of ghost cells to each sub-domain and to use the MPI library to exchange such cells at each time step. Depending on the type of mesh, the computation of layers of ghost cells is more or less complex and time-consuming. On regular structured meshes, their computation is quite straightforward \citep{halompifluids_struct, KOMA}. On unstructured meshes, however, it becomes much more complex and requires dedicated algorithms \citep{halompiswe_unstruct,DELAASUNCION,pat}.  The number of required ghost-layers depends on the discretization's stencil. When a simple first-order in space method is used \citep{toro}, only one layer of ghost cells is needed, however, when using higher-order methods the stencil widens \citep{weno4,wenoswe,riadh,louki} and more layers are required.\\

As the goal is to partition large meshes, the domain decomposition and the ghost-layers generation process both need to be performed in a parallel manner. A paper from \cite{Knepley} shows an implementation of an unstructured overlapping mesh distribution for the PETSc library \citep{petsc-web-page}, and in \cite{pat}, an algorithm to add multiple layers of ghost cells to sub-domains in parallel is developed for visualization purposes with Paraview \citep{para}. For our purposes, we will use ParMETIS \citep{parmetis} to perform the domain decomposition in parallel and then use a modified version of the algorithm presented in \cite{pat} to add multiple layers of ghost cells to the sub-domains. The main difference with previous works is that, whereas in \cite{pat} the interest was just in one single exchange of information between sub-domains, we want to generate send/receive information so that at each time step of the CFD simulation data be exchanged in the best way between processors. Moreover, we tackle the issue of storing the domain decomposition and the overlapping information by using the CFD general notation system  \citep{cgnsbase}, specifically using the PCGNS \citep{pcgns} library to perform parallel IO operations.\\

Concerning the resolution of the Shallow-Water equations using a multi-GPU solver, this paper follows from one of our previous works \citep{delmasvsoulaima} where a multi-GPU solver using OpenMPI \citep{openmpi}  and CUDA \citep{cudac,cudaf} was developed. One can find the use of GPUs in the context of the resolution of the Shallow-water equations in \cite{BRODTK, BRODTK2, SMITH, ESCALANTE,DELAASUNCION2, DELAASUNCION}. The key points of our previous work were the use of a CUDA-Aware version of OpenMPI that allowed for better performances of the memory exchange between GPUs at each time step, and, the overlap of the computation with communications that is further improved in this work. The numerical methods used are mostly based on \cite{toro,toro2,louki,DELAASUNCION,bristeau}. The first- and second-order resolution of the Shallow-Water equations has been studied in detail in \cite{toro,toro2,ZOKAGOA,riadh,multislopemuscl,secondorderbathyrecons,bristeau}. In this work, we specifically chose to use the Weighted Average Flux (WAF) \citep{toro2,louki} and the Monotone Upstream Scheme for Conservation Laws (MUSCL) \citep{toro2,multislopemuscl} second-order methods as they are the most widely known methods. Other methods like the Essentially Non-Oscillatory (ENO) and Weighted
Essentially Non-Oscillatory (WENO) schemes are studied in \cite{weno4,wenoswe}.

\section{Parallel algorithm to add multiple layers of ghost cells}
\label{sec:parghost}
In this section, we present the main steps of the process of performing the domain decomposition using ParMETIS and adding the multiple layers of ghost cells to each sub-domain. Figure \ref{mainalg} shows an overview of the algorithm; green blocks represent IO operations performed to read or write the CGNS mesh file, red blocks represent communication steps where the MPI library is used, and blue blocks represent the computation steps. The main goals of this algorithm are to use as little memory as possible, be fast, and scale well across multiple processors.  \\

\tikzstyle{blockcpu} = [rectangle, draw, fill=blue!20, text centered, rounded corners, minimum height=2em]
\tikzstyle{blockio} = [rectangle, draw, fill=green!20, text centered, rounded corners, minimum height=2em]
\tikzstyle{blockmpi} = [rectangle, draw, fill=red!20, text centered, rounded corners, minimum height=2em]
\tikzstyle{line} = [draw, -latex']
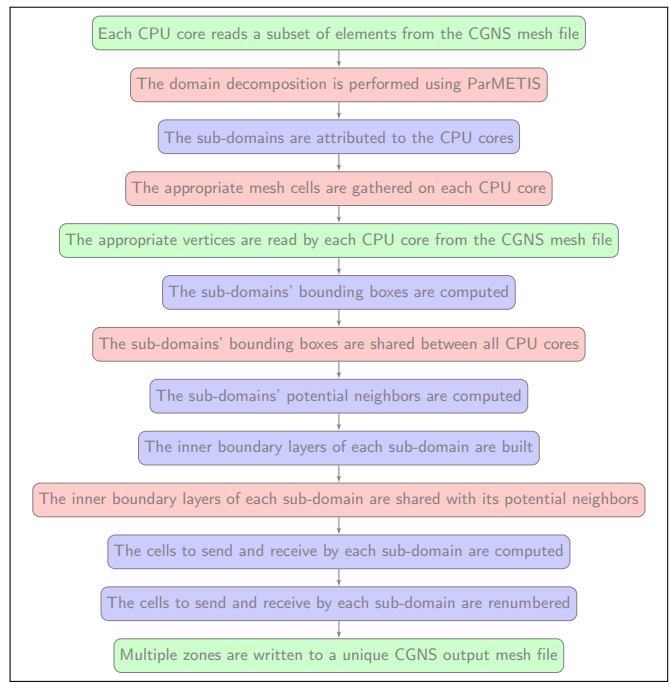
\begin{figure}[h]
\begin{center}
\fbox{
\resizebox{0.5\textwidth}{!}{
\begin{tikzpicture}[node distance = 1cm,auto]
    \node [blockio] (lect) {Each CPU core reads a subset of elements from the CGNS mesh file};
    \node [blockmpi, below of=lect] (ddec) {The domain decomposition is performed using ParMETIS};
    \node [blockcpu, below of=ddec] (supply) {The sub-domains are attributed to the CPU cores};
    \node [blockmpi, below of=supply] (gather) {The appropriate mesh cells are gathered on each CPU core};
    \node [blockio, below of=gather] (lectnod) {The appropriate vertices are read by each CPU core from the CGNS mesh file};
    \node [blockcpu, below of=lectnod] (subdo) {The sub-domains' bounding boxes are computed};
    \node [blockmpi, below of=subdo] (shareext) {The sub-domains' bounding boxes are shared between all CPU cores};
    \node [blockcpu, below of=shareext] (potnei) {The sub-domains' potential neighbors are computed};
    \node [blockcpu, below of=potnei] (sendrecv) {The inner boundary layers of each sub-domain are built};
    \node [blockmpi, below of=sendrecv] (share) {The inner boundary layers of each sub-domain are shared with its potential neighbors};
    \node [blockcpu, below of=share] (compute) {The cells to send and receive by each sub-domain are computed};
    \node [blockcpu, below of=compute] (renumber) {The cells to send and receive by each sub-domain are renumbered};
    \node [blockio, below of=renumber] (write) {Multiple zones are written to a unique CGNS output mesh file};
    
    \path [line] (lect) -- (ddec);
    \path [line] (ddec) -- (supply);
    \path [line] (supply) -- (gather);
    \path [line] (gather) -- (lectnod);
    \path [line] (lectnod) -- (subdo);
    \path [line] (subdo) -- (shareext);
    \path [line] (shareext) -- (potnei);
    \path [line] (potnei) -- (sendrecv);
    \path [line] (sendrecv) -- (share);
    \path [line] (share) -- (compute);
    \path [line] (compute) -- (renumber);
    \path [line] (renumber) -- (write);
    
\end{tikzpicture}
}
}
    \caption{Parallel algorithm to add layers of ghost cells on top of ParMETIS's decomposition. Green blocks represent IO, red blocks MPI communications, and, blue blocks CPU computations.}
\label{mainalg}
\end{center}
\vspace{-0.0\textwidth}
\end{figure}

The following three sections focus on the use of ParMETIS to perform the domain decomposition, the ghost-layers generation process, and, the renumbering of the send/receive cells. The writing of the mesh file and send/receive information is the subject of section \ref{sec:writedecomp}, along with the CGNS file format.

\subsection{Using ParMETIS to perform the domain decomposition}

ParMETIS \citep{parmetis} is the well-known parallel implementation of METIS \citep{metis}, a graph partitioning tool often used to generate the domain decomposition of a mesh. One interesting feature of ParMETIS is its ability to use a number of processes that differs from the number of sub-domains of the decomposition. This is significant as it allows, for example, to split a mesh into 1024 sub-domains using only 16 or 32 CPU cores on a single compute node. This feature is even more important when using computer clusters that have a small number of large memory nodes on which the aforementioned domain decomposition can take place. As a result, this is a feature we wanted to keep in our algorithm, hence, a process might well be in charge of multiple sub-domains during the decomposition and ghost-layers generation.\\

\begin{figure}[h]
\vspace{-1mm}
\begin{center}
\fbox{
\includegraphics[width=0.9\textwidth]{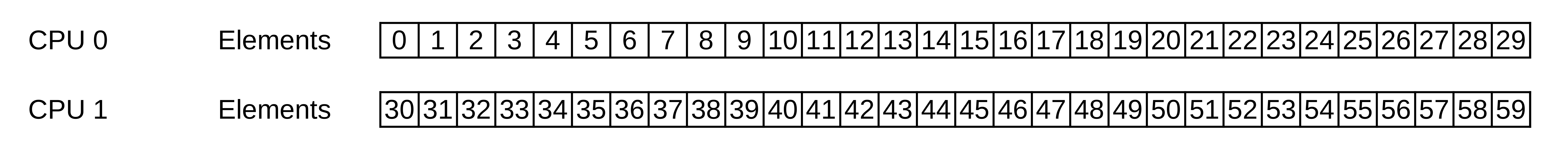}
}
\end{center}
\caption{Data owned by each CPU after reading the elements from the CGNS mesh file}
\label{figalg:fig1}
\vspace{-1mm}
\end{figure}

The way ParMETIS works is quite straightforward, at the beginning each process needs to own a subset of cells of the mesh. This is generally done by linearly partitioning the $n$ cells on the $p$ processors, i.e.,  the processor $k$ read the cells $k n/p$ to $(k+1) n/p$ as shown in Figure \ref{figalg:fig1}.  \\

\begin{figure}[h]
\vspace{-1mm}
\begin{center}
\fbox{
\includegraphics[width=0.9\textwidth]{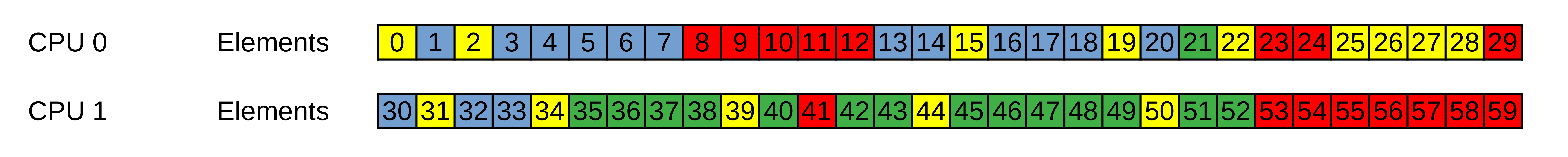}
}
\end{center}
\caption{Data owned by each CPU after using ParMETIS, each color corresponds to a sub-domain}
\label{figalg:fig2}
\vspace{-1mm}
\end{figure}

\textcolor{black}{ParMETIS needs the cell-to-cell adjacency graph which is not stored inside the CGNS mesh file. This graph is constructed by first looping through the cells and adding every edge/face to an unstructured map with the cell ID they are added from. We can then loop through the map to construct the cell-to-cell adjacency graph that will be used by ParMETIS.} Once each process has read a subset of the mesh's elements and the cell-to-cell adjacency graph has been computed, ParMETIS is used to partition the mesh into as many sub-domains as requested. ParMETIS will tag each element with the ID of the sub-domain it belongs to. This is shown by coloring the elements according to their sub-domain in Figure \ref{figalg:fig2}.\\

\begin{figure}[h]
\vspace{-1mm}
\begin{center}
\fbox{
\includegraphics[width=0.9\textwidth]{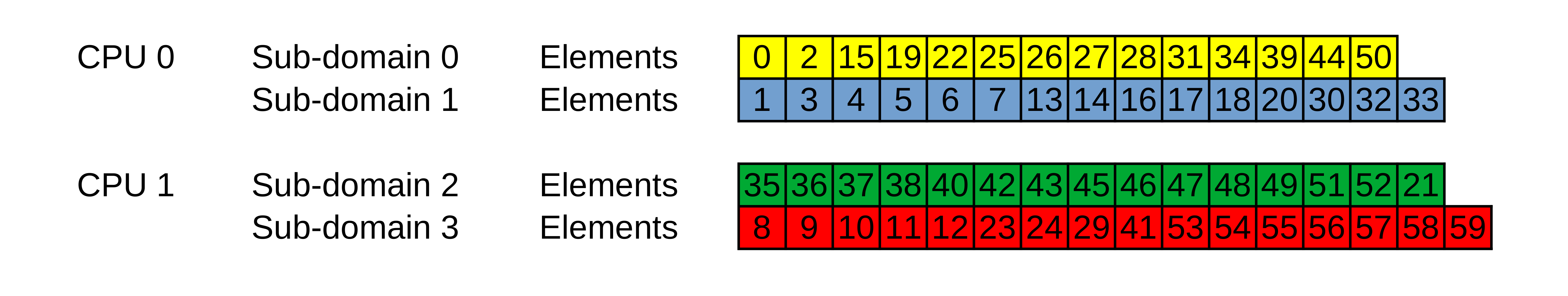}
}
\end{center}
\caption{Data owned by each CPU after the MPI memory exchange}
\label{figalg:fig3}
\vspace{-1mm}
\end{figure}

Each process owns at this point a subset of the cells that all belong to different sub-domains. We thus need to attribute a set of sub-domains to each process and exchange mesh cells so that each process owns every cell of the sub-domains it is in charge of. This is the first use of MPI to exchange data between processors and one major communication step as in the worst case each CPU core needs to send all its cells and receive every cell of the sub-domains it was chosen to handle. This is shown in Figure \ref{figalg:fig3}.\\

After each process gets all the elements of the sub-domains it owns, the vertices of the mesh are read. Each process will read the vertices from the mesh file and only keep the ones needed by the sub-domains it owns. To use as little memory as needed, the vertices are read in multiple batches rather than all in one step. Once the vertices have been read, the sub-domains are complete and the ghost cell layers can be added.

\subsection{Adding layers of ghost cells to each sub-domain}
To add multiple layers of ghost cells to each sub-domain, we use an algorithm derived from \cite{pat}. While the goal in the original paper was to exchange multiple layers of ghost cells once for visualization purposes in Paraview \citep{para}, we want to add the ghost cells to each sub-domain while keeping track of where they came from so that we can perform a memory exchange at each time step of the CFD simulations. \\

\begin{figure}[pos=htp]
\begin{center}
\fbox{
    \subfigure[One sub-domain in green]{
		\includegraphics[width=0.3\textwidth]{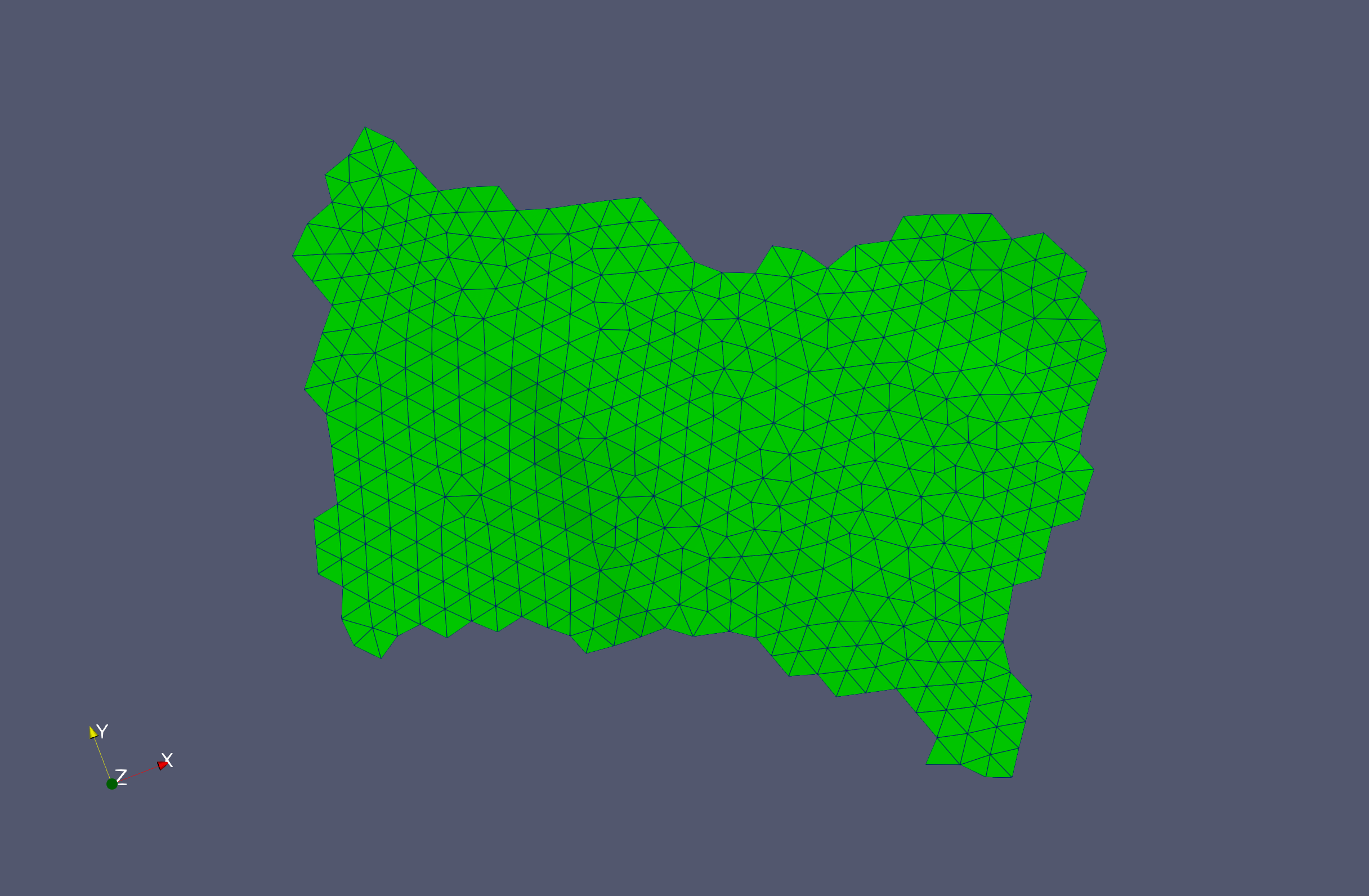}
		\label{figalg:subdo1}
    } 
    \subfigure[One inner layer in blue]{
		\includegraphics[width=0.3\textwidth]{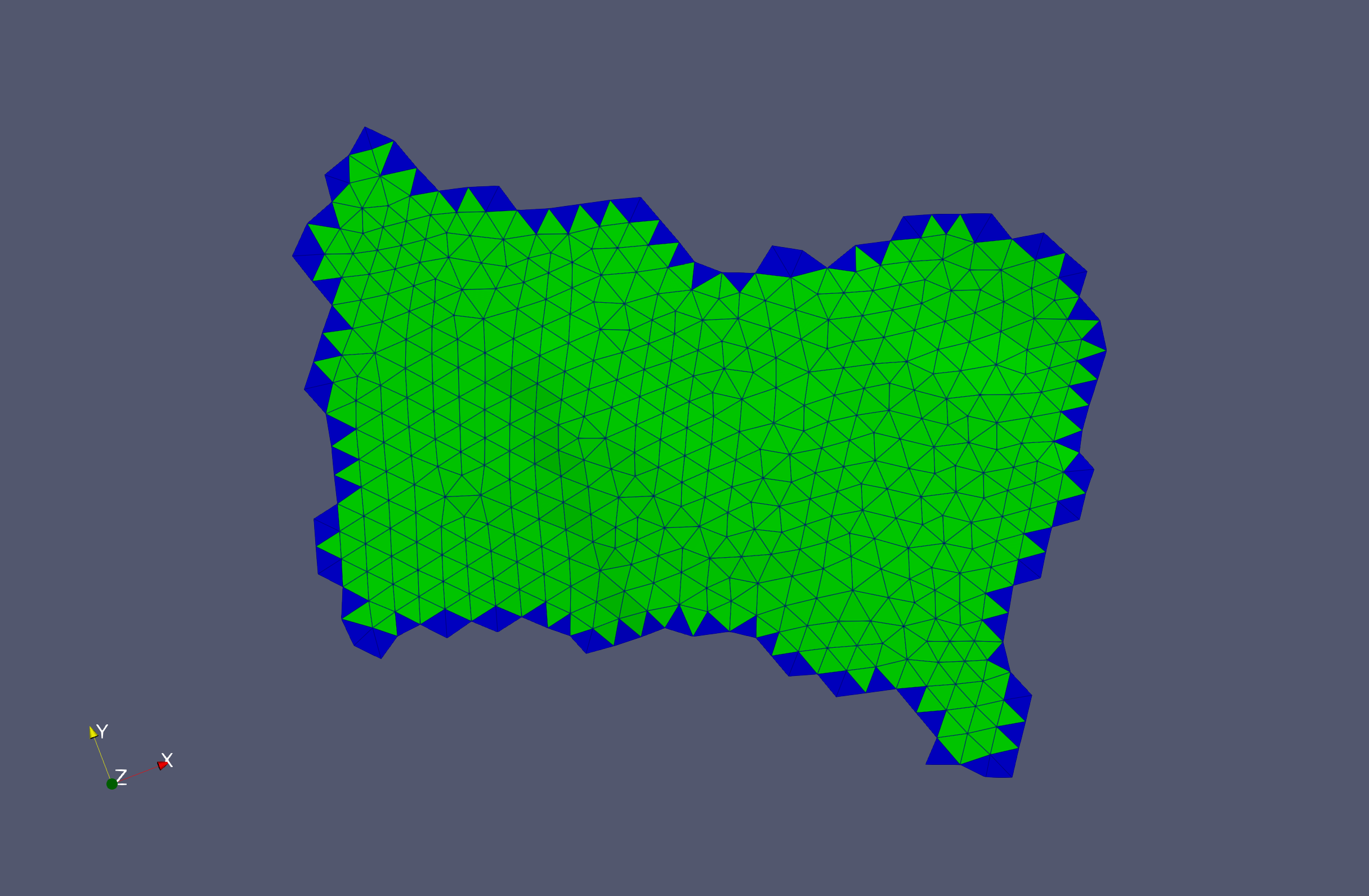}
		\label{figalg:subdo2}
    } 
    \subfigure[Three inner layers in blue]{
		\includegraphics[width=0.3\textwidth]{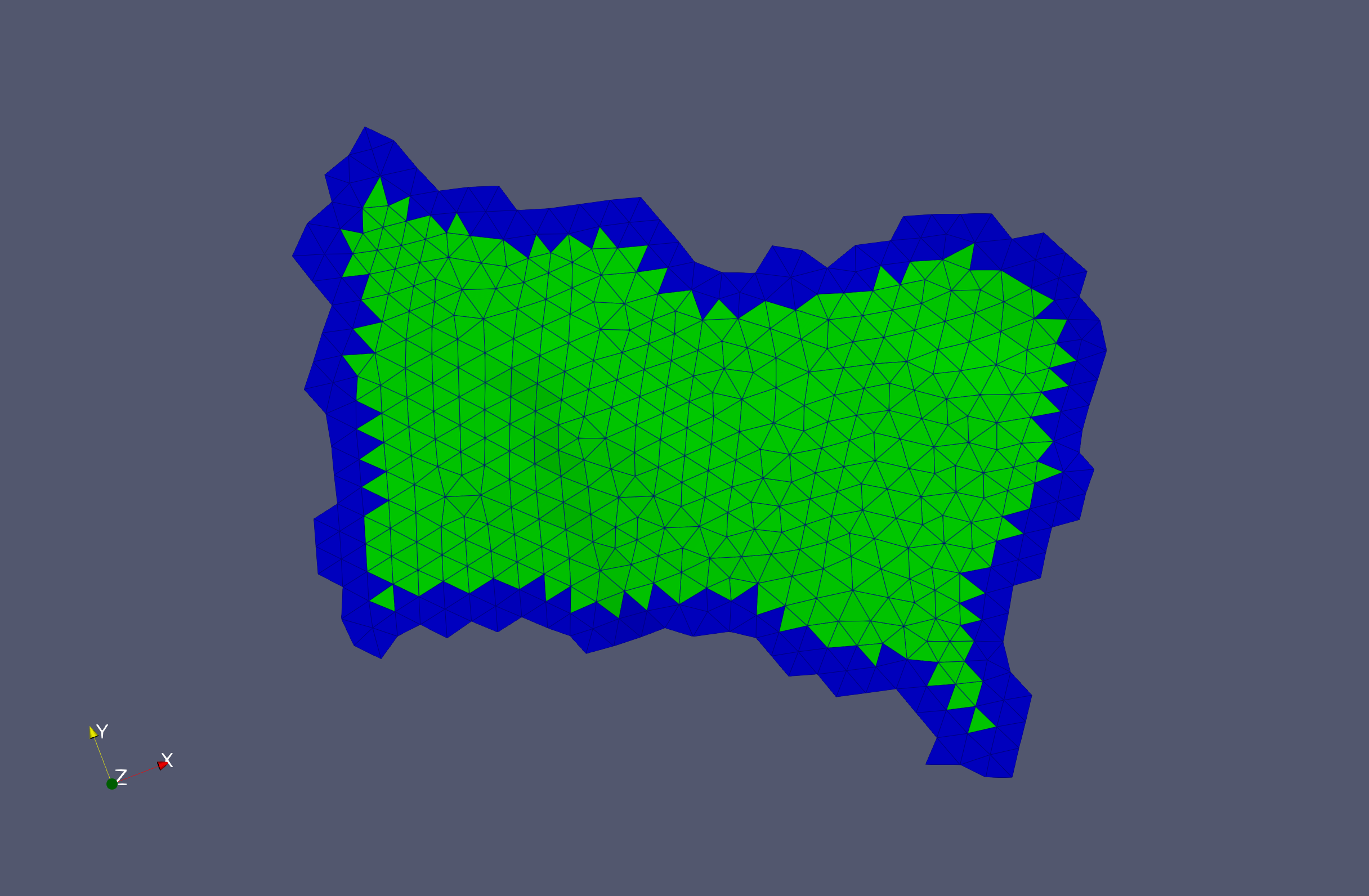}
		\label{figalg:subdo3}
    } 
    }
\end{center}
	 \caption{Inner boundary layers of a sub-domain}
	 \label{figalg:subdo}
\end{figure}

At the start of the algorithm, the bounding boxes of each sub-domain are computed and shared with every sub-domain. The goal is to create, for each sub-domain, a list of potential neighbors to reduce communications in future steps. The communication cost in this step is quite low as the extents are represented by six floating-point values for each sub-domain and are exchanged via an $MPI\_AllGather$. Once a list of potential neighbors has been computed for each sub-domain, the inner boundary layers of each sub-domain are built. This step is shown for a unique sub-domain in Figure \ref{figalg:subdo}.\\

\begin{figure}[pos=htp]
    \begin{center}
\fbox{
    \subfigure[Green sub-domain and it's neighbors]{
		\includegraphics[width=0.3\textwidth]{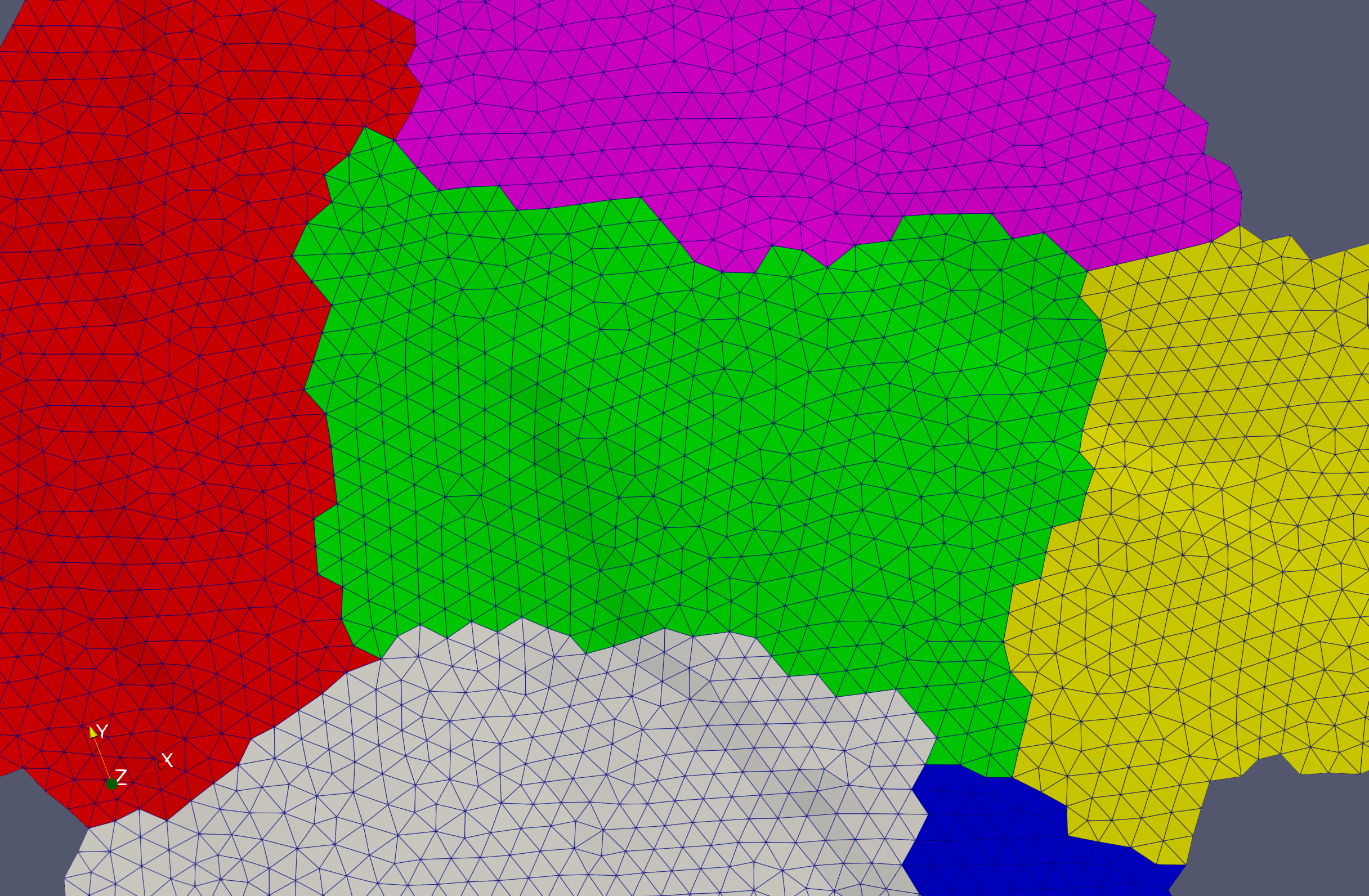}
		\label{figalg:multisubdo1}
    } 
    \subfigure[One inner boundary layer for the green sub-domain and it's neighbors]{
		\includegraphics[width=0.3\textwidth]{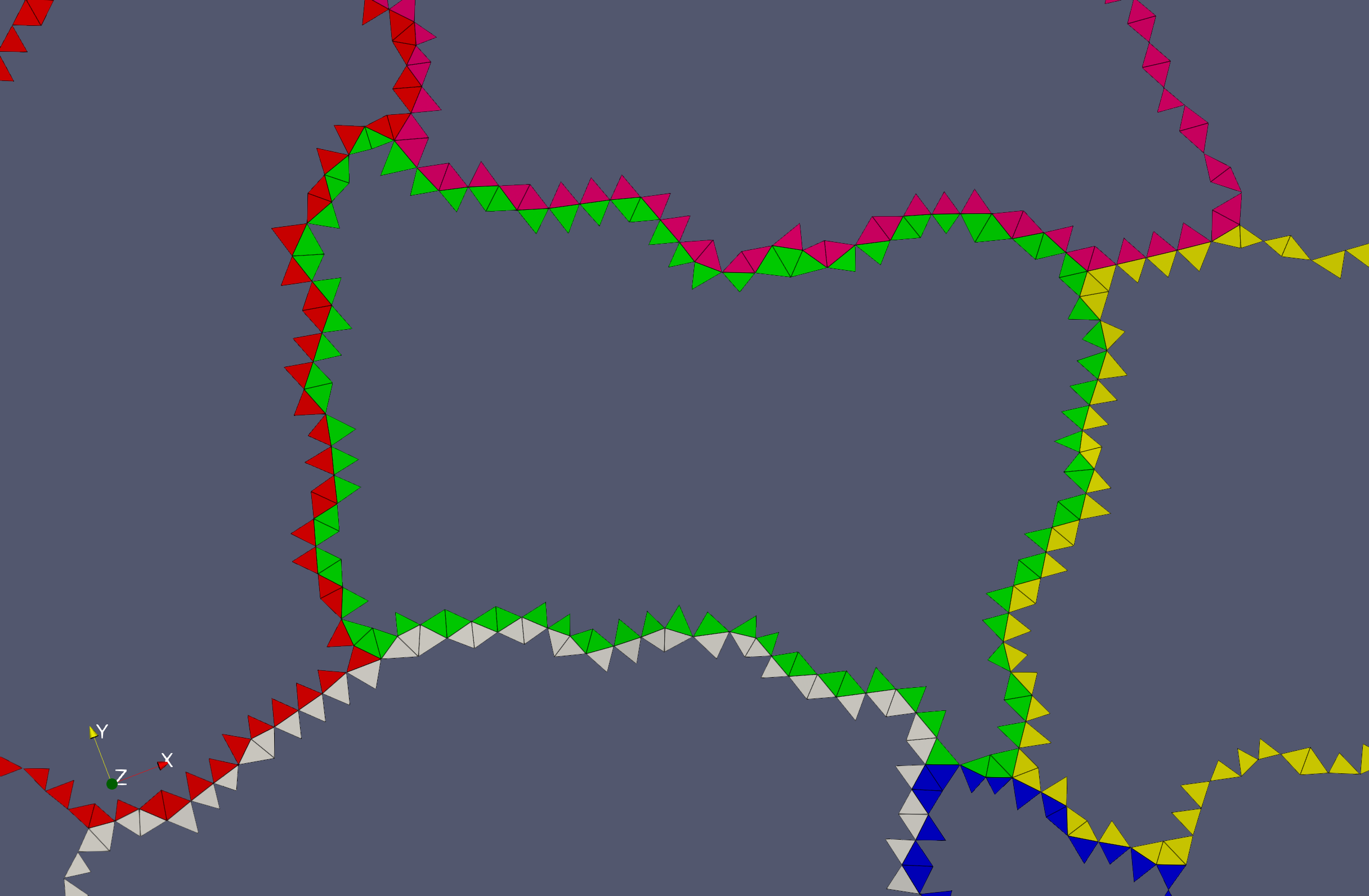}
		\label{figalg:multisubdo2}
    } 
    \subfigure[Three inner boundary layers for the green sub-domain and it's neighbors]{
		\includegraphics[width=0.3\textwidth]{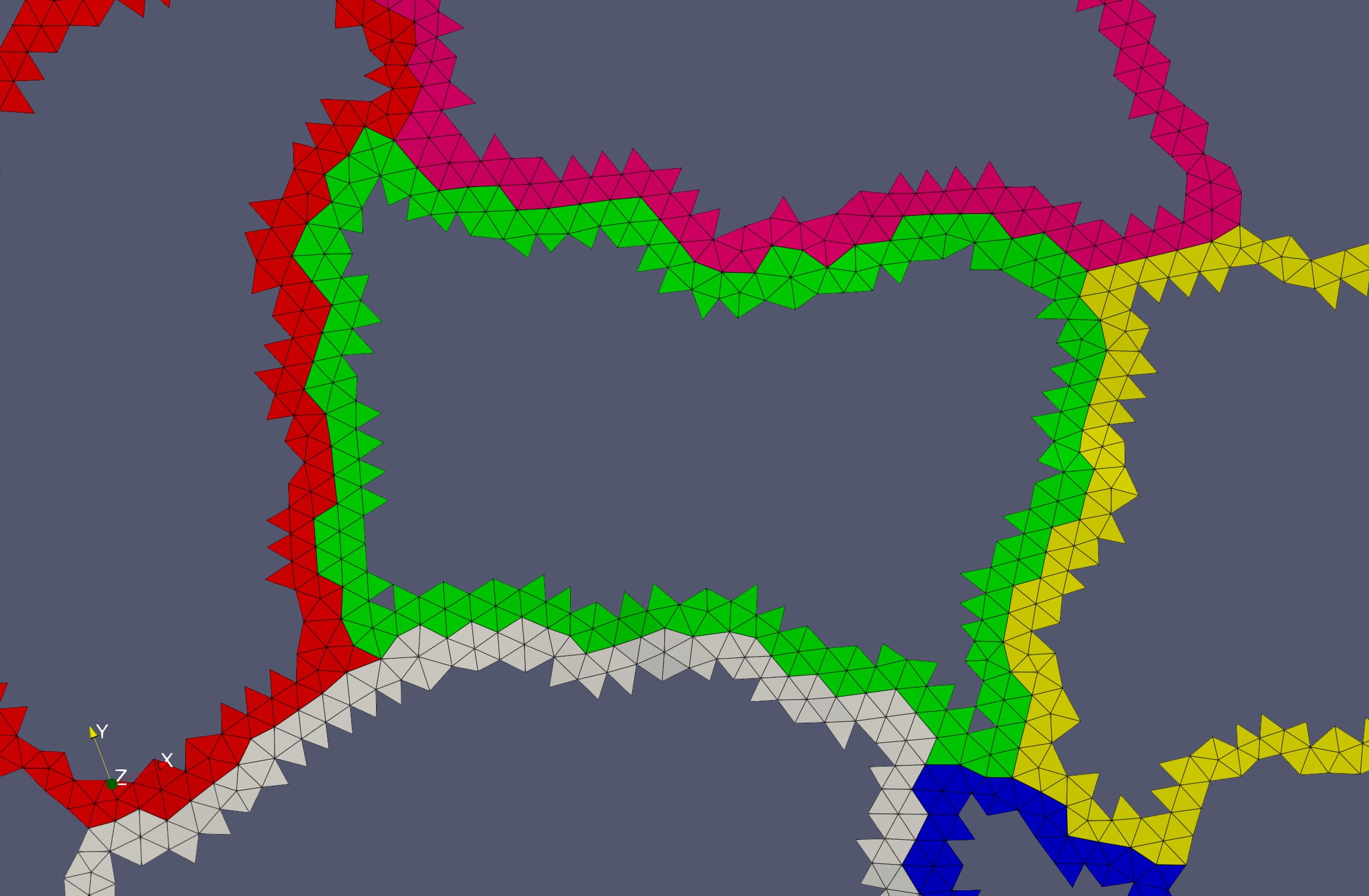}
		\label{figalg:multisubdo3}
    } 
    }
    \end{center}
	 \caption{Inner boundary layers for the green sub-domain and it's neighbors}
	 \label{figalg:multisubdo}
\end{figure}

These inner layers are then exchanged with potential neighbors of the sub-domain. Figure \ref{figalg:multisubdo} shows the inner boundary layers computed for the neighbors of the green sub-domain from Figure \ref{figalg:subdo1}.

As can be seen in Figure \ref{figalg:multisubdo}, the ghost cells that will be needed by the green sub-domain are all contained inside the inner boundary layers of its neighbors. This is the last large communication step as each sub-domain needs to send its inner boundary layers to its potential neighbors and receive their inner boundary layers. As mentioned earlier, as a process might own multiple sub-domains, there is sometimes no need for an MPI communication. As an example, in Figure \ref{figalg:multisubdo}, if the same process owns the green and the yellow sub-domains, it does not need to use MPI to communicate the inner boundary layers. \\

\begin{figure}[pos=htp]
    \begin{center}
\fbox{
    \subfigure[Three inner boundary layers for the green sub-domain and its neighbors]{
		\includegraphics[width=0.35\textwidth]{images/mulsubdo3.png}
		\label{figalg:green11}
    } 
        \subfigure[Green sub-domain with three added layers of ghost cells in yellow]{
		\includegraphics[width=0.35\textwidth]{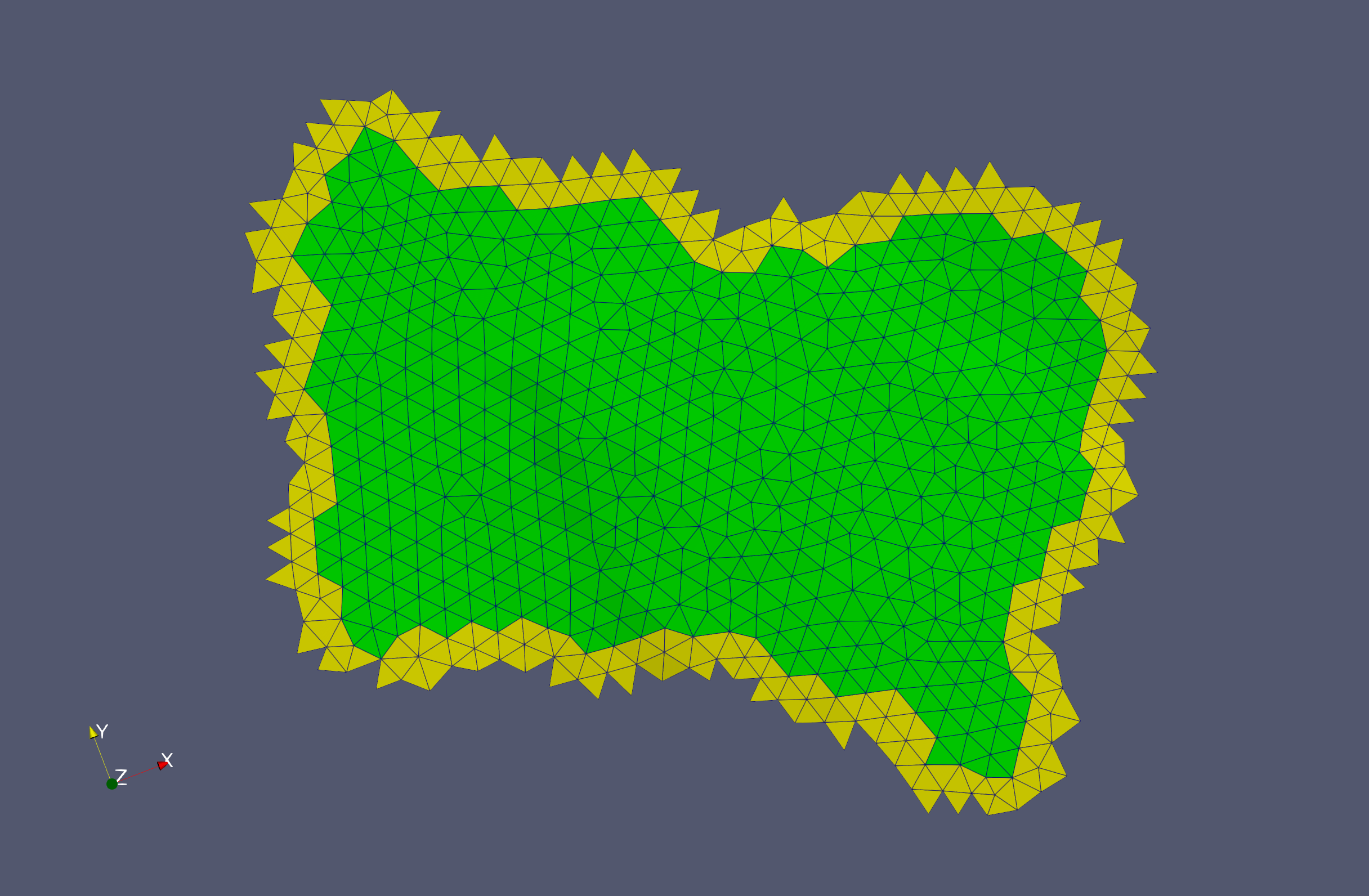}
		\label{figalg:green3}
    }
    }
    \end{center}
	 \caption{Inner boundary layers for the green sub-domain and it's neighbors}
	 \label{figalg:green10}
\end{figure}

Once each sub-domain can access its neighbors' inner boundary layers, we can easily find the cells to add. \textcolor{black}{The connectivity can be followed from the first layer of boundary elements to add a second layer as required by the methods in Section \ref{sec:application}}. Figure \ref{figalg:green3} shows in yellow the three layers of ghost cells that are added to the green sub-domain. \textcolor{black}{It should be noted that while this way of adding multiple layers of ghost-cells only requires a single exchange of the inner boundary layers it will miss ghost cells if the ghost layers are wider than the distance between non adjacent sub-domains. As the meshes resulting from this algorithm will be used to perform some high-order resolution, the number of overlapping layers will in most cases not be wider than the minimum distance between non adjacent sub-domains and there will be no problem.} We need to keep track of where the receive cells come from, as in the following CFD simulation each zone will need to receive the solution on these cells from the appropriate zone. To, speed up the MPI message exchange of these ghost cells during the simulation, they need to be contiguously numbered in the final mesh which is explained in the next section.\\

\subsection{Renumbering the cells}
\label{sec:sendrecv}
 The renumbering of the cells is the last step of the algorithm before the mesh is written to a file. This step is critical for the performance of the MPI memory exchange inside the CFD solver; the send and receive data must be treated as blocks of memory or as close as possible to blocks of memory. In this way blocks of memory containing multiple cells can be exchanged between processes efficiently, rather than sending and receiving every cell individually.\\

\begin{figure}[pos=htp]
    \begin{center}
\fbox{
    \subfigure[Green sub-domain with three inner layers in light blue]{
		\includegraphics[width=0.35\textwidth]{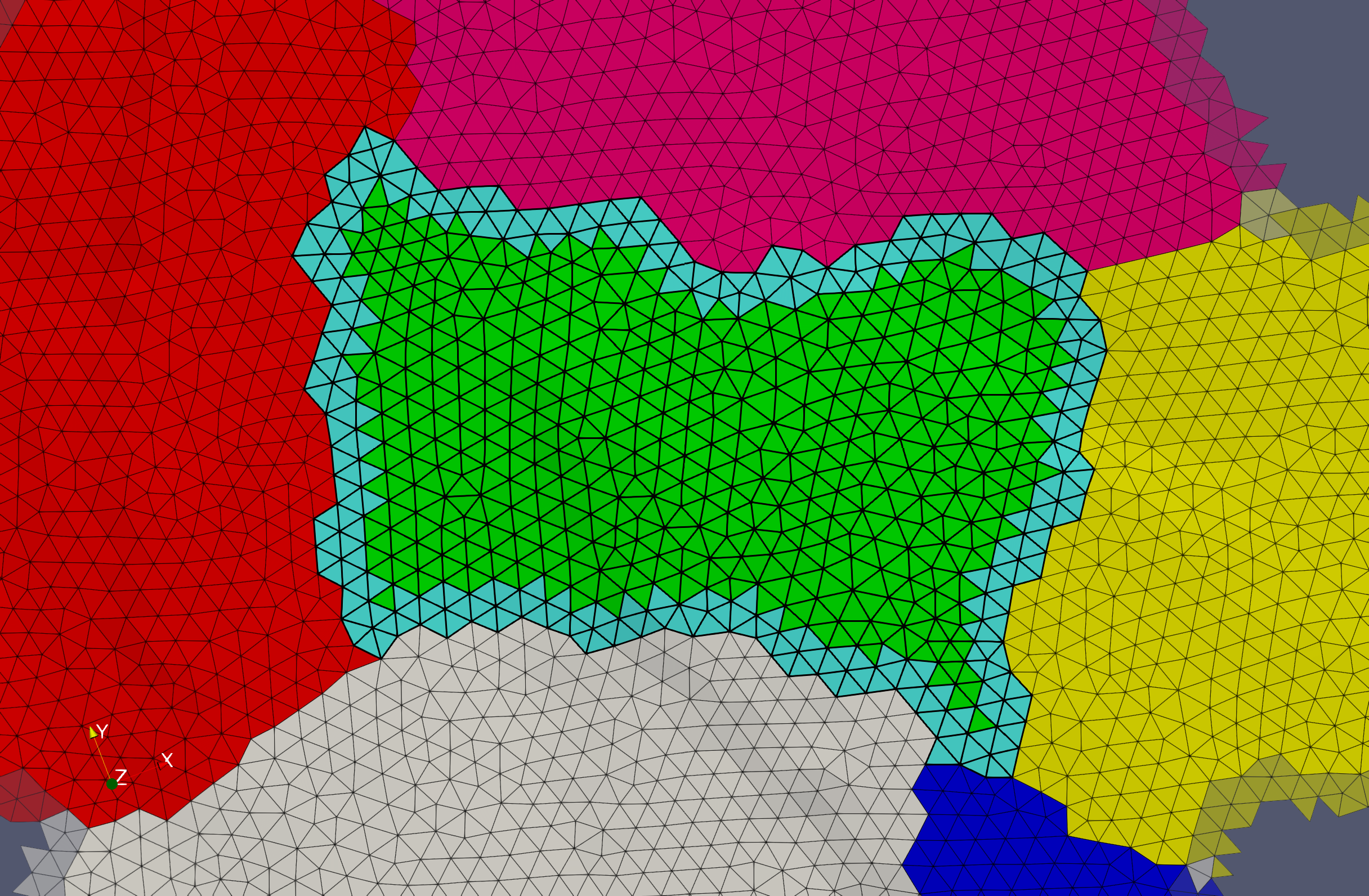}
		\label{figalg:161}
    } 
    \subfigure[Three layers of ghost cells overlapping the green sub-domain]{
		\includegraphics[width=0.35\textwidth]{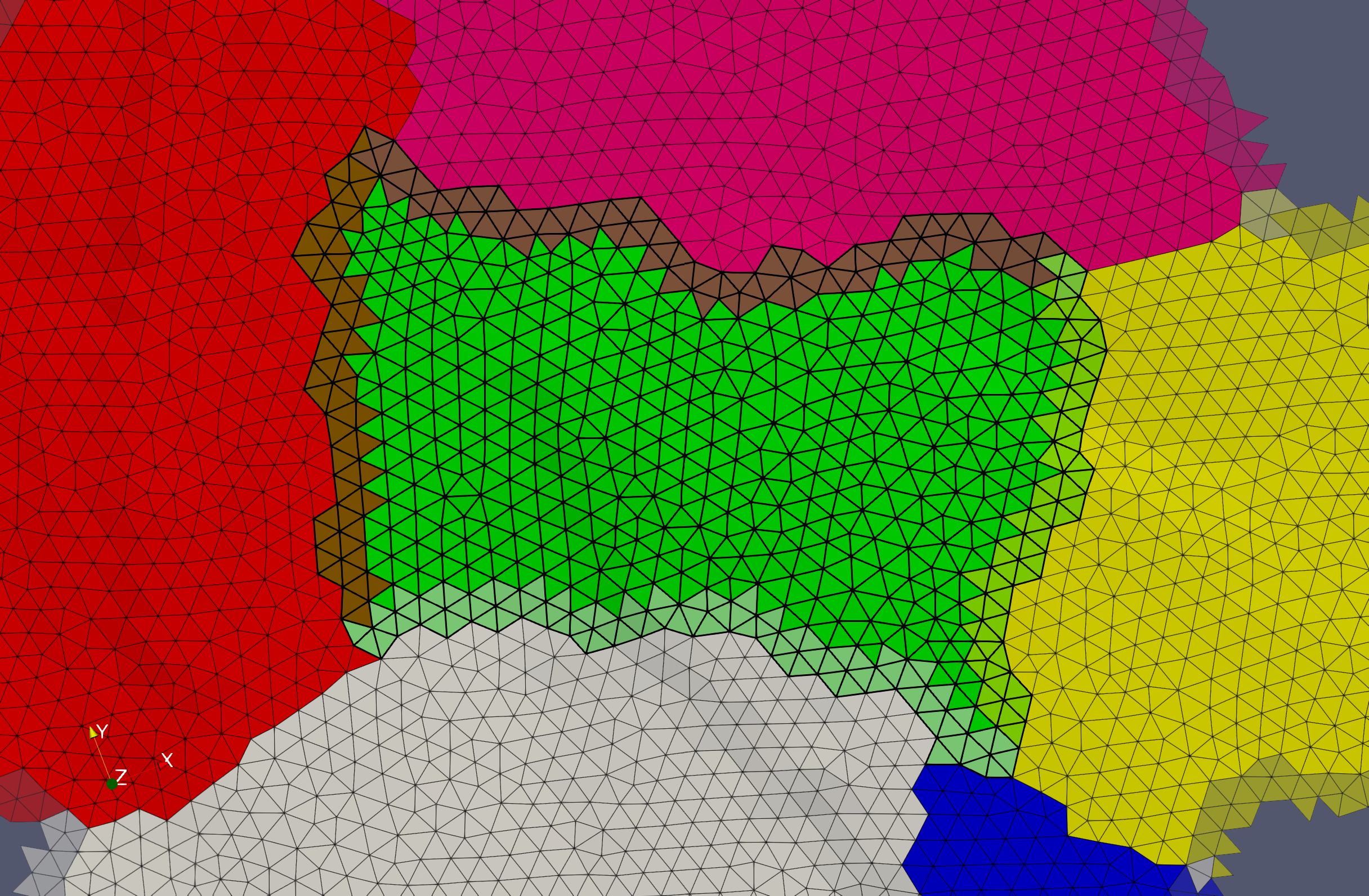}
		\label{figalg:162}
    } 
    }
    \end{center}
	 \caption{Overlapping of the neighbors' ghost cells on the inner boundary layer for three layers of ghost cells}
	 \label{figalg:green160}
\end{figure}

The first part of the renumbering process is to have the ghost cells that must be received contiguously numbered for each neighbor. Taking as example Figure \ref{figalg:green11}, from the point of view of the green sub-domain, we want to have contiguous blocks of cells coming from the red, purple, yellow, blue, and white sub-domains. This is quite easy to do, as each ghost cell is owned by only one other sub-domain. The second part is to contiguously number the cells that have to be sent to other sub-domains, which is more difficult and in most cases not possible. The issue is that one inner cell can be a ghost cell in multiple  sub-domains, which makes a strict contiguous renumbering by block infeasible. Figure \ref{figalg:162}  shows how the ghost cells of the adjacent domains overlap the green sub-domain and how some of the green cells are needed by multiple sub-domains. For example, in Figure \ref{figalg:162}, in the bottom right corner of the green sub-domain, some cells are needed by the blue, yellow, and white sub-domains. This circumstance means that we cannot number the inner cells so that we have a contiguous block of cells to send to each of the adjacent sub-domains. These cells are still renumbered in a manner as close to contiguously as possible; a send vector will need to be updated at each time step before being sent to the adjacent sub-domains at each step of the computation.

\section{Using the CFD General Notation System to write a domain decomposition}
\label{sec:writedecomp}
The CFD General Notation System (CGNS) provides a standard for storing computational fluid dynamics (CFD) meshes and data, as well as free and open software implementing this standard. It is administered by an international steering committee and is part of the American Institute of Aeronautics and Astronautics' (AIAA) recommended practices. The CGNS free and open software allows for reading and writing CGNS files from a variety of languages and platforms \citep{cgnsbase,pcgnsrecent}. A parallel implementation called PCGNS \citep{pcgns,pcgnsbenchmark1,pcgnsbenchmark2} is also available, allowing parallel IO operations using MPI. The CGNS file format and software allow for better compatibility between CFD solvers as well as between pre- and post-processing steps \citep{cgnsimpact}. In this paper, we use the CGNS file format and C++ library to read a large unstructured mesh on multiple processors and to store the domain decomposition. As the reading of the mesh file is quite straightforward, we focus here on describing how the domain decomposition is written.\\

The easiest way to store a domain decomposition is to write each zone in its own mesh file, as that will ensure fast and easy access by each processor during the CFD simulation. However, this does have some drawbacks, as the CFD solver will be storing the solution in multiple files, and at some point will have to reconstruct the solution on the global mesh or have a post-processing software do that. Furthermore, when storing a time-dependent solution, one file might be written for each time, which can lead to thousands of files generated by a single simulation. To avoid this problem, we want the whole domain decomposition to be written in a single CGNS mesh file, and the CFD solver to be capable of  writing the time-dependent solution alongside it. The task of writing the domain decomposition to a unique CGNS file can be quite complex, as there are multiple ways of tackling the problem, leading to trade-offs between communication, computation, memory, and IO time.\\

First and foremost, our main concern is that the IO operations inside the parallel CFD solver are as fast as possible, because the solution will need to be written for  numerous time steps. With this concern in mind, the CFD solver will have to use the  Parallel CGNS library to write the solutions in a parallel manner to the unique mesh file. \textcolor{black}{One of the advantages of the CGNS and PCGNS libraries is that they act in an asynchronous manner, meaning that inside the CFD solver, the solution will be written while the computations continue to be preformed. This saves time compared to classical methods of writing the solutions. We still need to chose the best configuration for the CGNS mesh file so that the PCGNS library is as fast as possible.} Benchmarks from \cite{pcgnsbenchmark2} show the performances of the PCGNS library on multiple mesh configurations and it is shown that the best configuration is for the CGNS file to contain multiple independent zones. This is the configuration we choose to write at the end of our pre-processing step and this is how the domain decomposition will be stored in a single mesh file. Each sub-domain will be stored as a zone in the mesh file, containing nodes, elements, and boundary conditions. The task is then to find a way to store the send/receive information for each zone. At first, the \textit{Rind layers} feature from the CGNS file format seems to fit the description, as it is designed to store ghost cells for each zone. The issue is that the \textit{Rind layers} are only a mechanism to store the external cells, that is to say, the cells each zone needs to receive from other sub-domains. We cannot use \textit{Rind layers} to store the internal cells that each zone needs to send to other sub-domains. As the \textit{Rind layers} are not well-adapted to what is needed here, we will use the simple \textit{PointSet} node of the CGNS library, which allows us to store either a list or a range of elements. We make use of the \textit{PointRange} feature to store the indices of the elements to receive from other partitions. Since they are contiguous in memory, we only have to store the index of the first element to receive and the number of elements to receive. As mentioned in section \ref{sec:sendrecv}, this is not possible for the elements to send; in that case, we use the \textit{PointList} feature which allows us to explicitly write every cell index that we need to send to a specific sub-domain. At the end, we use the ordinal feature to specify the sender or receiver of each set of cells. In this way, each zone will contain a \textit{PointSet} of the cells it needs to send and receive from each of its neighbors. \\

\begin{figure}[pos=H]
\begin{adjustbox}{varwidth=0.6\textwidth,fbox,center}
    \centering
    \subfigure[CGNS mesh file \textbf{before} the domain decomposition]{
		\includegraphics[width=0.3\textwidth]{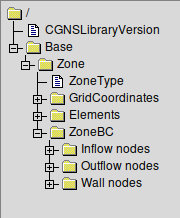}
    } 
    \subfigure[CGNS mesh file \textbf{after} the domain decomposition]{
		\includegraphics[width=0.3\textwidth]{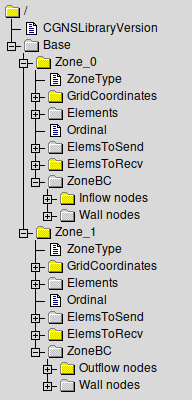}
    } 
    \subfigure[Close-up view of the send/receive information]{
		\includegraphics[width=0.3\textwidth]{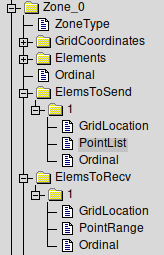}
    } 
    \end{adjustbox}
	 \caption{View of the structure of the mesh file before and after the domain decomposition is performed}
	 \label{fig:cgnsview}
\end{figure}

A typical structure for a mesh file is shown in Figure \ref{fig:cgnsview}. Figure \ref{fig:cgnsview} (a) shows  the structure of the mesh file before the domain decomposition. The original mesh is composed of: the coordinates of the nodes under \textit{GridCoordinates}, the connectivity of the elements under \textit{Elements}, and the boundary conditions under \textit{ZoneBC}, which contains three types of boundaries, inflow, outflow and wall. Figure \ref{fig:cgnsview}  (b) shows the mesh file after the mesh has been decomposed in two sub-domains. Two zones are now present, \textit{Zone\_0} and \textit{Zone\_1}. Each zone contains its own node coordinates and elements as well as its own boundary conditions. For this specific mesh the entirety of the inflow nodes are contained in \textit{Zone\_0} and the outflow nodes in \textit{Zone\_1}. Figure \ref{fig:cgnsview} (c) shows how the send/receive information is stored for \textit{Zone\_0}. Under \textit{ElemsToSend}, for each adjacent sub-domain (here only for \textit{Zone\_1}), a list of the send cells is stored. A range of elements to receive is stored under \textit{ElemsToRecv}. The equivalent send/receive information is stored for \textit{Zone\_1} as well.\\

The goal at the end of our algorithm is to write the mesh file shown in Figure \ref{fig:cgnsview} (b) in an efficient manner. The simplest way would be to use the CGNS library to make each process write the sub-domains it owns to the mesh file sequentially. This has the advantages of being very easy to implement and of having no restriction on the kind of information to be written for each zone. The obvious drawback is that at any given time, only one process would be writing in the file, which would be a big bottleneck. To overcome this issue, we want to use the PCGNS library to  allow parallel IO operations to the file. The problem is that the PCGNS library was not designed to create a mesh file from scratch; instead, it was designed to be able to write solutions to an already-created mesh file in parallel. As such, it lacks some features, such as the capability of writing boundary conditions or point lists to the mesh file in parallel. To use the PCGNS library to create a usable mesh file, we need to use the classic CGNS library to write the missing data to the file. Doing so without the CGNS library hanging requires that each process has the information we want to write to the file, and consequently requires data to be communicated between the processes. The performances of different ways of writing the mesh file using a combination of the CGNS and PCGNS libraries are studied in section \ref{sec:perfio}.

\section{Performance analysis}
\label{sec:perf}
In the first part of this section, we analyze the time taken by the algorithm and the memory it uses. The scaling of these two metrics across multiple MPI processes is displayed and commented upon in multiple scenarios. In the second part of this section, we present the performances of multiple ways of writing the mesh file using a combination of the CGNS and PCGNS libraries. All the results presented in this section were generated on Graham, one of the \textit{Compute Canada} clusters \url{https://docs.computecanada.ca/wiki/Graham}. We used the most common CPU nodes on the cluster, consisting of two Intel E5-2683 v4 Broadwell @ 2.1GHz for a total of 32 CPU cores and 125 GB of RAM per node. The complete program is available at \url{https://github.com/ETS-GRANIT/parmetis-ghostlayers}.

\subsection{Time and memory usage}
The main focus of this section is to assess the scaling of the ghost layers' generation process. We analyze the performances of the algorithm on multiple test cases of increasing difficulty; first on 2-dimensional triangular meshes, then on 3-dimensional tetrahedral meshes. The time taken by the ghost layers' generation process, ParMETIS, and the file operations are logged using the C++ standard library \textit{chrono}. The memory (resident set size) used by the program is logged by checking the proc filesystem from within the C++ program in a manner similar to  \url{https://hpcf.umbc.edu/general-productivity/checking-memory-usage/} . The results were compared with the $sacct$ command from the \textit{slurm workload manager} (\url{https://slurm.schedmd.com/sacct.html}) to ensure consistency. Both time and memory are reported for multiple layers of ghost cells on log-log plots, as the ideal scaling would be a straight line of slope $-1$. The ideal time and memory are calculated by fitting a function of the form $\textit{Ideal}(n) = C\frac{1}{n}$ to the result using the lowest number of MPI processes for the one-layer case to determine the constant.\\

\subsubsection{2-dimensional unstructured triangular mesh cases}

The first mesh used in this section is the smallest used in these tests with its 480k triangular cells. It is decomposed into 16 sub-domains, each containing around 30k elements. The number of MPI processes used is increased until 16 processes are used, at which time each MPI process owns only one sub-domain and the number of processes cannot be increased any further. \\

\begin{figure}[pos=H]
\begin{adjustbox}{varwidth=\textwidth,fbox,center}
    \centering
    \subfigure[Execution time of the ghost-layers generation process]{
		\includegraphics[width=0.4\textwidth]{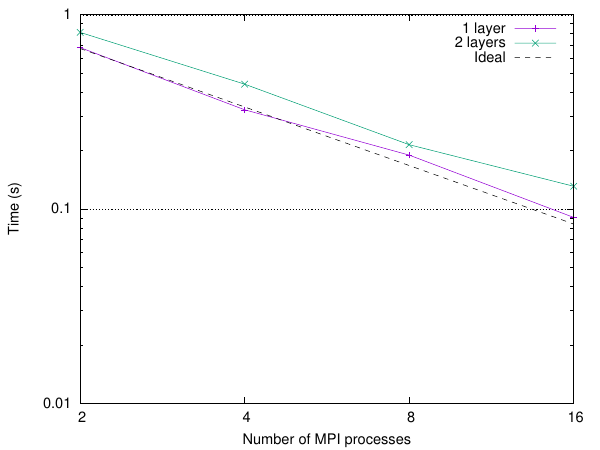}
    } 
    \subfigure[Average memory (RSS) used per MPI process]{
		\includegraphics[width=0.4\textwidth]{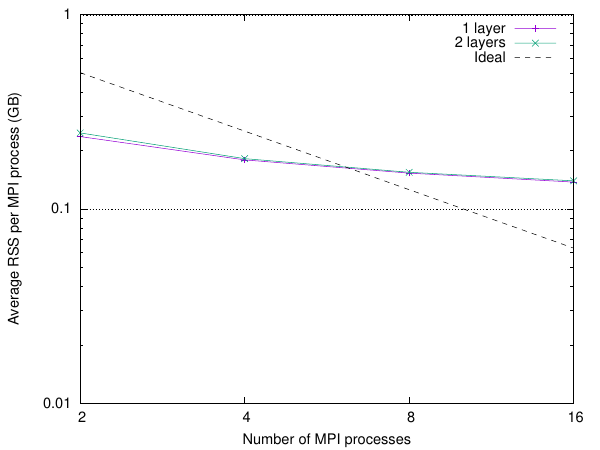}
    } 
    \end{adjustbox}
	 \caption{Time taken and memory usage for a 480k 2D triangular cell mesh as a function of the number of MPI processes when decomposed in 16 sub-domains}
	 \label{graph:48}
\end{figure}

Figure \ref{graph:48} shows the time taken and the memory used by the algorithm for a varying number of MPI processes. It can be seen that the time taken is close to ideal when increasing the number of processes, whereas the memory used per MPI process is not. The memory scaling that we see is due to the fact that, as more and more MPI processes are used, more data needs to be duplicated across the processes. This is to be expected and shows the benefit of assigning multiple sub-domains to a single MPI process, as data can be shared between all the sub-domains, thereby  reducing the memory usage.\\

\begin{figure}[pos=H]
\begin{adjustbox}{varwidth=\textwidth,fbox,center}
    \centering
    \subfigure[Execution time of the ghost layers' generation process]{
		\includegraphics[width=0.4\textwidth]{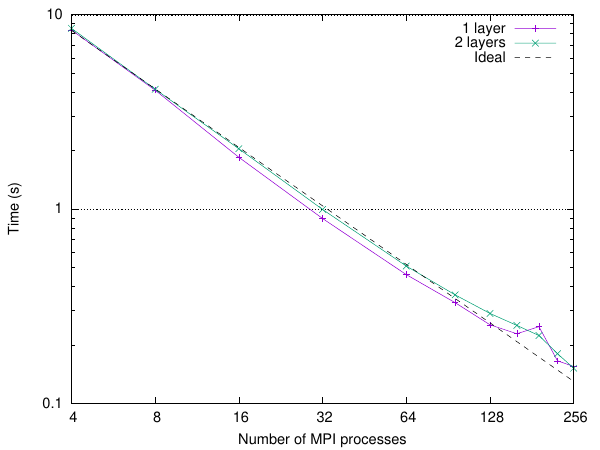}
    } 
    \subfigure[Average memory (RSS) used per MPI process]{
		\includegraphics[width=0.4\textwidth]{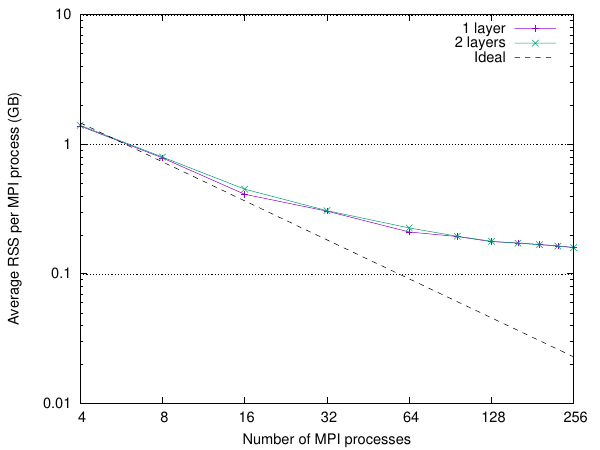}
    } 
    \end{adjustbox}
	 \caption{Time taken and memory usage for a 13 million 2D triangular cell mesh as a function of the number of MPI processes when decomposed into 256 sub-domains}
	 \label{graph:11M}
\end{figure}

The second mesh contains 13 million triangular cells and is decomposed into 256 sub-domains, each containing around 42k elements. The number of MPI processes is increased until 256 processes are used. Figure \ref{graph:11M} shows the time and memory scaling. When compared to Figure \ref{graph:48}, the results are similar for the time scaling and better for the memory usage, as the memory usage stays close to ideal until 16 processes are being used (16 sub-domains per process).

\subsubsection{3-dimensional unstructured tetrahedral mesh cases}
This section presents the results for 3-dimensional unstructured tetrahedral meshes. The main difference between 2-dimensional and 3-dimensional meshes is the scaling of the number of ghost cells per sub-domain. In a 3-dimensional case, the number of ghost cells is proportional to the surface of each sub-domain, whereas in the 2-dimensional case it is proportional to the boundary length of the sub-domain. This means that the ghost cell layers will use much more memory in the 3-d case than in the 2-d case, especially when using multiple layers. \\

\begin{figure}[pos=htp]
\begin{adjustbox}{varwidth=\textwidth,fbox,center}
    \centering
    \subfigure[Execution time of the ghost layers' generation process]{
		\includegraphics[width=0.4\textwidth]{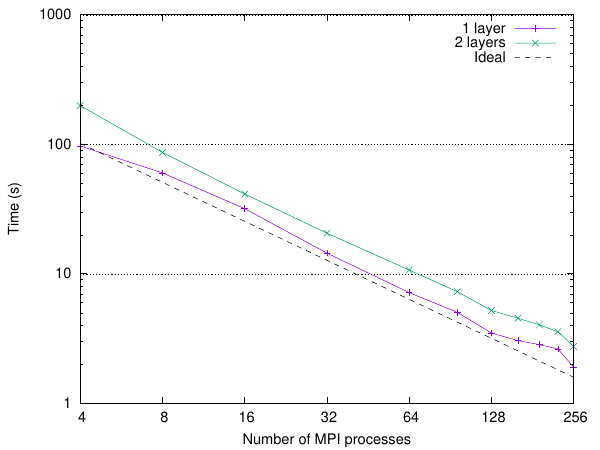}
    } 
    \subfigure[Average memory (RSS) per MPI process]{
		\includegraphics[width=0.4\textwidth]{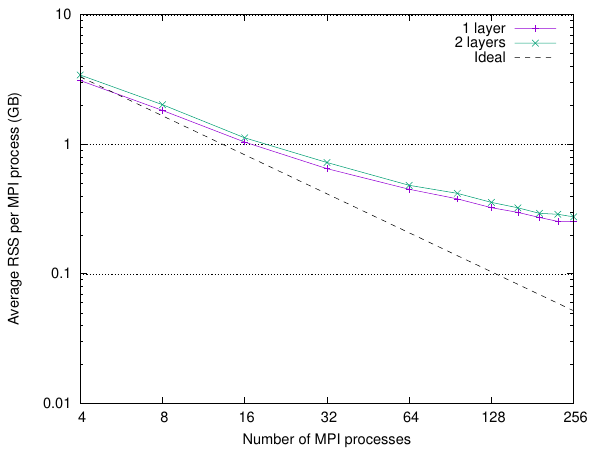}
    } 
    \end{adjustbox}
	 \caption{Time taken and memory usage for a 27 million 3D tetrahedral cell mesh as a function of the number of MPI processes when decomposed in 256 sub-domains}
	 \label{graph:27}
\end{figure}

The first mesh is a simple box containing 27 million tetrahedral cells. It was generated using GMSH \citep{gmsh}, because it supports the CGNS file format in its last version. Figure \ref{graph:27} shows the time and memory scaling when decomposing this mesh into 256 sub-domains. The results are quite similar to the 2-dimensional cases, the main difference being the greater impact of the number of ghost layers used.\\

\begin{figure}[pos=htp]
\begin{adjustbox}{varwidth=\textwidth,fbox,center}
    \centering
    \subfigure[Execution time of the ghost layers' generation process]{
		\includegraphics[width=0.4\textwidth]{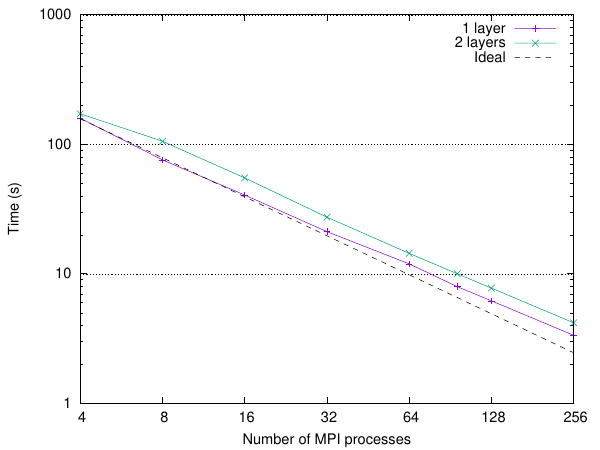}
    } 
    \subfigure[Average RSS per MPI process]{
		\includegraphics[width=0.4\textwidth]{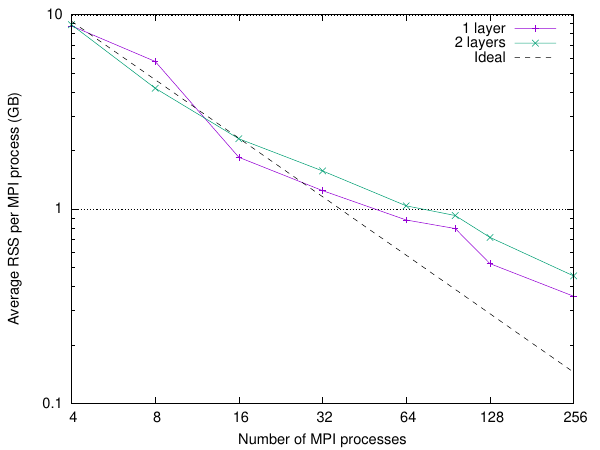}
    } 
    \end{adjustbox}
	 \caption{Time taken and memory usage for a 70 million 3D tetrahedral cell mesh as a function of the number of MPI processes when decomposed into 2048 sub-domains}
	 \label{graph:70}
\end{figure}

Finally, the last mesh is taken from \url{https://hiliftpw-ftp.larc.nasa.gov/HLPW4/CRM-HL_Grids/Committee_Grids/1-Pointwise-Unstructured/1-Config_CRM-HL_40-37_Nominal/1.1-Pointwise-Unstr-Tet-V2/1.1.A/}. It is a mesh of a plane containing 70 million tetrahedral cells, and we choose to decompose it into 2048 sub-domains. The results are shown in Figure \ref{graph:70}; the time scaling is similar to previous cases, and the memory scaling is closer to the ideal than for the other cases. The fact that the memory scaling becomes better for the larger meshes is as expected, and is an indication that the program is performing as expected.\\

This section presented the performances of the ghost layers' generation process and showed how both time and memory scale when using up to 256 processes. The time scaling is always close to ideal for the number of processes tested, which highlights the fact that the communication steps are not a bottleneck for the algorithm. The memory usage is however more difficult to manage due to the inner boundary layers often needed by multiple processes and being duplicated in memory. Nevertheless, when using sufficiently large meshes, the memory used per MPI process scaled close to ideally up to a certain point, depending on the specific mesh used.\\

\subsection{Time taken by the IO operations}
\label{sec:perfio}
 This section reports the time taken to write the domain decomposition in a CGNS mesh file. \textcolor{black}{All the results of this section were obtained using the \textit{scratch} space of \textit{Graham} a computer cluster that is part of the \textit{Digital Research Alliance of Canada}. More information on the high performance lustre-based file system (3.2 PB Lustre over 100 Gb/s InfiniBand with flash-based MDS and 72 OSTs across 12 OSS's) can be found on \url{https://docs.alliancecan.ca/wiki/Graham/en}}. To use the PCGNS library, the more recent HDF5 CGNS file format has been used instead of the older ADF format. As discussed in section \ref{sec:writedecomp}, there are multiple ways of writing the domain decomposition:
\begin{enumerate}
    \item \textbf{CGNS sequential single file}: Each process writes its sub-domains in a common CGNS mesh file in a sequential manner. This is the worst case for writing in a single file, as each MPI process has to wait its turn to write its sub-domains into the file. The upside is that the original mesh's boundary conditions and the new ghost cells' send/receive information can easily be written to the mesh file.
    
    \item \textbf{PCGNS without send/receive}: This approach uses the PCGNS library exclusively to write the mesh file. However, some missing features in the PCGNS library prohibit us from writing the boundary conditions and the send/receive information, which makes the mesh files \textbf{unusable in practice}. This is still shown here to highlight the best speed achievable for a single-file writing approach in parallel.
    
    \item \textbf{PCGNS/CGNS with send/receive}: This is a process that uses the PCGNS library to write most of the mesh file. Communications are then performed between the MPI processes so that the CGNS library can be used to write the boundary conditions as well as the send/receive information. This type of writing is slower than the pure PCGNS version, however, it is the only way to write all the information so that the meshes are usable in practice.  
\end{enumerate}    

\begin{table}[pos=H]
    \centering
    \caption{Time in seconds taken by the IO steps using a 13 million 2D triangular cell mesh}
\begin{adjustbox}{max width=\textwidth}
\begin{tabular}{|l|c|c|c|c|c|c|c|c|c|c|c|c|} 
\hline
Number of MPI processes & 2 & 4 & 8 & 16 & 32 & 64 & 128 & 160 & 192 & 224 & 256 \\
\hline
CGNS sequential (278M) & 10.5898 & 12.7135 & 21.5934 & 28.6133 & 60.3067 & 145.157 & 249.996 & 384.383 & 467.244 & 530.731 & 543.769 \\
\hline
PCGNS without ghost layers (278M) & 2.84463 & 1.81206 & 1.46898 & 1.10387 & 1.08244 & 1.93859 & 2.31247 & 2.9228  & 3.12405 & 3.07873 & 4.52603 \\
\hline
PCGNS/CGNS with ghost layers (286M) & 7.8463 & 8.29877 & 7.87522 & 6.38313 & 7.26209 & 14.6417 & 15.1863 & 15.9122 & 16.867 & 21.8684 & 18.8944 \\
\hline
\end{tabular}
\end{adjustbox}
    \label{tab:11M}
\end{table}

Table \ref{tab:11M} shows the write times for the 13 million 2D cell mesh. The results indicate that using the simple CGNS sequential version is quite slow compared to the PCGNS versions, especially when a large number of processes are used. The PCGNS version without ghost layers shows the best times achievable in parallel, however, as the PCGNS library does not support writing the boundary conditions and the send/receive data to the file, a hybrid version must be used. The hybrid version is slower than the PCGNS version but is still at least one order of magnitude faster than the CGNS sequential version. Similar results are shown in Figure \ref{tab:av70M} for a 70 million 3D tetrahedral cell mesh. 

\begin{table}[pos=H]
    \centering
    \caption{Time in seconds taken by the IO steps using a 70 million 3D tetrahedral cell mesh}
\begin{adjustbox}{max width=\textwidth}
\begin{tabular}{|l|c|c|c|c|c|c|c|c|c|c|c|c|} 
\hline
Number of MPI processes & 32 & 64 & 96 & 128 & 160 & 192 & 224 & 256 \\
\hline
CGNS sequential (1.7GB) & 1267.76 & 2847.47 & 4257.12 & 5695.69 & 7159.36 & 8522.84 & 9751.57 & 11140.9 \\
\hline
PCGNS without ghost layers (1.6GB) & 8.24 & 16.23 & 20.53 & 22.14 & 26.28 & 24.47 & 28.51 & 36.27 \\
\hline
PCGNS/CGNS with ghost layers (1.7GB) & 96.62 & 182.82 & 197.12 & 211.57 & 216.308 & 220.54 & 222.475 & 254.70\\
\hline
\end{tabular}
\end{adjustbox}
    \label{tab:av70M}
\end{table}

The results presented in this section show that writing the domain decomposition in a single file is possible in an efficient manner by using both the PCGNS and CGNS libraries. It is 10 to 100 times faster to use the hybrid PCGNS/CGNS version when compared to the sequential CGNS version. Better times could be achieved if the PCGNS library was extended to support writing boundary conditions and send/receive data to the file in an efficient way.\\

\textcolor{black}{We now want to show the performance of the I/O operations when the CFD solver writes multiple solutions to the mesh file. We show three ways of writing solutions with the CFD solver:}
\begin{enumerate}
    \item \textbf{VTK separate files}: \textcolor{black}{This is the old way our in-house solver saved the solutions using the VTK file format (ASCII). Each MPI process writes the solution on its sub-domain to a file. This allows the processes to be independent of each other at the writing step and is the typical way solutions are written.}
    
    \item \textbf{CGNS single file}: \textcolor{black}{In this method, each process opens the common mesh file to store the solution on the zone it is assigned to before closing it. This is very slow as only one process is writing the solution at any given time.}
    
    \item \textbf{PCGNS}: \textcolor{black}{This method uses the PCGNS library to write the solution on all the zones at the same time. This is the most efficient use of the PCGNS library as was demonstrated in \cite{pcgnsbenchmark2}. This method specifically leverages the fact that the PCGNS library works in an asynchronous manner that allows solutions to be written to the mesh file all the while the solution is being computed.} 
\end{enumerate} 

\begin{table}[pos=H]
    \centering
    \caption{\textcolor{black}{Time in seconds taken by the CFD solver to complete a simulation on a 13 million 2D mesh while using different solution write modes.}}
\begin{adjustbox}{max width=\textwidth}
\begin{tabular}{|l|c|c|c|c|c|c|c|c|c|c|c|c|} 
\hline
Number of MPI processes & 8 & 12 & 16 & 32 \\
\hline
No solution file (s) & 663.29 & 449.28 & 345.75 & 233.94\\
\hline
VTK separate files (80GB total) (s) & 1116.81 & 750.41 & 602.42 & 393.34 \\
\hline
CGNS single file (80GB total) (s) & 2027.63 & 3951.04 & 5904.30 & 25979.66\\
\hline
PCGNS single file (80GB total) (s) & 746.22 & 539.82 & 449.98 & 409.85\\
\hline

\end{tabular}
\end{adjustbox}
    \label{tab:solGB}
\end{table}

\textcolor{black}{Table \ref{tab:solGB} shows the time taken by the CFD solver to complete a simulation on a 13 million 2D mesh while writing the solution in multiple ways. All the writing modes have been adjusted so that the same amount of data (80GB) is written for each case. From these figures, it can be seen that using the CGNS library to sequentially write to a single file is by a large margin the slowest method. This is the expected result of accessing in a sequential manner a single common mesh file where all the solutions are stored. In contrast, it can be seen that using the PCGNS library to write the solutions to the mesh file is comparable to writing the solutions to separate files as in the VTK case. This is caused by the asynchronous way the PCGNS library works that allows the solution to be cached for the write operations. This means that even if communication steps slow down the PCGNS file write when compared to the VTK separate file write, it still can catch up by writing the solution in an asynchronous way while the solver continues computing the solution. These results confirm that choosing a single zone for each sub-domain leads to a very good performance of the PCGNS library \citep{pcgnsbenchmark2}.}

\section{Application to the parallel high-order resolution of the Shallow-water equations with a multi-GPU solver}
\label{sec:application}
In this second part of the paper, we use the preceding work to perform the parallel high order resolution of the Shallow-water equations using the multi-CPU multi-GPU solver developed in \cite{delmasvsoulaima}. We start by briefly describing the Shallow-water equations and their resolution using a finite volume method. The classic first-order in space and time HLLC method is recalled before presenting the second-order WAF and MUSCL methods. The accuracy of the resolution is studied on a test case of an idealized dam break on a wet bottom. The need for multiple layers of ghost cells for the second-order resolution is demonstrated by comparing the second-order solutions computed on a mesh decomposed into 8 sub-domains using one and two layers of ghost cells. Finally, the first- and second-order solutions are compared on two real domains with complex bathymetries using meshes of up to 13 million cells and 32 GPUs to perform the resolution.

\subsection{Numerical resolution of the Shallow-water equations}

\begin{wrapfigure}{r}{0.5\textwidth}
\centering
	\fbox{
		\includegraphics[width=0.45\textwidth]{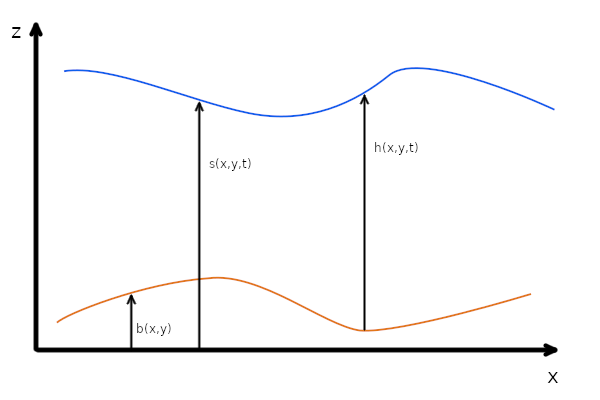}
	}
	 \caption{Illustration of the notations}
	 \label{fig:notations}
	 \vspace{-0.8cm}
\end{wrapfigure}

The Shallow-water equations and their numerical resolution using a time-explicit finite volumes method are presented in this section. The notations used are represented in Figure \ref{fig:notations}.
\bigbreak

The Shallow-water equations \citep{toro} can be written as:
 \begin{equation}
     U_t + G(U)_x + H(U)_y = S(U),
     \label{eqswefinal}
 \end{equation}
 with
 \begin{equation}
     U = \left[ \begin{array}{c}
          h  \\
          h\bar{u} \\
          h\bar{v}
     \end{array}\right],
     \hspace{0.5cm}
     G(U) = \left[ \begin{array}{c}
          h\bar{u}  \\
          h\bar{u}^2 + \frac{1}{2}gh^2 \\
          h\bar{u}\bar{v}
     \end{array}\right],
     \notag
 \end{equation}
  \begin{equation}
     H(U) = \left[ \begin{array}{c}
          h\bar{v}  \\
          h\bar{u}\bar{v}\\
          h\bar{v}^2 + \frac{1}{2}gh^2
     \end{array}\right],  
     \hspace{0.5cm}
     S(U) = \left[ \begin{array}{c}
          s_1  \\
          s_2 \\
          s_3
     \end{array}\right].
     \notag
 \end{equation}
 
where $\bar{u}$ and $\bar{v}$ are the depth-averaged velocities in the x and y direction, $h$ is the height of the water column as defined in Figure \ref{fig:notations}, and g is the gravitational acceleration.
\bigbreak

\textcolor{black}{By integrating \eqref{eqswefinal} on triangular volumes, using the appropriate cell averages and using the rotational invariance between $G$ and $H$ \citep{toro}}

\begin{equation}
    F(U).n_{ij} = T_{n_{ij}}^{-1}G(T_{n_{ij}}U), \hspace{0.02\textwidth}T_{n_{ij}} = \begin{bmatrix}
1 & 0 & 0\\
0 & n_{ij}^1 & n_{ij}^2\\
0 & -n_{ij}^2 & n_{ij}^1\\
\end{bmatrix}
\end{equation}

\textcolor{black}{we get a discrete form of the Shallow-water equations \cite{toro, louki, delmasvsoulaima}}

\begin{equation}
    \abs{\Omega_i} \frac{dU_i}{dt} = -\sum_{j=1}^3 L_{ij}T_{n_{ij}}^{-1}\Tilde{G}(T_{n_{ij}}U_i,T_{n_{ij}}U_j) + \abs{\Omega_i} S_{O_i}(U) + \abs{\Omega_i} S_{f_i}(U),
    \label{vlap2}
\end{equation}
with \textcolor{black}{$S_{O_i}$ and $S_{f_i}$ respectively the bathymetric and friction source terms, $\abs{\Omega_i}$ the area of $\Omega_i$ (triangular), $L_{ij}$ the length of the side $j$ of $\Omega_i$, $n_{ij}$ the unit normal of the side $j$ outwards of $\Omega_i$, and } $\Tilde{G}(T_{n_{ij}}U_i,T_{n_{ij}}U_j)=\Tilde{G}(U_L,U_R)$ a discrete flux computed by solving a Riemann problem with $U_L=T_{n_{ij}}U_i$ and $U_R=T_{n_{ij}}U_j$ as the initial states:

\begin{equation}
\begin{dcases}
    \frac{\partial U}{\partial t} + \frac{\partial G(U)}{\partial x_n} = 0,\\
    U(x,0) = \left\{\begin{array}{ll}U_L \text{ if } x_n < 0,\\ U_R \text{ if } x_n > 0.\\\end{array} \right.
    \label{riem}
\end{dcases}
\end{equation}

For the boundary conditions, we choose to proceed as in \cite{louki}. A transmissive condition is solved by assuming a state $U_R = U_L$ in the resolution of the Riemann problem. For a condition with an incoming flow $Q$ we calculate the flow directly with $\Tilde{G} (U_L) = (Q,~Q^2/h_l + gh_l^2/2,~0)^T$. For a non-transmissive wall condition, we set $Q=0$ in the preceding calculation, leading to $\Tilde{G} (U_L) = (0,~gh_l^2/2,~0)^T$.\\

Temporal discretization is done using an explicit Euler method, which makes it possible to avoid having to solve a linear system at the cost of a time step constrained by a stability condition. The stability analysis from \cite{louki} gives the following CFL condition:

\begin{equation}
    CFL = \Delta t \frac{max(\sqrt{gh}+\sqrt{u^2+v^2})}{min(d_{L,LR})},
\end{equation}
with $d_{L,LR}$ the distance between the cell center and the L/R interface. However, for simplicity we choose to take $d_{L,LR} = R_{L} $, the radius of the circle inscribed in cell L. This is a conservative condition and has proven to result in great stability for $\text{CFL}=0.9$. \\

More complex time discretization can be found in \cite{riadh}, and details concerning the bathymetric and friction source terms can be found in \cite{toro,louki,bristeau}.\\

\subsection{Resolution of the 1-dimensional Riemann problem}
\label{sec:riemprob}

To compute the inter-cell flux $\Tilde{G}(U_L,U_R)$, the 1-dimensional Riemann problem \eqref{riem} has to be solved either exactly or approximately. The typical way of computing this flux in the context of the resolution of the Shallow-water equations is to use an approximate Riemann solver. In the next sections, we briefly recall the classic first-order HLLC scheme and the second-order WAF and MUSCL schemes. As we are only interested here in demonstrating how multiple layers of ghost cells are used to ensure the quality of the second-order solution across multiple sub-domains, we only present these methods in their most basic form. More detailed studies of the schemes presented hereafter can be found in \cite{hllc,toro,toro2,louki,multislopemuscl,riadh,secondorderbathyrecons,bristeau}.

\begin{figure}[pos=htp]
\begin{center}
\fbox{
\parbox{.5\columnwidth} {
  \includegraphics[width=.98\linewidth]{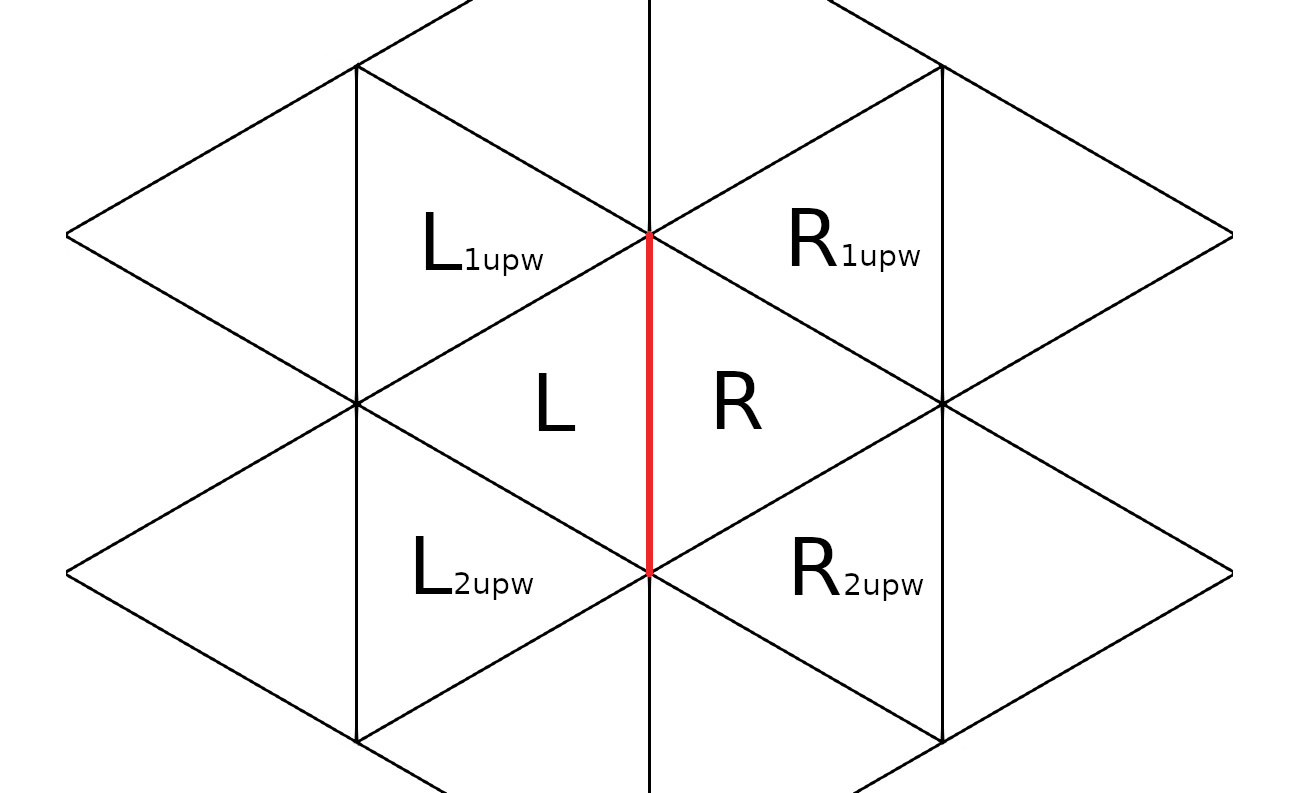}
    } 
}
\end{center}
\caption{Notations used for identifying the left and right cells of the 1-dimensional Riemann problem.}
\label{fig:fsev}
\end{figure}

\subsubsection{HLL and HLLC scheme}
\label{sec:preshllc}
We recall here the flux of the HLL scheme \citep{hll,toro}
\begin{equation}
\Tilde{G}^{HLL} =\begin{dcases}
G(U_L),~S_L \geq 0,\\G(U_R),~S_R \leq 0,\\\Tilde{G}(U_*),~ S_L \leq 0 \text{ and } S_R \geq 0,
\label{hll}
\end{dcases}
\end{equation}

with $\Tilde{G}(U_*)$ the flux in the so-called star region given by: 

\begin{equation}
    \Tilde{G}(U_*) = \left[S_R G(U_L) - S_L G(U_R) + S_R S_L (U_R-U_L)\right]/(S_R - S_L).
\end{equation}

The right and left wave speeds, $S_R$ and $S_L$, are estimated as follows:
\begin{equation}
S_L = u_L - a_L p_L,~~~ S_R = u_R + a_R p_R,
\end{equation}

where $k=L,R$, $a_k=\sqrt{gh_k}$, and
\begin{equation}
    p_k = \begin{dcases}[h^*(h^*+h_k)/2)]^{1/2},& h^* > h_k,\\1,& h^* \leq h_k.\end{dcases}
\end{equation}

The water height $h^*$ is evaluated in multiple steps. A first approximation indicates if this is a shock wave or a rarefaction wave
\begin{equation}
    h^*_0 = \frac{h_L+h_R}{2} - \frac{(u_R-u_L)(h_L+h_R)}{4(a_R+a_L)}.
\end{equation}

If $h^*_0 \leq min(h_L,h_R) $, then it is a rarefaction wave and we have
\begin{equation}
    h^* = [(a_R+a_L)/2+(u_L-u_R)/4]^2/g,
\end{equation}

whereas, if $h^*_0 > min(h_L,h_R) $, then it is a shock wave and we have 
\begin{equation}\begin{aligned}
    h^* &= (h_L g_L + h_R g_R +u_L-u_R)/(g_L+g_R),\\
    g_k &= \left[ \frac{g(h^*_0+h_k)}{2h^*_0h_k}\right]^{1/2}, k=L,R .
    \end{aligned}
\end{equation}

We can then modify this scheme to obtain the HLLC scheme \citep{hllc,toro} by accounting for the speed $S^*$ in the star region. This step only modifies the last component of the flux, as follows 
\begin{equation}
    \Tilde{G}_3^{HLLC} = \begin{dcases}\Tilde{G}_1^{HLL}(U_L,U_R)v_L,~S^* \leq 0, \\\Tilde{G}_1^{HLL}(U_L,U_R)v_R,~S^* \geq 0, \end{dcases}
\end{equation}

with 
\begin{equation}
    S^* = \frac{s_L h_R(u_R-S_R) - s_R h_L(u_L-S_L)}{h_R(u_R-S_R) - h_L(u_L-S_L)}.
\end{equation}

\subsubsection{WAF scheme}
The Weighted Average Flux (WAF) method is derived from a space-time integral average of the flux. Using a weighted average of the HLLC fluxes for the first two components and an average of the WAF flux itself for the third component, we obtain \citep{toro}:

\begin{equation}
   \Tilde{G}_i^{WAF}(U_L,U_R) = w_L G_i(U_L) + w_{LR} \Tilde{G}^{HLL} + w_R G_i(U_R), ~~\text{i=1,2}
\end{equation}

\begin{equation}
   \Tilde{G}_3^{WAF} = (w_{L*}v_L + w_{R*}v_R)\Tilde{G}^{WAF}_1
\end{equation}

where the weights are given by

\begin{equation}
\begin{aligned}
    &w_L~ = (1+c_L)/2, ~~~~ w_R~ = (1-c_R)/2, ~~~~ w_{LR} = (c_R-c_L)/2,\\
    &w_{L*} = (1+c^*)/2, ~~~~ w_{R*} = (1-c^*)/2,
    \end{aligned}
\end{equation}

in which $c_L=S_L\frac{\Delta t}{\Delta x}$, $c_R=S_R\frac{\Delta t}{\Delta x}$ and $c_*=S_*\frac{\Delta t}{\Delta x}$ are the Courant numbers for the right left and middle wave speeds, respectively.\\

As the WAF scheme is second-order in space and time, it produces spurious oscillations near steep gradients and thus requires Total Variation Diminishing (TVD) stabilization. This is a well-known phenomenon, often addressed by using limiter functions derived from classical flux limiters $\phi(r)$ as follows: $A(r) = 1 - (1-|c|)\phi(r)$. The weights take a new expression using these limiter functions:

\begin{equation}
\begin{aligned}
    &w_L = (1+c_L A(r_L,c_L))/2, ~~~~ w_{LR} = (c_R A(r_R,c_R)-c_L A(r_R,c_L))/2, ~~~~ w_R = (1-c_R A(r_R,c_R))/2,\\
     &w_{L*} = (1+c^* A(r_*,c_*))/2, ~~~~ w_{R*} = (1-c^* A(r_*,c^*))/2,
\end{aligned}
\end{equation}

where for $k=L,R, and~ *$, $r_k$ is the ratio of the upwind change $\Delta_{upw} q^{(k)}$ to the local change $\Delta_{loc} q^{(k)}$, defined by

\begin{equation}
    \Delta_{loc} q^{(k)} = q_R^{(k)} - q_L^{(k)}
\end{equation}

\begin{equation}
    \Delta_{upw} q^{(k)} = \begin{dcases}
    q_L^{(k)} - q_{L_{upw}}^{(k)} , &S_k > 0, \\
    q_{R_{upw}}^{(k)} - q_R^{(k)} , &S_k < 0 .
    \end{dcases}
\end{equation}

Following \cite{louki}, $q^{(k)} = h$  is chosen for $k=L,R$ and $q^{(*)}=v$. The upwind values are obtained by averaging the solution on the cells $L_{1_{upw}}$ and $L_{2_{upw}}$ for $q_{L_{upw}}$, and, $R_{1_{upw}}$ and $R_{2_{upw}}$ for $q_{R_{upw}}$.\\

The results presented in the following sections mainly use the SUPERBEE limiter \eqref{eq:superbee}, as it has been shown to give the sharpest wave pattern.
\begin{equation}
    A(r,c) = \begin{dcases}
        1, &r \leq 0,\\
        1-2(1-|c|)r, & 0 < r \leq \frac{1}{2},\\
        |c|, & \frac{1}{2} < r \leq 1,\\
        1-(1-|c|)r, & 1 < r \leq 2,\\
        2|c|-1 & r > 2.
    \end{dcases}
    \label{eq:superbee}
\end{equation}

\subsubsection{MUSCL scheme}
The Monotone Upstream Scheme for Conservation Law (MUSCL) method was introduced by Van Leer \citep{VANLEER}. The idea is to perform a piecewise linear reconstruction of the solution to achieve high order. The method presented in this paper can be found in \cite{toro2}. A more complete description of the MUSCL method in a 2-dimensional context can be found in \cite{monomultimuscl,multislopemuscl}. Other methods based on the MUSCL idea, such as the MUSCL-Hancock Method (MHM) can be found in \cite{toro2}.\\

The left and right states of the Riemann problem \eqref{riem} are modified using a piecewise linear reconstruction of the solution. That is,

\begin{equation}
    \Tilde{U}_L = U_L + |x_{RL}-x_L|\Delta U_L, ~~~~ \Tilde{U}_R = U_R - |x_R-x_{RL}|\Delta U_R
\end{equation}

where $\Delta U_L = \frac{U_R-U_{L_{upw}}}{|x_R-x_{L_{upw}}|}$ and  $\Delta U_R = \frac{U_{R_{upw}}-U_L}{|x_{R_{upw}}-x_L|}$ are the approximated slopes of the solution on the left and right cells, respectively. The upwind values $U_{L_{upw}}$ and $U_{R_{upw}}$ are chosen in a similar manner as for the WAF scheme by averaging the solution on the cells $L_{1_{upw}}$ and $L_{2_{upw}}$, and $R_{1_{upw}}$ and $R_{2_{upw}}$, respectively. The Riemann problem \eqref{riem} is transformed using these new initial conditions:
 
\begin{equation}
\begin{dcases}
    \frac{\partial U}{\partial t} + \frac{\partial G(U)}{\partial x_n} = 0,\\
    U(x,0) = \left\{\begin{array}{ll}\Tilde{U}_L \text{ if } x_n < 0,\\ \Tilde{U}_R \text{ if } x_n > 0.\\\end{array} \right.
    \label{riemnew}
\end{dcases}
\end{equation}

This new Riemann problem is then solved using the HLLC scheme presented in section \ref{sec:preshllc}. Similar to the WAF scheme, the MUSCL scheme, being second-order in space and time produces spurious oscillations near steep gradients and thus needs a TVD stabilization. The TVD condition is achieved by using slope limiters that play the same role as the flux limiters in the WAF scheme. The limited slopes can, for example, be found with:\\
\begin{equation}
    \bar{\Delta U_L} =  
    \begin{dcases}
        max[0,min(\beta \tfrac{U_L-U_{L_{upw}}}{|x_L-x_{L_{upw}}|},\tfrac{U_R-U_L}{|x_R-x_L|}),min(\tfrac{U_L-U_{L_{upw}}}{|x_L-x_{L_{upw}}|},\beta \tfrac{U_R-U_L}{|x_R-x_L|})],~~~U_R-U_L > 0,\\ 
        min[0,max(\beta \tfrac{U_L-U_{L_{upw}}}{|x_L-x_{L_{upw}}|},\tfrac{U_R-U_L}{|x_R-x_L|}),max(\tfrac{U_L-U_{L_{upw}}}{|x_L-x_{L_{upw}}|},\beta \tfrac{U_R-U_L}{|x_R-x_L|})],~~~U_R-U_L < 0,\\ 
    \end{dcases}
\end{equation}
and
\begin{equation}
    \bar{\Delta U_R} =  
    \begin{dcases}
        max[0,min(\beta \tfrac{U_R-U_L}{|x_R-x_L|},\tfrac{U_{R_{upw}}-U_R}{|x_{R_{upw}}-x_R|}),min(\tfrac{U_R-U_L}{|x_R-x_L|},\beta \tfrac{U_{R_{upw}}-U_R}{|x_{R_{upw}}-x_R|})],~~~U_{R_{upw}}-U_R > 0,\\ 
        min[0,max(\beta \tfrac{U_R-U_L}{|x_R-x_L|},\tfrac{U_{R_{upw}}-U_R}{|x_{R_{upw}}-x_R|}),max(\tfrac{U_R-U_L}{|x_R-x_L|},\beta \tfrac{U_{R_{upw}}-U_R}{|x_{R_{upw}}-x_R|})],~~~U_{R_{upw}}-U_R < 0,\\ 
    \end{dcases}
\end{equation}

where for particular values of the parameter $\beta$, traditional flux limiters are reproduced. For example, $\beta = 1$ reproduces the MINBEE flux limiter and $\beta = 2$ reproduces the SUPERBEE flux limiter \citep{toro2}.\\

The MUSCL method may be advantageous over the WAF method for the resolution of the Shallow-water equations because it is easier to combine with the hydrostatic reconstruction, a step aimed at improving stability for complex bathymetries \citep{bristeau,secondorderbathyrecons}. \\

\subsection{Case of a one-dimensional dam break on a wet bottom}
We present the resolution of an idealized (no friction) one-dimensional dam break on a wet bottom. This problem has an analytical solution that can be found in \cite{toro}. \\

To simulate this problem, we take a rectangular domain $ \Omega = [0,10]*[0,1] $ and choose the following initialization values

\begin{equation}
\begin{dcases}
    h &= \left\{\begin{array}{ll}1 \text{ if } x < 5,\\ 0.5 \text{ if } x > 5.\\\end{array} \right.\\
    \bar{u} &= 0 , \\
    \bar{v} &= 0 .
\end{dcases}
\notag
\end{equation}

\begin{figure}[pos=htp]
	\centering
	\fbox{
		\includegraphics[width=0.95\textwidth]{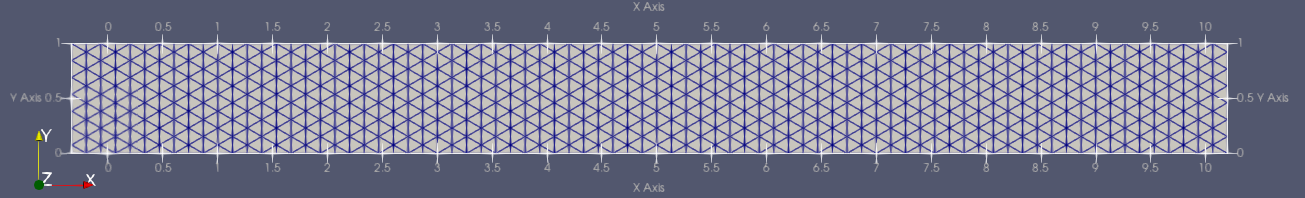}
	}
	 \caption{Coarse mesh containing 1000 triangular cells}
	 \label{fig:meshused}
\end{figure}

A coarse mesh is used in this section, as shown in Figure \ref{fig:meshused}.\\

\begin{figure}[pos=htp]
\begin{center}
\fbox{
\parbox{.96\columnwidth} {
    \subfigure[]{
  \includegraphics[width=.49\linewidth]{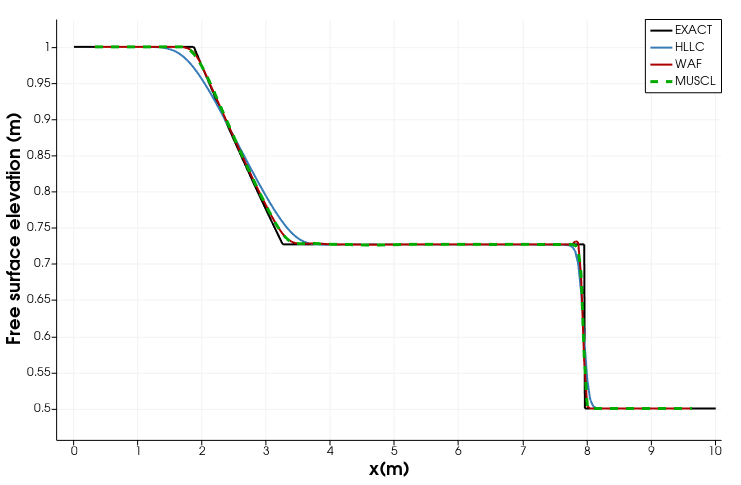}
    } 
    \subfigure[]{
  \includegraphics[width=.49\linewidth]{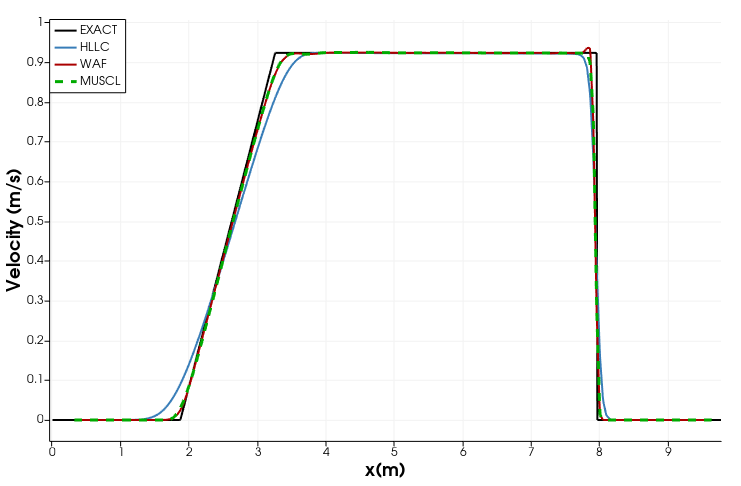}
    \label{fig:fseb}
    } 
    }
}
\end{center}
\caption{Projected solutions on the y-axis for the one-dimensional dam-break case on a 2470-cell mesh at $t=1$s. A SUPERBEE style limiter is used both for the WAF and MUSCL methods.}
\label{fig:compar3w}
\end{figure}

Figure \ref{fig:compar3w} shows the results for the HLLC, WAF, and MUSCL methods. As expected, the WAF and MUSCL methods give better results than the simple HLLC method. While both the WAF and MUSCL solutions are almost perfectly identical, a slight difference can be noted near the steepest part of the solution, where the WAF solution presents a slight oscillation. This kind of oscillation can be smoothed by using other limiting functions or by slightly modifying how the upwind values are averaged. Precise numerical computation of the order of the methods presented here can be found in \cite{riadh,secondorderbathyrecons}.

\subsubsection{Comparison of the second-order solutions using one and two layers of ghost cells}
\begin{figure}[pos=htp]
\begin{center}
\fbox{
\parbox{.98\columnwidth} {
\subfigure[]{
  \includegraphics[width=.31\linewidth]{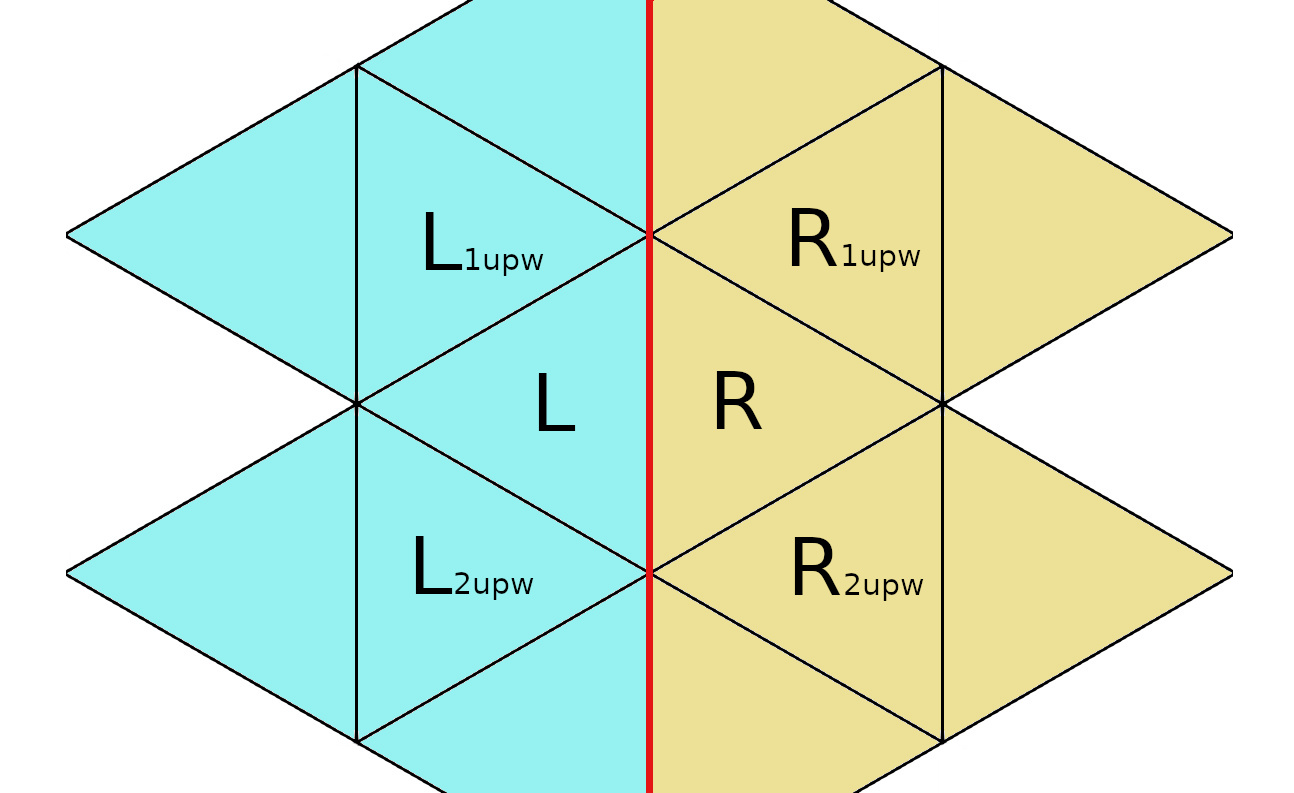}
  }
  \subfigure[]{
      \includegraphics[width=.31\linewidth]{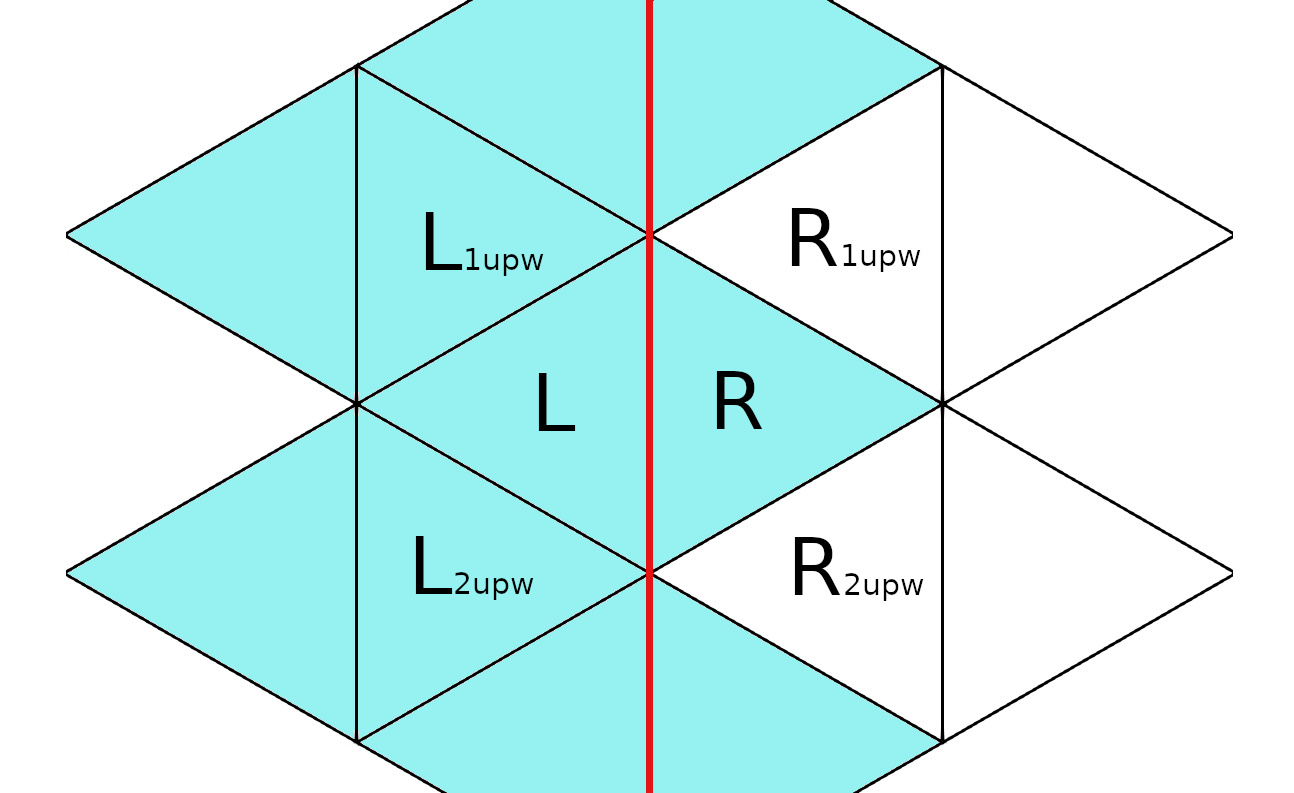}
      \label{fig:fsev2b}
      }
      \subfigure[]{
          \includegraphics[width=.31\linewidth]{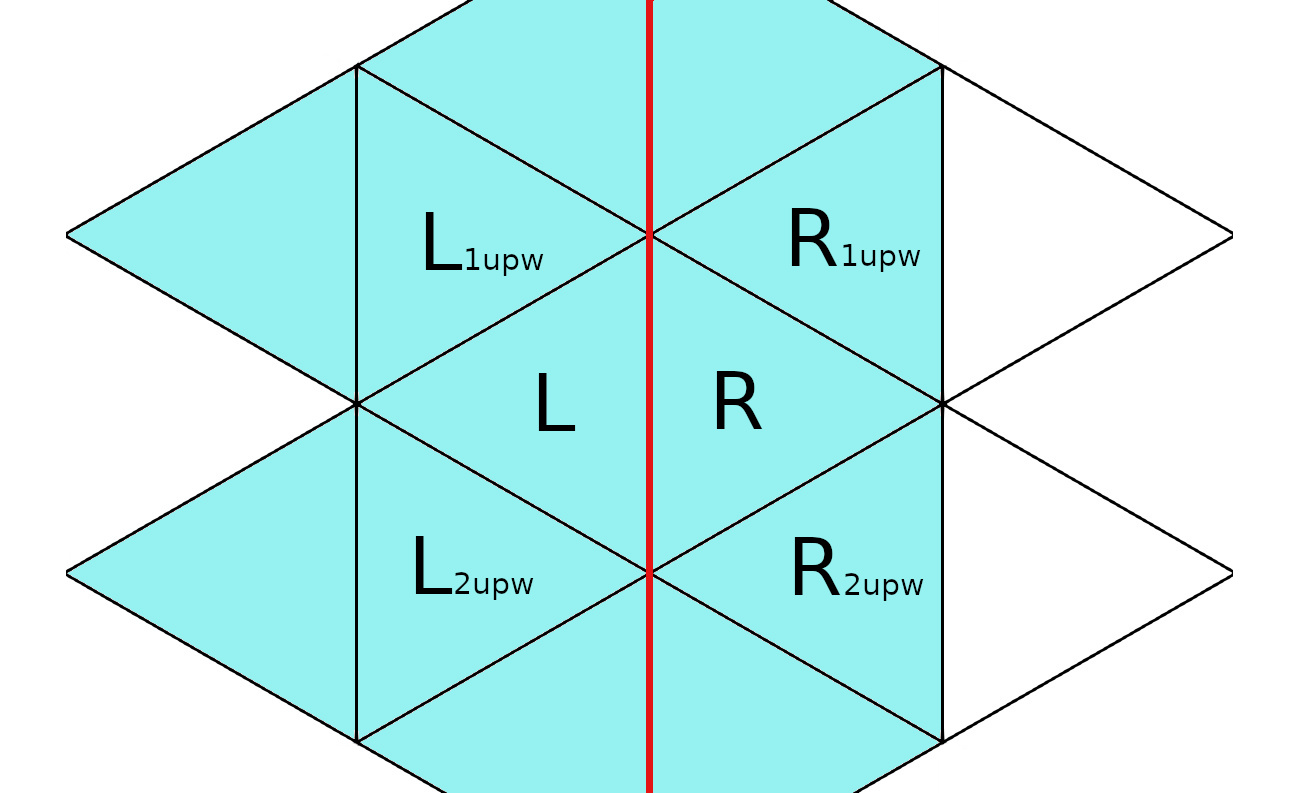}
          }
    }
}
\end{center}
\caption{Left and right cells of the Riemann problem at the interface between two sub-domains (a). Left sub-domain with one layer of ghost cells (b). Left sub-domain with two layers of ghost cells (c).}
\label{fig:fsev2}
\end{figure}

\begin{figure}[pos=htp]
\begin{center}
\fbox{
\parbox{.96\columnwidth} {
    \subfigure[]{
  \includegraphics[width=.95\linewidth]{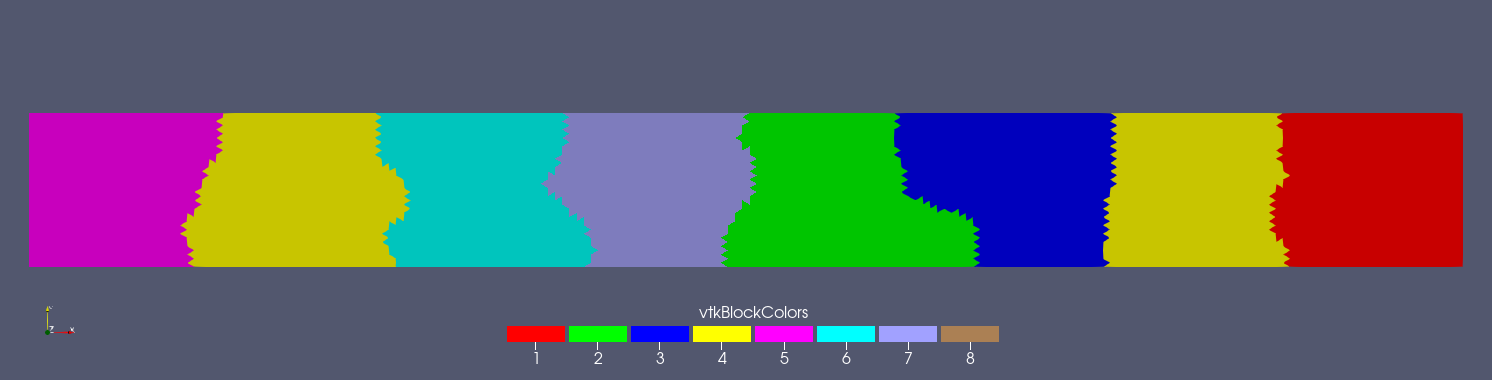}
  \label{fig:diffmesh}
    } 
    \subfigure[]{
  \includegraphics[width=.98\linewidth]{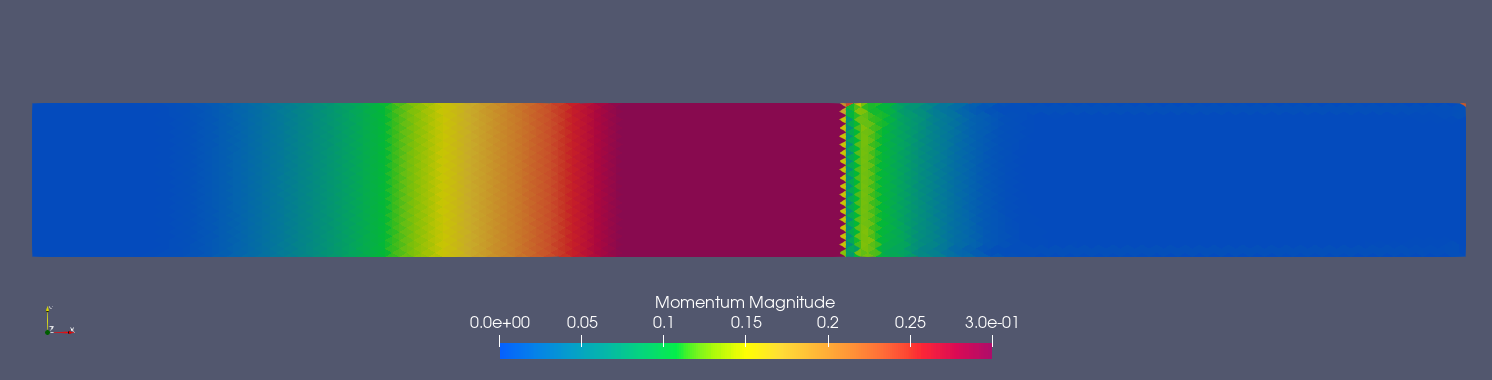}
  \label{fig:diffa}
    } 
    \subfigure[]{
  \includegraphics[width=.98\linewidth]{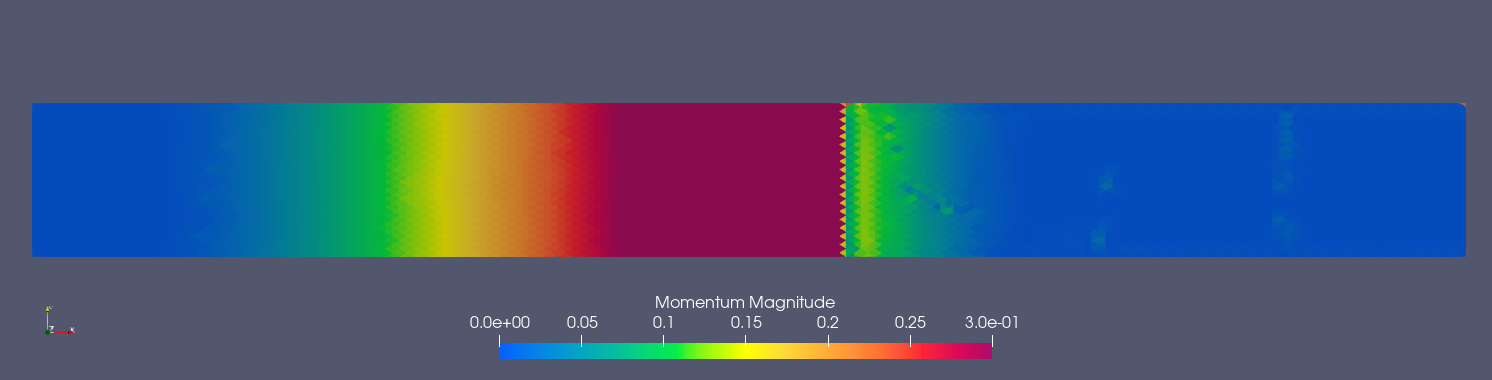}
    \label{fig:diffb}
    } 
    \subfigure[]{
  \includegraphics[width=.98\linewidth]{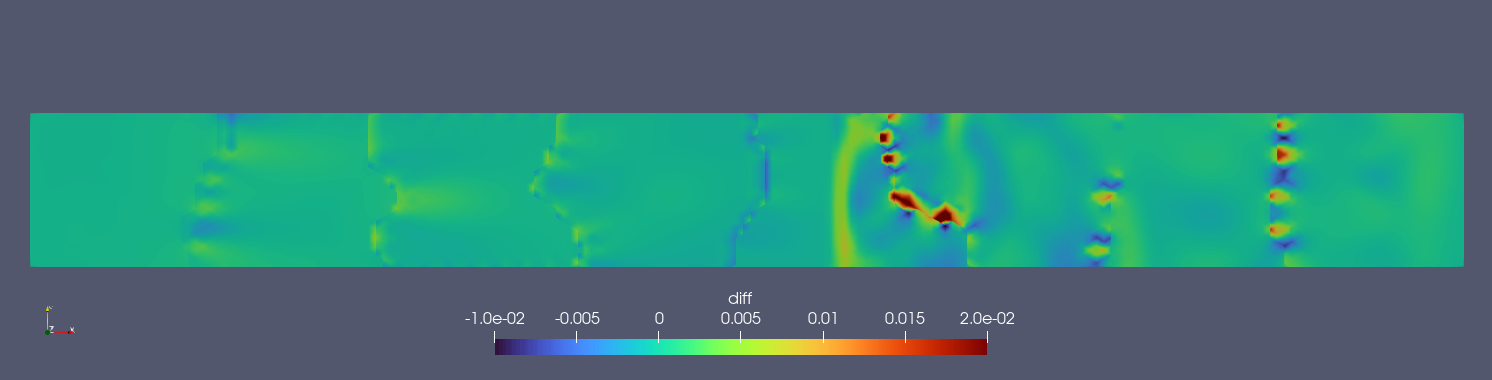}
    \label{fig:diffc}
    } 
    }
}
\end{center}
\caption{Mesh decomposed into 8 sub-domains (a). Solutions computed with the second order MUSCL method, using 8 GPUs on the 8 sub-domains' mesh, using 2 layers of ghost cells (b), 1 layer (c) and their difference (d) at time $t=3.2s$ for the dam break on a wet bottom.}
\label{fig:diff}
\end{figure}

As shown in Figure \ref{fig:fsev}, a second-order method will need to know the state of the upwind cells, and so it cannot be used on the boundaries between sub-domains with only one layer of ghost cells. To make this aspect clearer, Figure \ref{fig:fsev2} shows what would happen at the interface between two sub-domains for the resolution of the Riemann problem. It can be seen in Figure \ref{fig:fsev2b} that with only one layer of ghost cells, the states $R_{1upw}$ and $R_{2upw}$ are not known by the left sub-domain. In that case, we could still use a second-order method on the interior of the sub-domains, and revert to a first-order method on their boundaries. The goal of this section is to show the problems of this sort of combination of a first- and second-order resolutions. We show the second-order solutions computed using 8 GPUs on a mesh decomposed into 8 sub-domains for one and two layers of ghost cells. The mesh decomposed into 8 sub-domains is shown in Figure \ref{fig:diffmesh}. We compute the solution with the same initial condition as in the preceding section.  \\

Figure \ref{fig:diff} shows the solutions obtained by using one (\ref{fig:diffa}) and two (\ref{fig:diffb}) layers of ghost cells and their difference (\ref{fig:diffc}). It is clear in Figure \ref{fig:diffc} that the difference between the solutions is most pronounced at the interfaces between sub-domains. Even if an order one method is used on the boundaries of every sub-domain, mixing a first-order solution with a second-order solution may lead to inaccurate solutions with waves reflecting off the sub-domains boundaries. The inaccuracy in the solution can be seen in Figure \ref{fig:diffb} where only one layer of ghost cells is used. In contrast, the solution generated when using two layers of ghost cells in Figure \ref{fig:diffa} is exactly the same as the solution generated on the original mesh before the decomposition. 

\subsection{Performance of the solver for a varying number of ghost cell layers}
In this section, we show the solver's performance for varying numbers of ghost cell layers for the HLLC, WAF, and MUSCL methods. The goal is to evaluate the impact of increasing the amount of the communications on the total computation time. A complete description of the parallel solver and of the specific way the memory exchange is performed using a CUDA-Aware version of OpenMPI can be found in \cite{delmasvsoulaima}. \textcolor{black}{All the results shown in this section have been computed on \textit{Graham}, a computer cluster that is part of the \textit{Digital Research Alliance of Canada}. \textcolor{black}{The GPU nodes used consist of 2 \textit{NVIDIA P100 Pascal} per node, connected by a 56Gb/s Mellanox FDR InfiniBand interconnect}. A complete description of the cluster can be found on \url{https://docs.alliancecan.ca/wiki/Graham}.}\\

The parallel solver works by assigning one MPI process to each of the sub-domains. Each MPI process uses one GPU to compute the solution on the inner cells of the sub-domain it has been assigned to, before it is sent to the adjacent sub-domains. The whole process of generating the send/receive information needed in this step was presented in section \ref{sec:parghost}. Depending on the method, multiple layers of ghost cells might be needed so that the solution is exactly the same no matter the number of sub-domains used (see Figure \ref{fig:diff}). The downside of needing multiple layers of ghost cells is that more data must be exchanged at each time-step, which slows down the solver. \textcolor{black}{The memory exchange is performed using a send buffer and an in-place reception on the GPU side using a CUDA-Aware version of OpenMPI. A send buffer is updated on the GPU for each of the neighbors using multiple streams and then sent to the appropriate MPI process where it is received in-place, that is to say, directly inside the solution vector. This is possible because the cells to receive have been renumbered in a contiguous way. More information about the CUDA-Aware version of OpenMPI that is used in the solver can be found in \cite{delmasvsoulaima}.}\\

In order to get the best performance, the memory exchange is overlapped with the computations so that it has as little impact as possible on the overall execution time. Next, we show the results for a low overlapped version of the solver and for a high overlap version. In the low overlap version of the solver, the memory exchange is overlapped with the computation of the next iteration's time step. This is a notable improvement over a non-overlapping version, as demonstrated in \cite{delmasvsoulaima}. The issue is that using more complex methods like WAF or MUSCL does not increase the time taken to compute the next iteration's time step, and so there is no improvement of the scaling of the algorithm with more complex methods. To take advantage of the fact that the WAF and MUSCL methods take more time to compute than the HLLC method, we want to overlap the memory exchange with part of the solution's computation. This is done by first computing the solution only on the cells that need to be sent to other sub-domains, and to then begin the  asynchronous memory exchange, all while the rest of the solution is being computed. This results in a high overlap of the communications and computations and takes advantage of the slower methods.

\begin{table}[pos=H]
    \centering
    \caption{Execution times for the HLLC, WAF and MUSCL methods using one and two layers of ghost cells for a varying number of GPUs on a 13M cell mesh of the Mille Îles River, Montréal, with the \textbf{low overlap version} of the solver.}
\begin{tabular}{|l|c|c|c|c|c|c|} 
\hline
Number of MPI processes & 8 & 12 & 16 & 20 & 24 & 32 \\
\hline
HLLC  1 layer (s) & 205.37 & 148.24 & 117.28 & 102.38 & 91.98& 74.62 \\
\hline
HLLC  2 layers (s) & 218.02 &166.79 &136.41 &120.50 &111.71 &94.54 \\
\hline
MUSCL 2 layers (s) & 336.09 &241.66 &192.59 &165.13 &151.37 &124.69 \\
\hline
WAF   2 layers (s) & 337.41 &243.10 &192.54 &169.61 &151.48 &125.53 \\
\hline

\end{tabular}
    \label{tab:tempscomparlow}
\end{table}

\begin{table}[pos=H]
    \centering
    \caption{Execution times for the HLLC, WAF and MUSCL methods using one and two layers of ghost cells for a varying number of GPUs on a 13M cell mesh of the Mille Îles River, Montréal, with the \textbf{high overlap version} of the solver.}
\begin{tabular}{|l|c|c|c|c|c|c|} 
\hline
Number of MPI processes & 8 & 12 & 16 & 20 & 24 & 32 \\
\hline
HLLC 1 layer (s) & 205.26 &138.37 &106.86 &87.38 &75.58 &60.63126\\
\hline
HLLC 2 layers (s) & 204.65 &139.36 &107.36 &88.27 &76.24 &69.39 \\
\hline
MUSCL 2 layers (s) & 326.38 &217.51 &168.44 &134.42 &115.82 &90.08\\
\hline
WAF 2 layers (s) & 329.56 &219.82 &167.94 &135.68 &116.57 &90.82\\
\hline

\end{tabular}
    \label{tab:tempscomparhigh}
\end{table}

Tables \ref{tab:tempscomparlow} and \ref{tab:tempscomparhigh} show the execution times of the solver using the methods described earlier on a 13 million cell mesh of the Mille Île River, Montréal. First, the execution times of the HLLC method using one and two layers of ghost cells are reported. These results show how using two layers of ghost cells increases the amount of memory exchanged at each time step and its impact on the total execution time. Using two layers of ghost cells instead of one for the HLLC method results in a $7$ to $20\%$ longer total execution time, depending on the number of sub-domains for the low overlap version, whereas for the high overlap version, there is no notable difference until 32 GPUs are used. This highlights the advantage of using the high overlap version of the solver, especially when dealing with a large number of sub-domains. The low overlap version using the HLLC method is $23\%$ slower than the high overlap method using a single layer of ghost cells, and $36\%$ slower when using two layers.\\

Tables \ref{tab:tempscomparlow} and \ref{tab:tempscomparhigh} also show how using a second-order method like WAF or MUSCL slows down the computations. These second-order methods both need to use the upwind values to compute a local gradient of the solution, which takes time. In our solver, the WAF and MUSCL methods take approximately the same time to compute the solution. Both methods take around $30$ to $50\%$ more time than the simple HLLC method. These figures are in line with \cite{riadh}, where, on a specific test case of the Malpasset dam break, CPU times of $46s$ and $1m08s$ are reported for the HLLC and WAF methods, respectively, which means the WAF method takes around $47.8\%$ more time than the HLLC method. It should be noted that more complex MUSCL methods usually take longer to compute than the WAF method \citep{riadh}. Both these methods need to use two layers of ghost cells to ensure the quality of the solution across sub-domains.\\

\begin{figure}[pos=htp]
\begin{center}
\fbox{
    \subfigure[Element time step per second]{
		\includegraphics[width=0.41\textwidth]{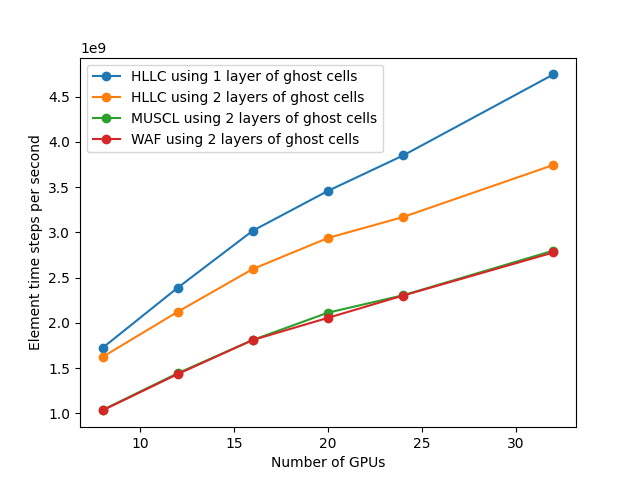}
    } 
    \subfigure[Element time step per second per process]{
		\includegraphics[width=0.41\textwidth]{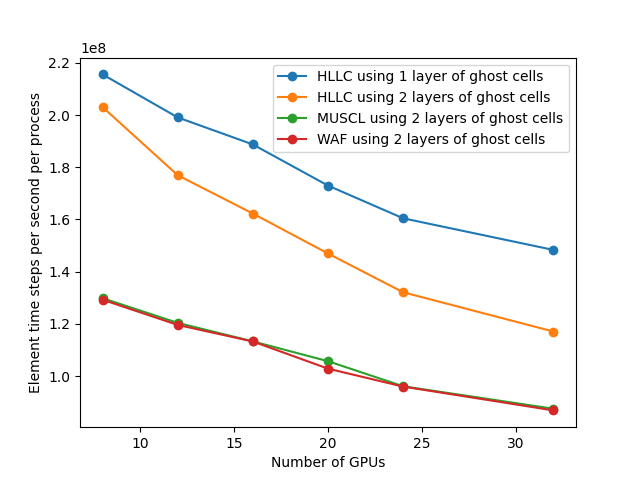}
    }
}
\end{center}
	 \caption{Element time step per second (a) and Element time step per second per process (b) of the \textbf{low overlap version} of the solver with the HLLC, WAF and MUSCL methods using one and two layers of ghost cells on a 13M cell mesh of the Mille Îles River, Montréal.}
	 \label{fig:etspslow}
\end{figure}

\begin{figure}[pos=htp]
\begin{center}
\fbox{
    \subfigure[Element time step per second]{
		\includegraphics[width=0.41\textwidth]{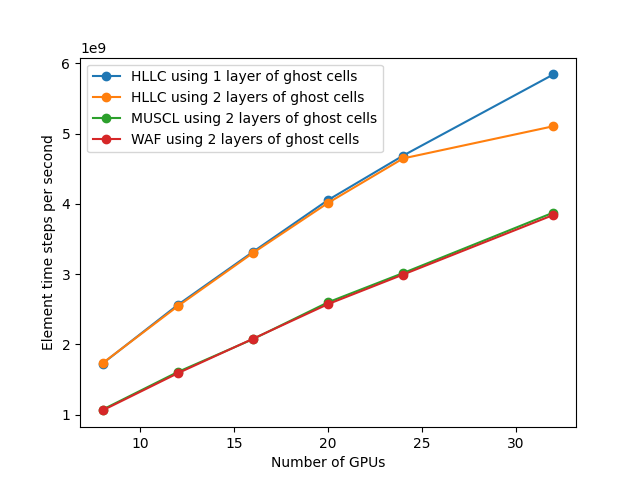}
    } 
    \subfigure[Element time step per second per process]{
		\includegraphics[width=0.41\textwidth]{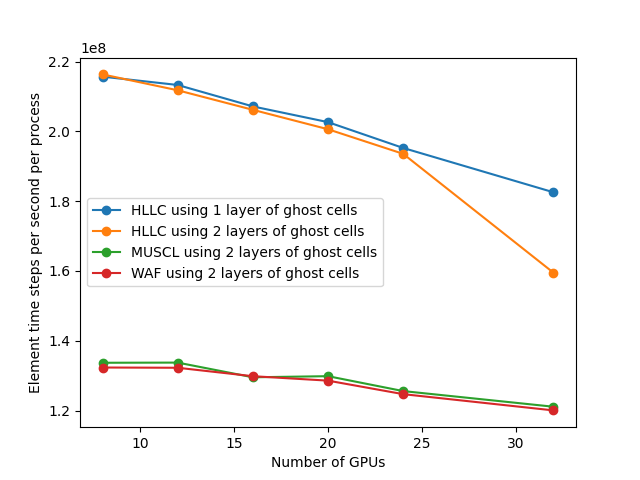}
    }
}
\end{center}
	 \caption{Element time step per second (a) and Element time step per second per process (b) of the \textbf{high overlap version} of the solver with the HLLC, WAF and MUSCL methods using one and two layers of ghost cells on a 13M cell mesh of the Mille Îles River, Montréal.}
	 \label{fig:etspshigh}
\end{figure}

\begin{figure}[pos=htp]
\begin{center}
\fbox{
		\includegraphics[width=0.51\textwidth]{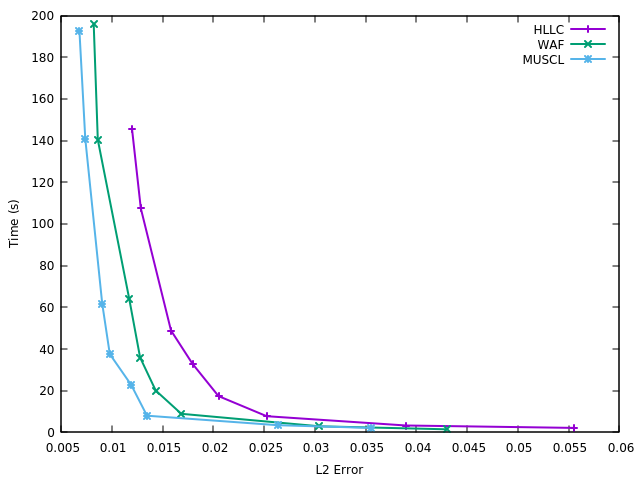}
}
\end{center}
	 \caption{\textcolor{black}{Time-Error graph computed on the classic Riemann dam break problem on a wet bottom for the HLLC, WAF and MUSCL methods using a single GPU with meshes ranging from 20 000 to 600 000 éléments. A similar SUPERBEE style limiting function was used for both the WAF and MUSCL methods.}}
	 \label{fig:timee}
\end{figure}

\textcolor{black}{To give a better idea of the absolute performance of the MUSCL and WAF methods when compared to HLLC, we use Tables \ref{tab:tempscomparlow} and \ref{tab:tempscomparhigh} to compute the element time step per second and element time step per second per process for each method. Figures \ref{fig:etspslow} and \ref{fig:etspshigh} show respectively the performance of the low and high overlap versions. We can clearly see in Figure \ref{fig:etspslow} the performance penalty caused by using two layers of ghost cells instead of one with the HLLC method. This results from the increase in communication while the computations remain the same. When using the high overlap version, we can see in Figure \ref{fig:etspshigh} that the performance of the HLLC method when using two layers of ghost cells is similar to using only one layer until 32 GPUs are used. This highlights the better overlap between computations and communications in the high overlap version. Furthermore, the high overlap version outperforms the low overlap version in every single scenario. Because the benefit of the second-order methods is to produce a more accurate solution, a time-error graph computed on the classic Riemann dam break on a wet bottom problem is presented in Figure \ref{fig:timee}.}\\

\noindent\textcolor{black}{In Figure \ref{fig:timee} we plot the time taken to compute the solution against the error in L2 norm for the HLLC, WAF and MUSCL methods. The results of Figure \ref{fig:timee} were produced using a single GPU on meshes ranging from 20 000 to 600 000 elements. It can be seen that the second-order WAF and MUSCL methods produce a more accurate solution than the first-order HLLC method for the same computation time. In our implementation, the MUSCL method produces slightly more accurate solution than the WAF method. A comparative study of TVD-Limiters can be found in \cite{limitkemm}.}

\newpage
\subsection{Application to real domains with complex bathymetries}
The Shallow-water equations are solved here for real domains with complex bathymetries. We show how the second-order solution impacts the flow patterns on real rivers near Montréal, QC, Canada.\\

\subsubsection{The archipelago of Montréal}

\begin{figure}[pos=htp]
\begin{center}
\fbox{
\parbox{0.95\linewidth} {
    \subfigure[]{
  \includegraphics[width=.23\linewidth]{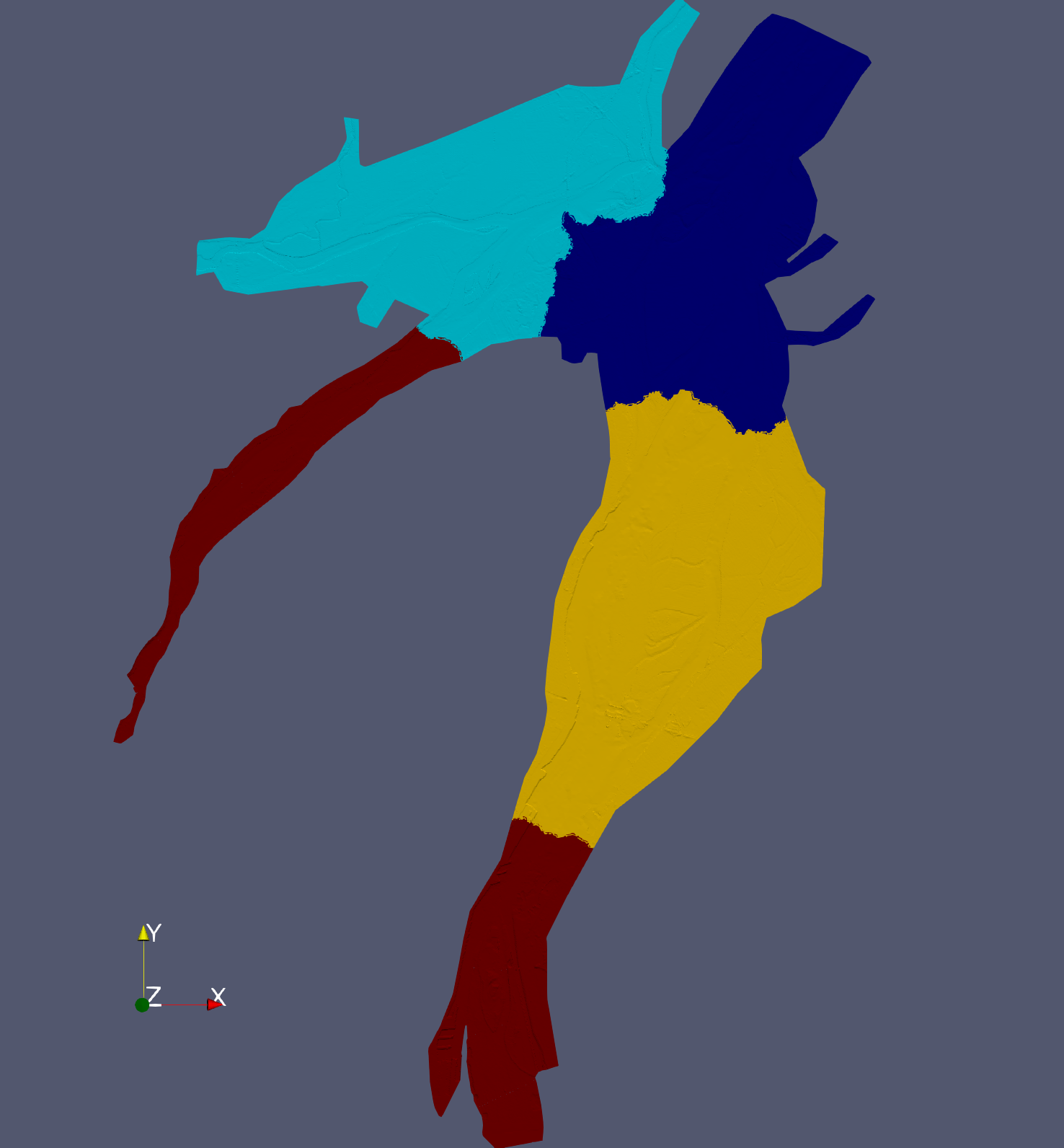}
  \label{fig:descarchia}
    } 
    \subfigure[]{
  \includegraphics[width=.23\linewidth]{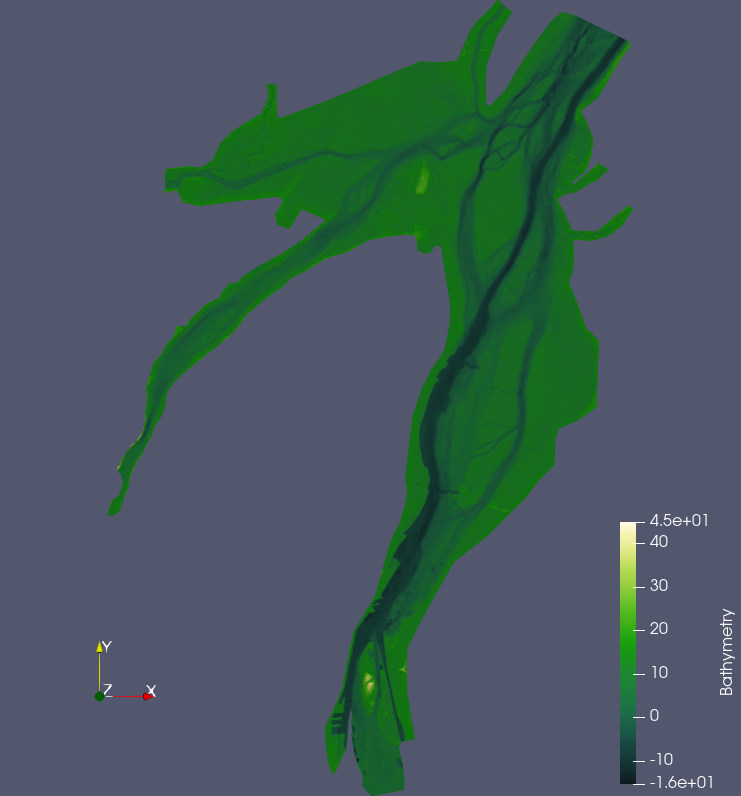}
  \label{fig:descarchib}
    } 
    \subfigure[]{
  \includegraphics[width=.23\linewidth]{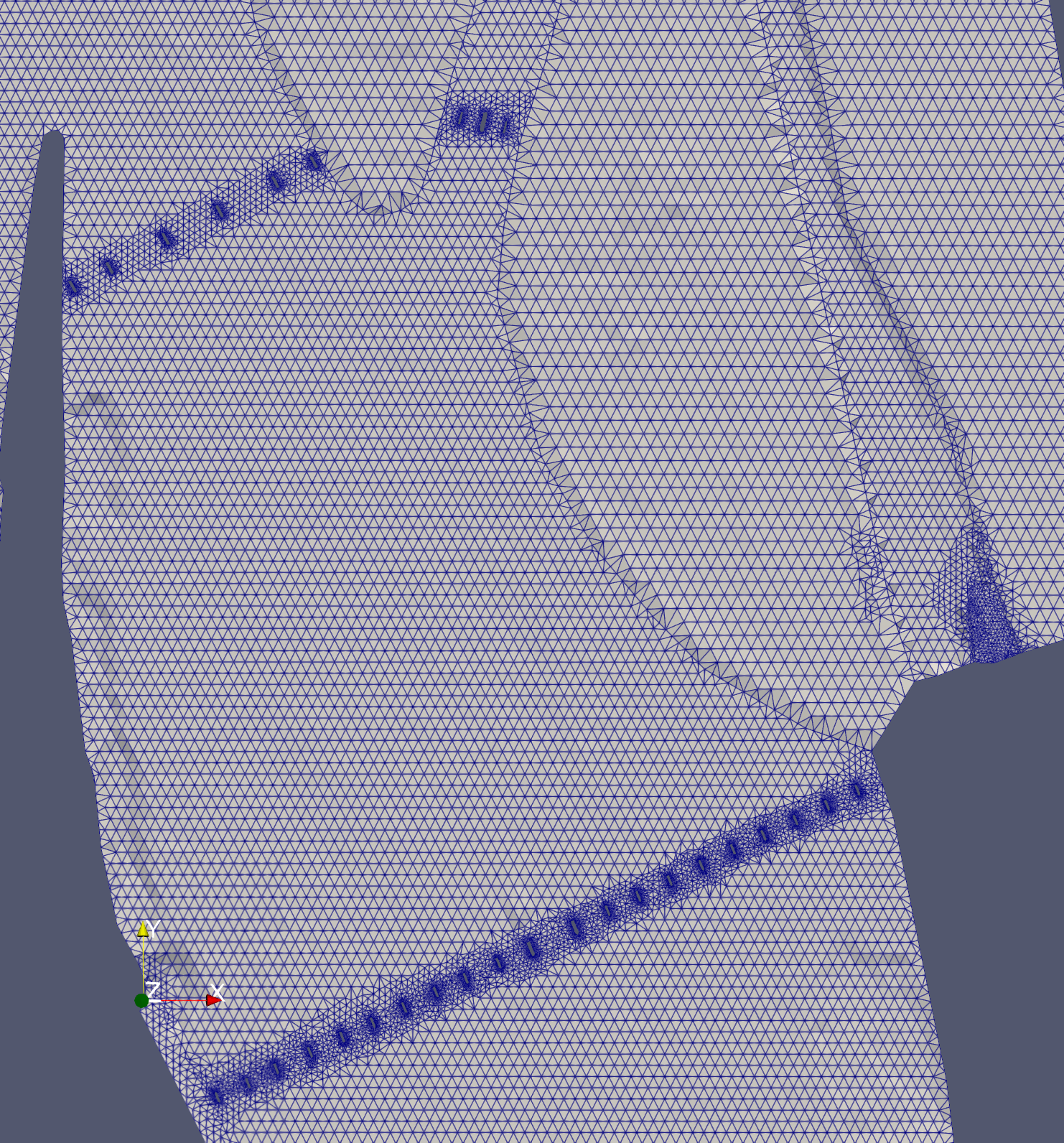}
  \label{fig:descarchic}
    } 
    \subfigure[]{
  \includegraphics[width=.23\linewidth]{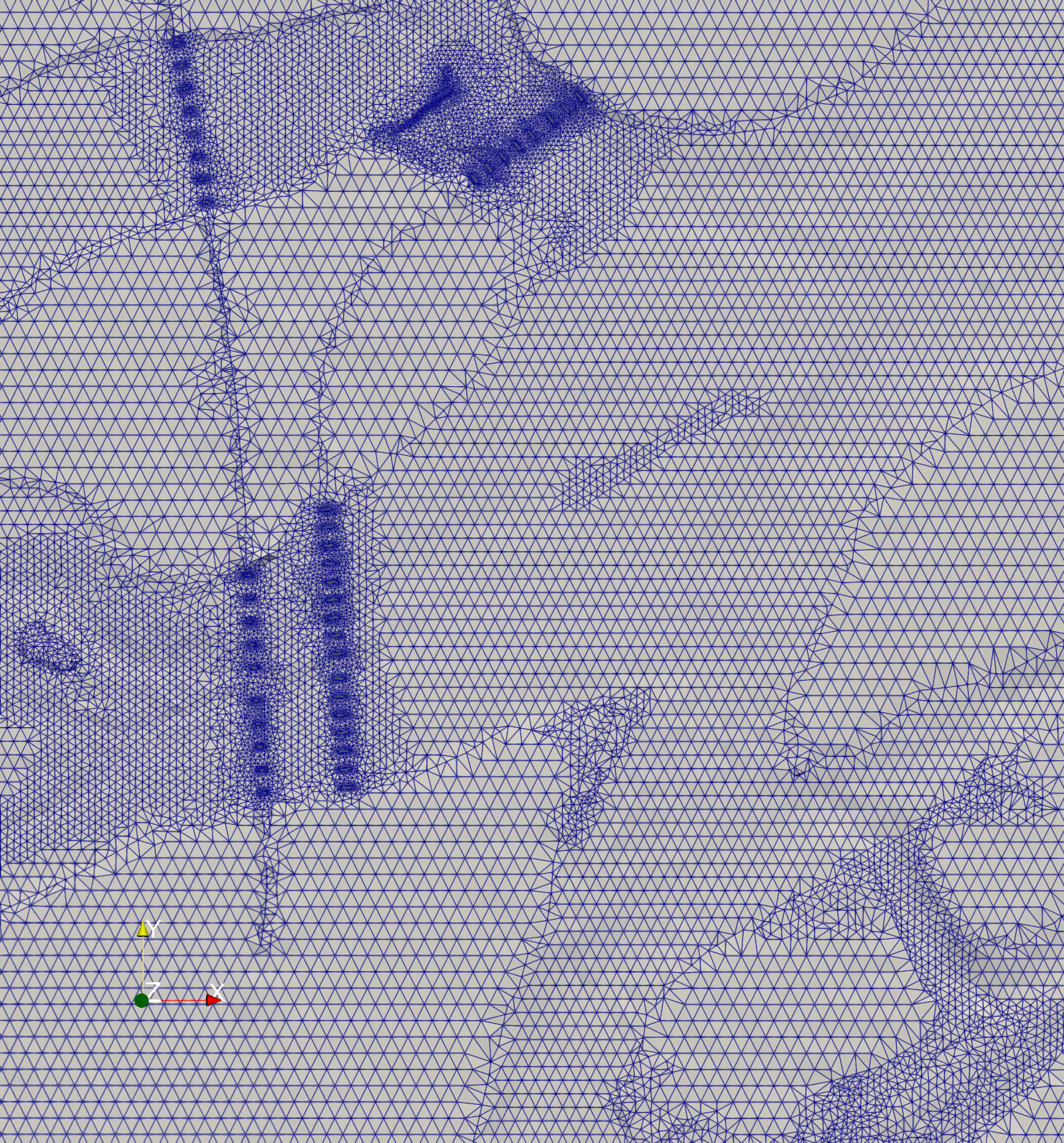}
  \label{fig:descarchid}
    } 
    }
}
\end{center}
\caption{Mesh of the archipelago of Montréal decomposed into 4 sub-domains (a), colored according to the bathymetry (b). North is up, the output of the domain is at the top right of the image. Close-up view of the entrance of the St Lawrence River near the south of the domain (c), and close-up view of a confluence near the top of the domain (d).}
\label{fig:descarchi}
\end{figure}

The first domain studied is a large domain containing multiple rivers that are part of the Montréal archipelago. The domain consists of three major rivers, the Mile Îles, the Prairies, and the St Lawrence, along with a few small streams. The domain encompasses around 30km of the St Lawrence River, and the mesh consists of 700K triangular cells. The domain is shown in Figure \ref{fig:descarchib}, colored according to the bathymetric data. The 2D triangular mesh is shown in Figure \ref{fig:descarchic} and \ref{fig:descarchid}, in which close-ups of two areas of the mesh show the refinement around bridge piers.\\

\begin{figure}[pos=htp]
\begin{center}
\fbox{
\parbox{0.95\linewidth} {
    \subfigure[$t=0s$]{
  \includegraphics[width=.23\linewidth]{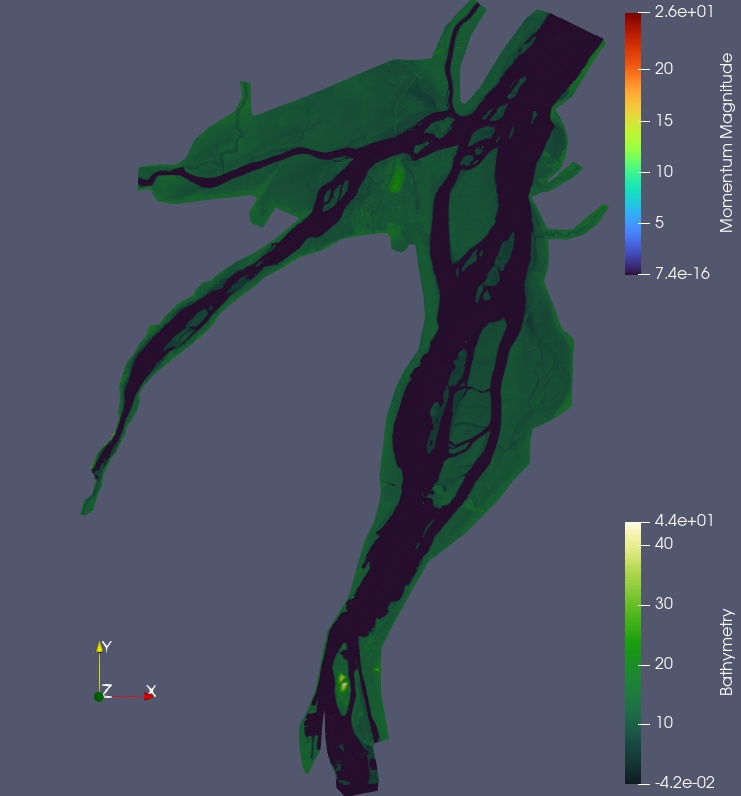}
    } 
    \subfigure[$t=500s$]{
  \includegraphics[width=.23\linewidth]{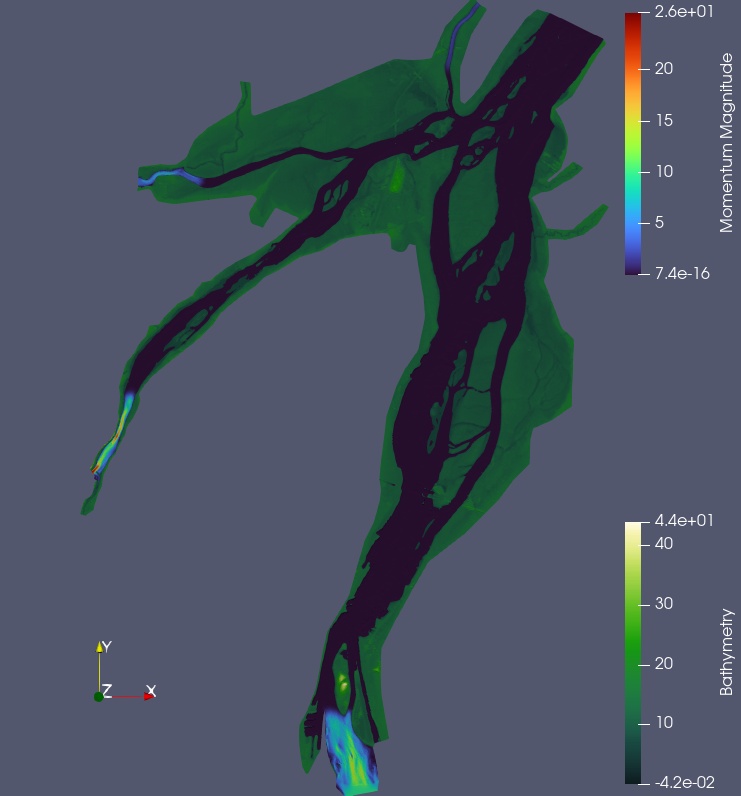}
    } 
    \subfigure[$t=1000s$]{
  \includegraphics[width=.23\linewidth]{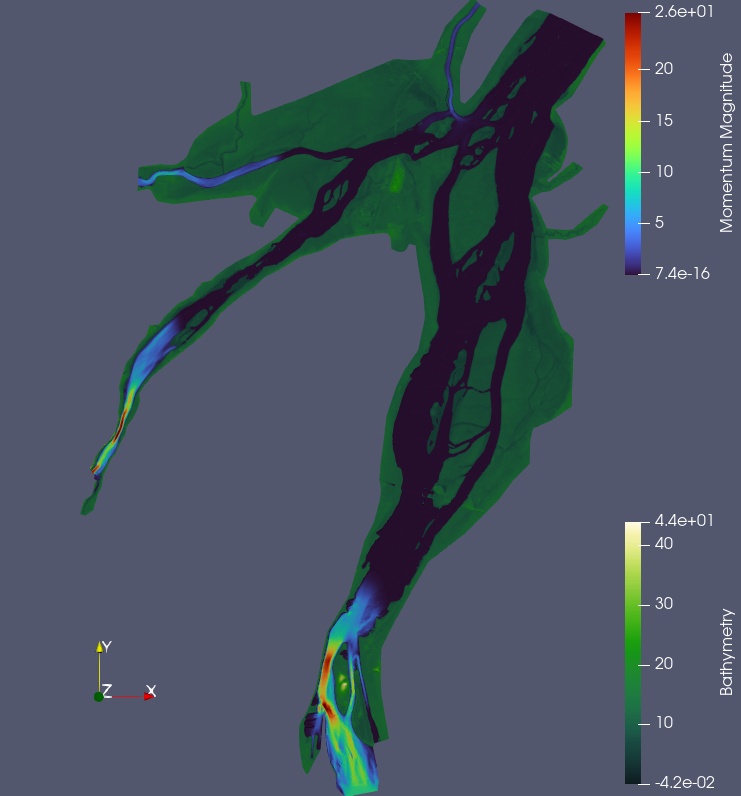}
    } 
    \subfigure[$t=1500s$]{
  \includegraphics[width=.23\linewidth]{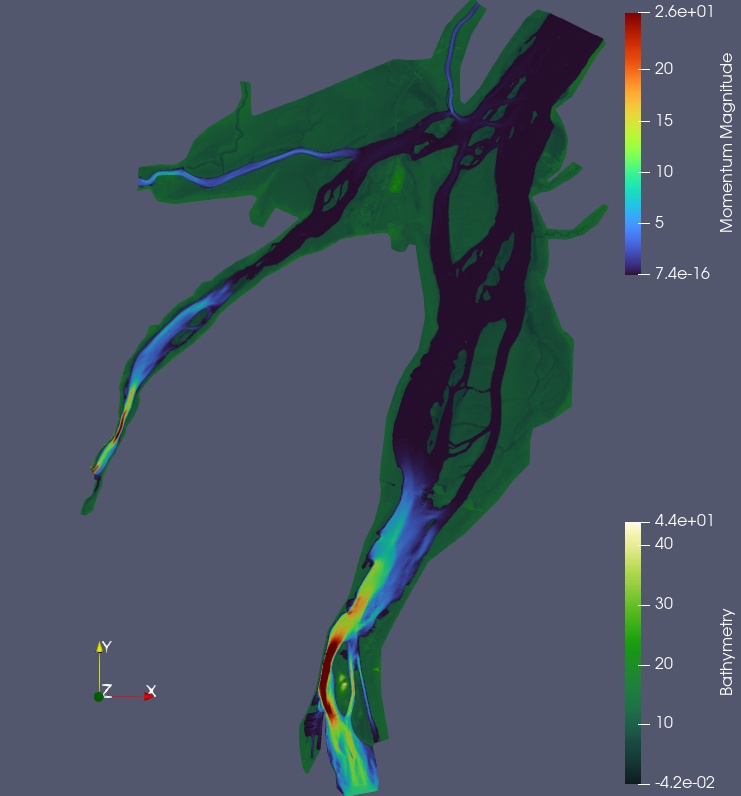}
    } 
    \subfigure[$t=2000s$]{
  \includegraphics[width=.23\linewidth]{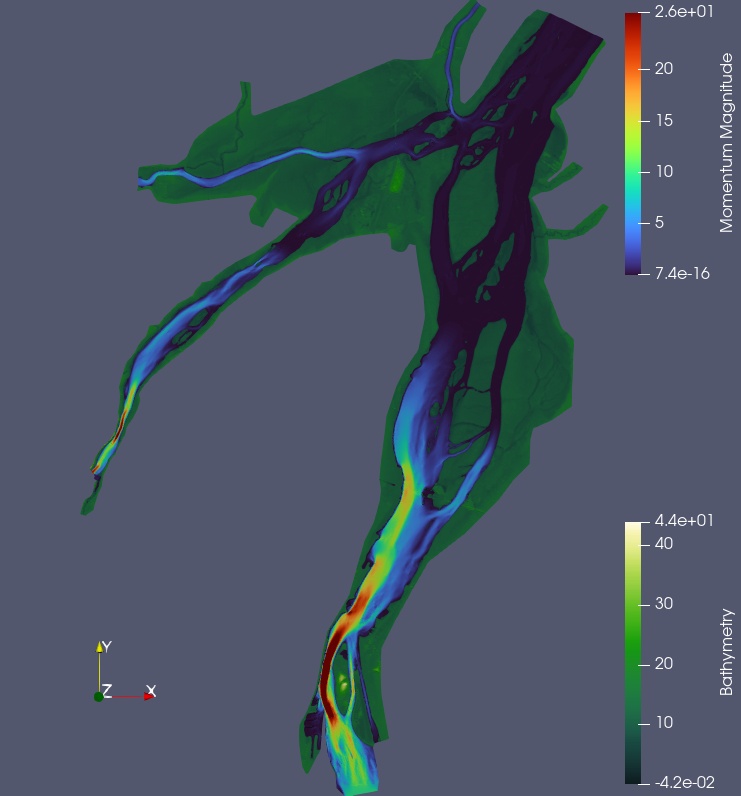}
    } 
    \subfigure[$t=2500s$]{
  \includegraphics[width=.23\linewidth]{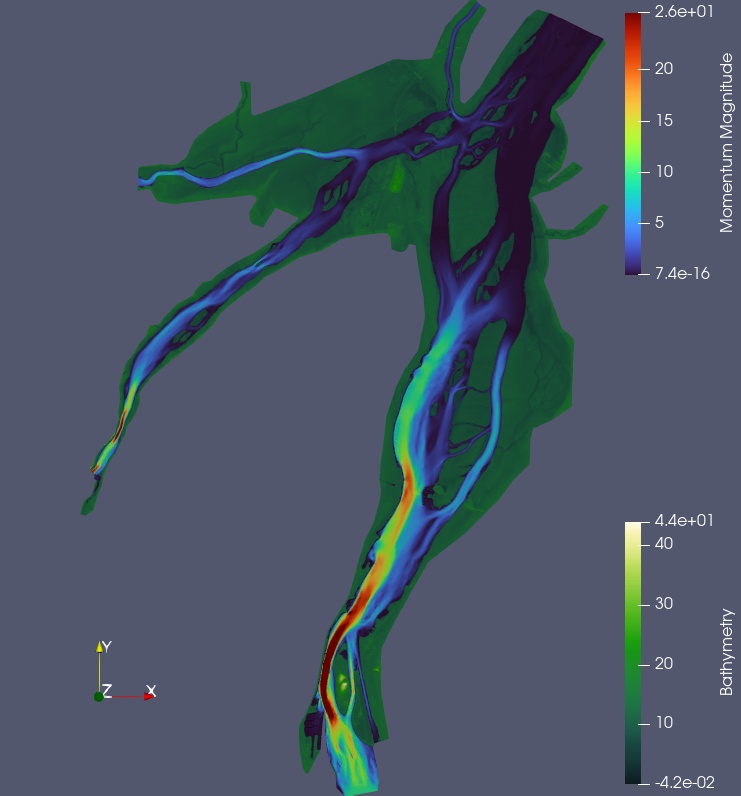}
    } 
    \subfigure[$t=3000s$]{
  \includegraphics[width=.23\linewidth]{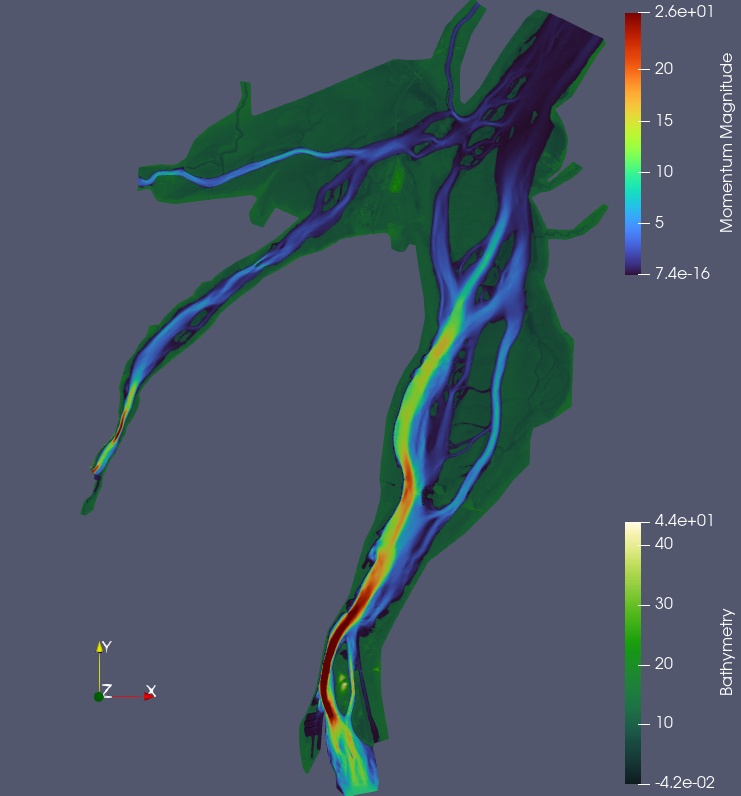}
    } 
    \subfigure[$t=3500s$]{
  \includegraphics[width=.23\linewidth]{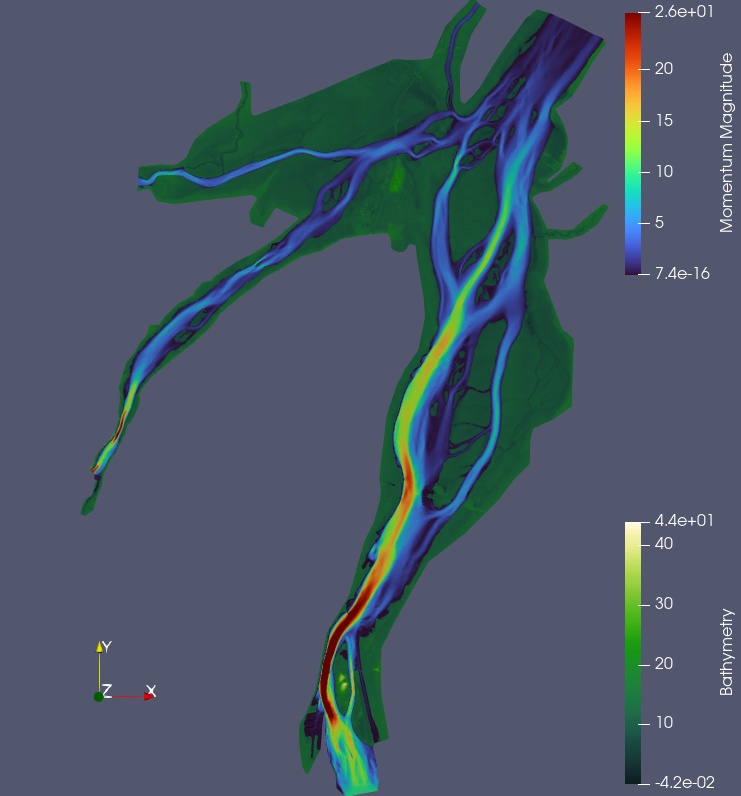}
    } 
    }
}
\end{center}
\caption{Second order solution using the MUSCL method with two layers of ghost cells with 4 sub-domains. The dry part of the domain is colored according to the bathymetry whereas the wet part of the domain is colored according to the momentum magnitude.}
\label{fig:compar_archi}
\end{figure}

The mesh was decomposed into 4 sub-domains, as the best performances with our in-house solver were achieved with 4 GPUs. The domain decomposition is shown in Figure \ref{fig:descarchia}. As explained in the preceding sections, one layer of ghost cells will be used with the first-order method and two layers of ghost cells with the second-order method. The results generated in this way are independent of the number of sub-domains used. The simulation is initialized with a flat free surface with a level corresponding to the output level, with the initial velocity set to zero, and a typical inflow imposed at each of the inputs of the domain. We show snapshots of the second-order solution in Figure \ref{fig:compar_archi}, where the dry part of the domain is colored according to the bathymetry and the wet part is colored according to the momentum magnitude ($h \sqrt{u^2+v^2}$).

\begin{figure}[pos=htp]
\begin{center}
\fbox{
\parbox{0.7\linewidth} {
    \subfigure[]{
  \includegraphics[width=.48\linewidth]{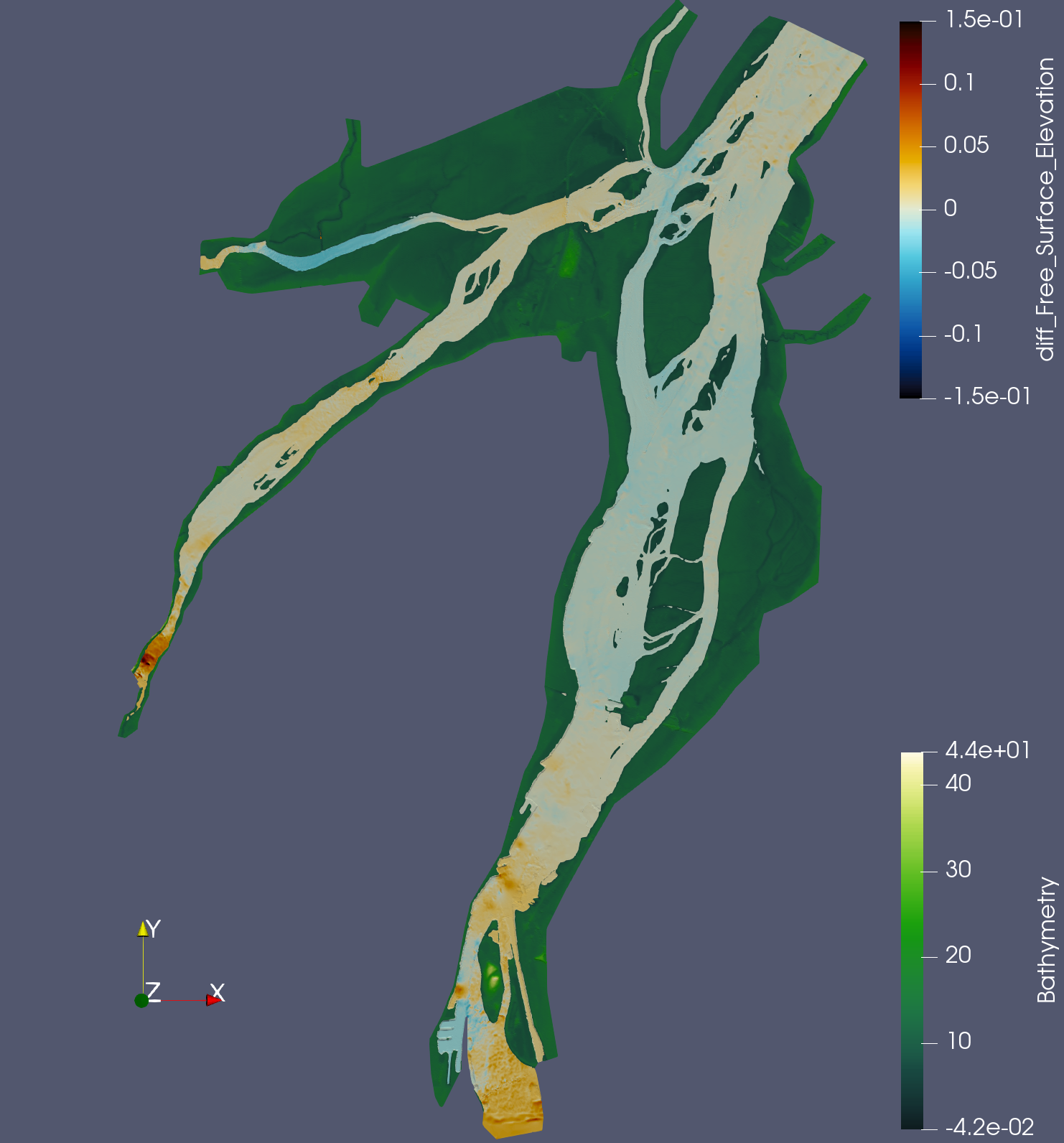}
\label{fig:diff_archia}
    } 
    \subfigure[]{
  \includegraphics[width=.48\linewidth]{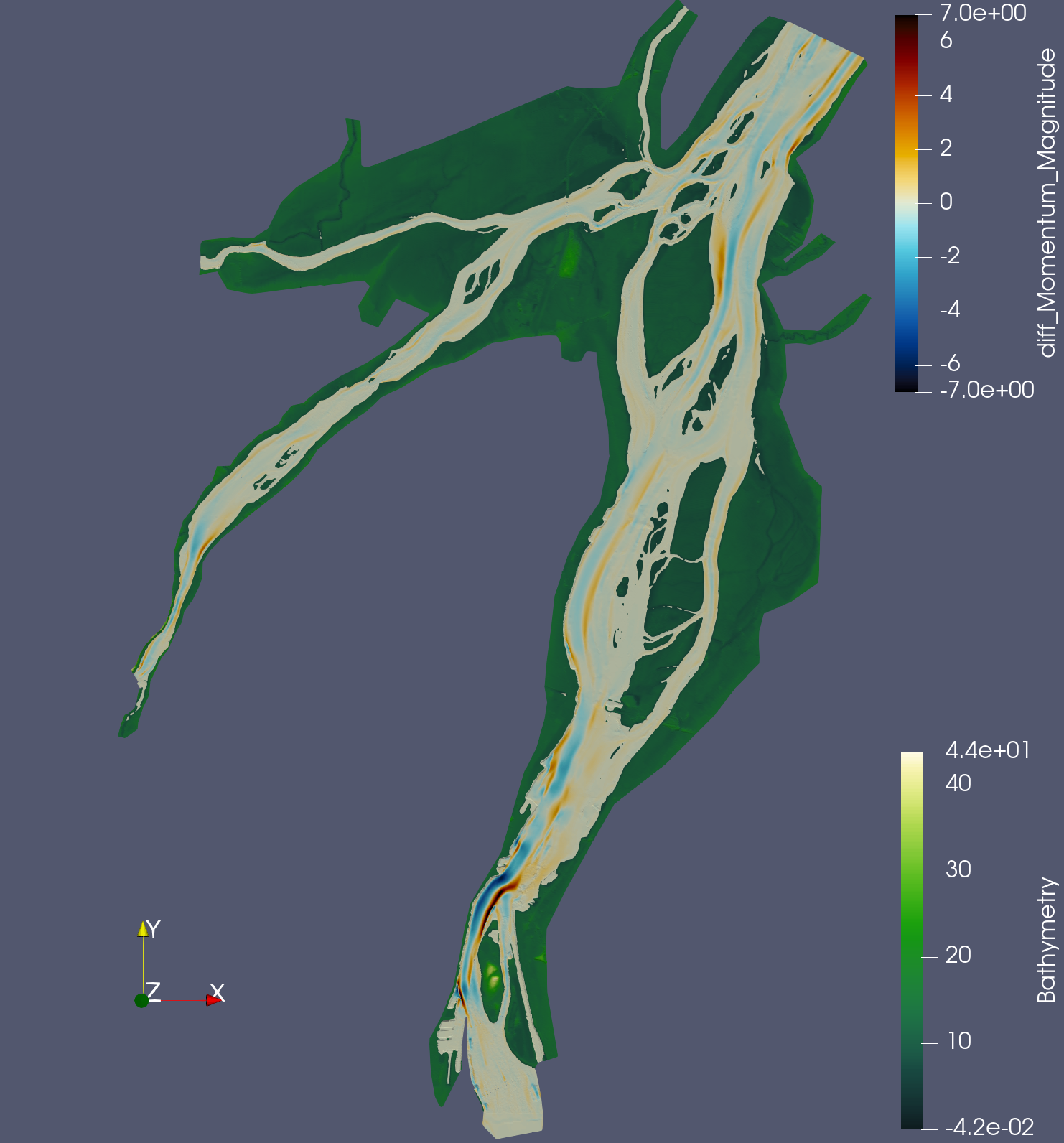}
\label{fig:diff_archib}
    } 
    }
}
\end{center}
\caption{Difference between the second and first-order solutions at $t=10000s$; the dry part of the domain is colored according to the bathymetry, the wet part of the domain is colored according to the difference in free surface elevation (a) and momentum magnitude (b).}
\label{fig:diff_archi}
\end{figure}

As it is difficult to see the differences between the first- and second-order solutions just by displaying them side by side, we chose to compute the difference between the solutions and display it in Figure \ref{fig:diff_archi}. It can be seen that the majority of the differences in free surface elevation in Figure \ref{fig:diff_archia} are in zones near the inputs, where the free surface gradient is the most pronounced. The differences in momentum magnitude in Figure \ref{fig:diff_archib} show that the second-order solution reshapes the flow patterns in the river, especially near the entrance of the St Lawrence river at the bottom of Figure \ref{fig:diff_archib}.

\subsubsection{The Mille Îles River}
\begin{figure}[pos=htp]
\begin{center}
\fbox{
\parbox{0.9\linewidth} {
    \subfigure[]{
  \includegraphics[width=.31\linewidth]{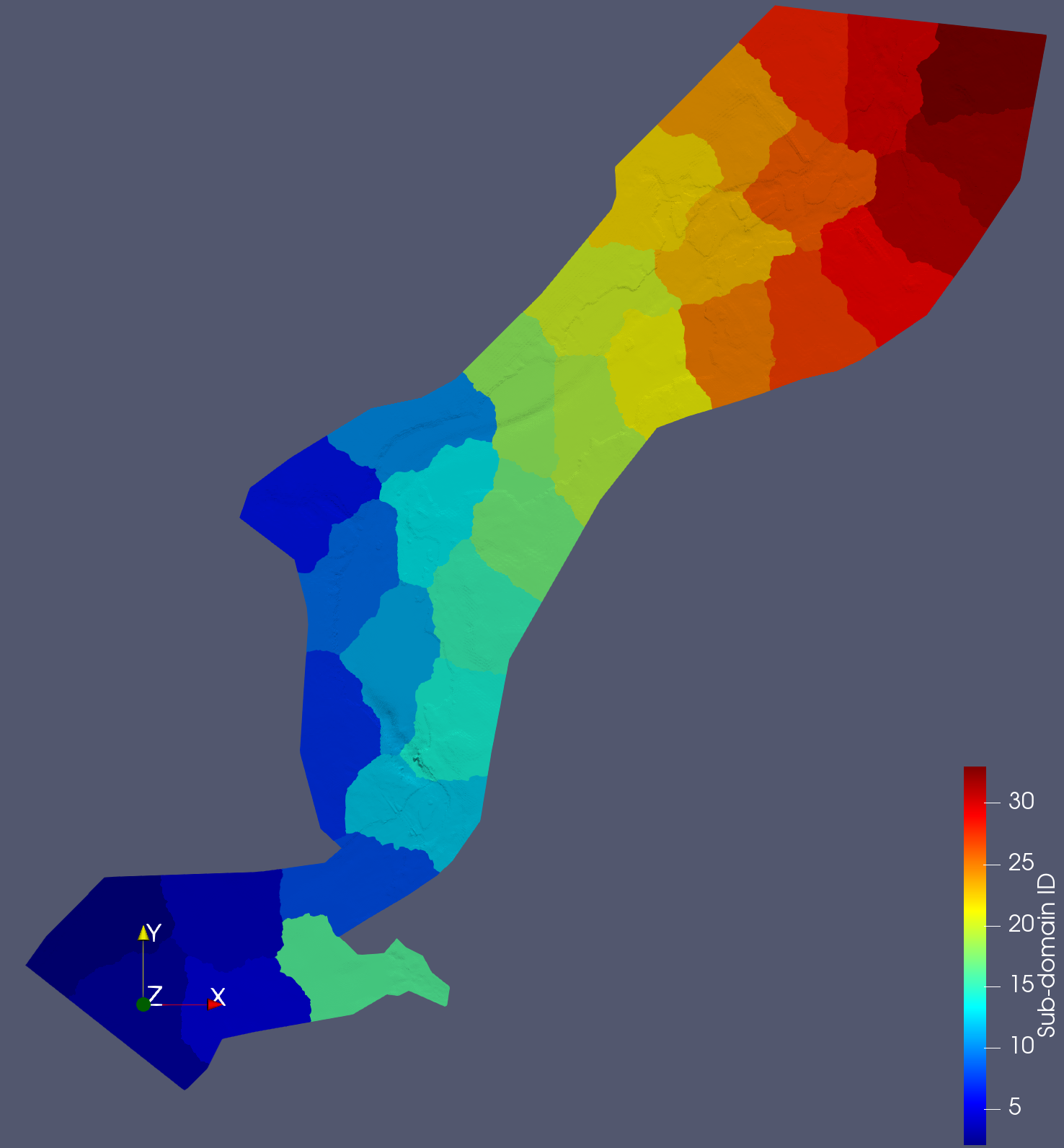}
  \label{fig:descmillea}
    } 
    \subfigure[]{
  \includegraphics[width=.31\linewidth]{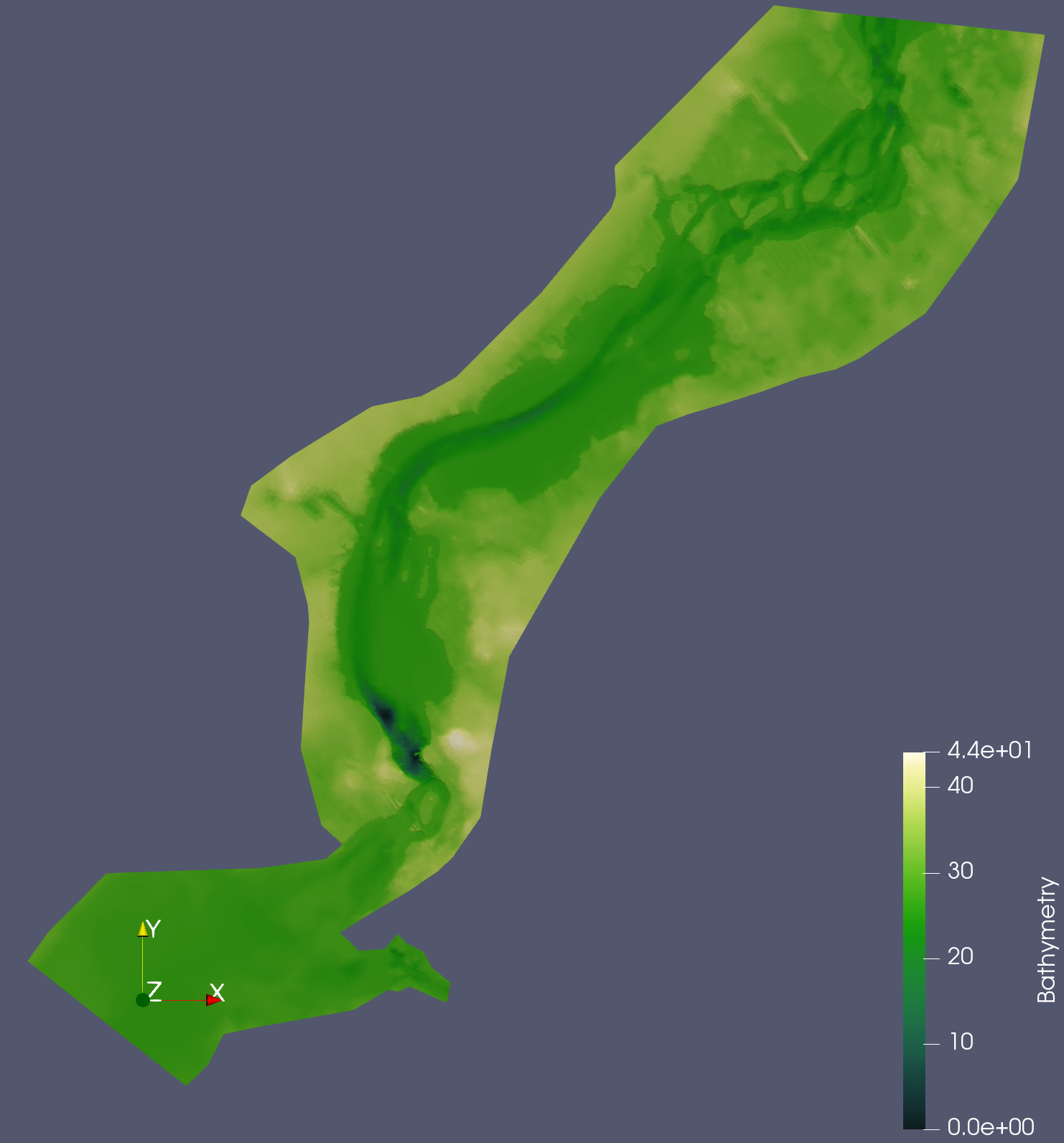}
  \label{fig:descmilleb}
    } 
    \subfigure[]{
  \includegraphics[width=.31\linewidth]{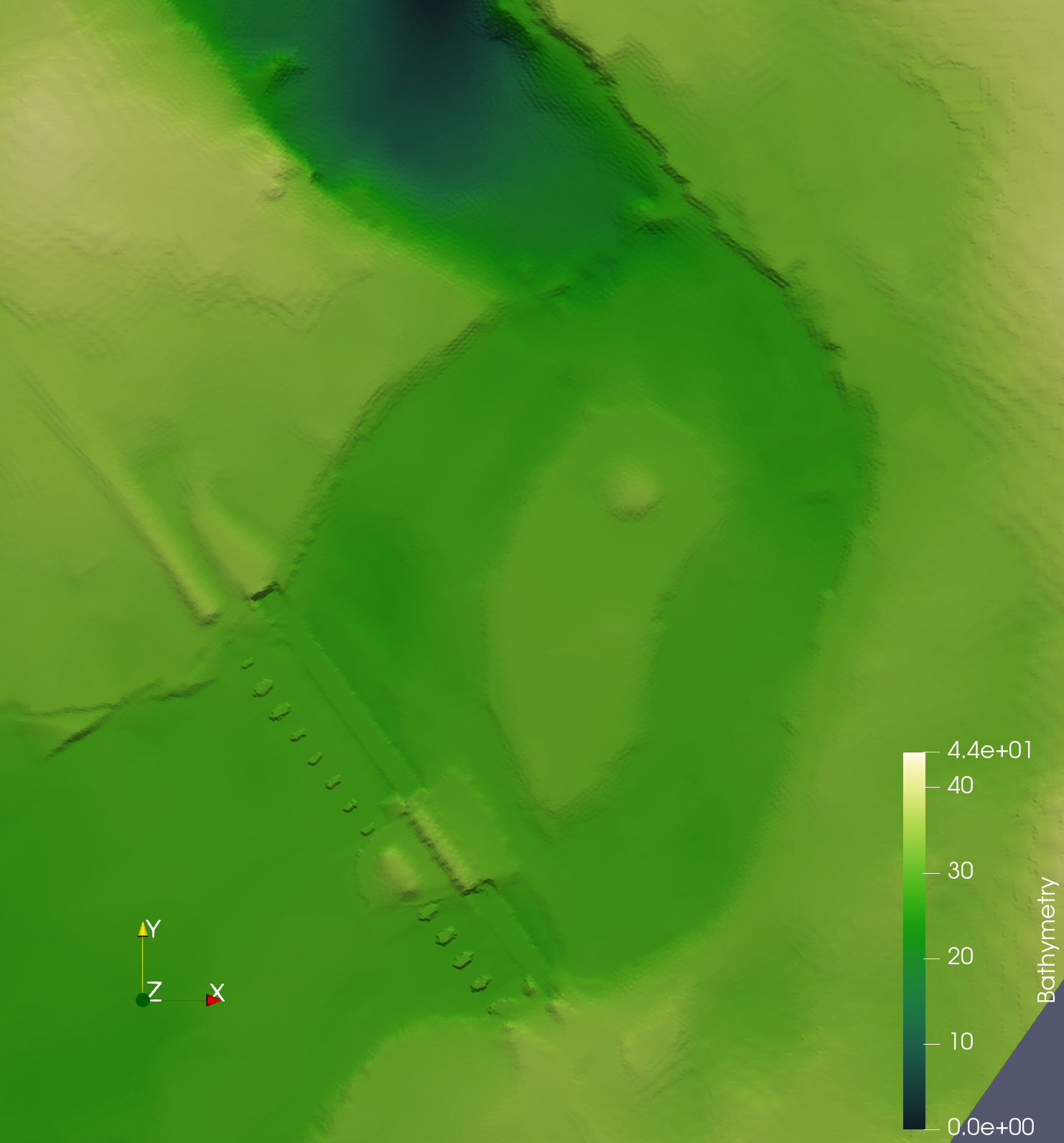}
  \label{fig:descmillec}
    } 
    }
}
\end{center}
\caption{Mille Îles River domain decomposed into 32 sub-domains (a) and colored according to bathymetry (b). North is up, the output of the domain is at the top right of the image. Close-up on a zone near the south of the domain where the river flows between the piers of a bridge, around an island, before taking a sharp left turn (c).}
\label{fig:descmille}
\end{figure}

This section shows a comparison of the first- and second-order solutions on a 13 million cell mesh of the Mille Îles River. First, a fictitious dam break is shown on a complex part of the domain where the river flows under a bridge and around an island, before taking a sharp left turn. The best performances for this mesh are achieved with 32 GPUs with our in-house solver. The domain decomposition is shown in Figure \ref{fig:descmillea}.\\

\begin{figure}[pos=htp]
\begin{center}
\fbox{
\parbox{0.9\linewidth} {
    \subfigure[$t=0s$]{
  \includegraphics[width=.23\linewidth]{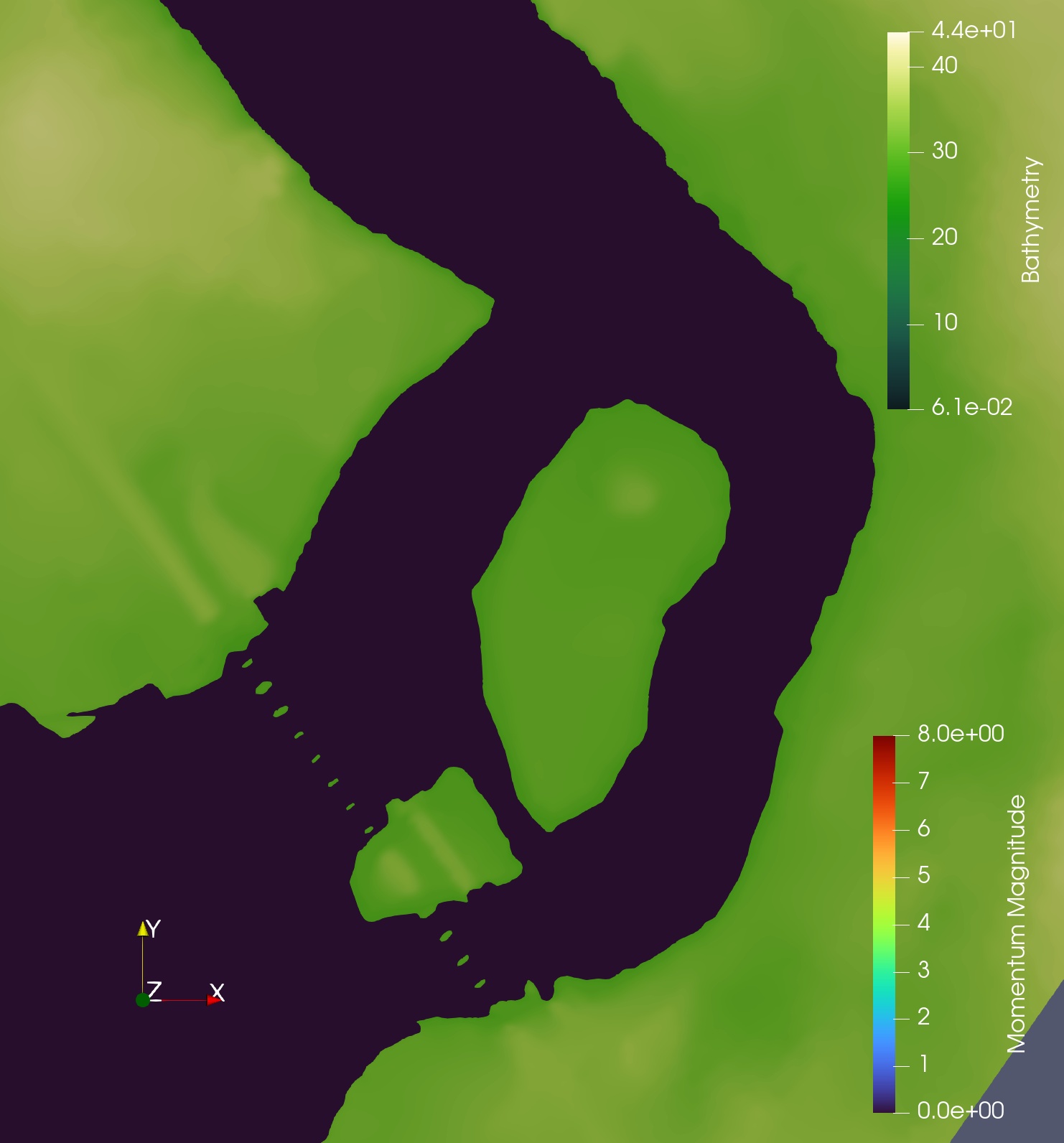}
    } 
    \subfigure[$t=5s$]{
  \includegraphics[width=.23\linewidth]{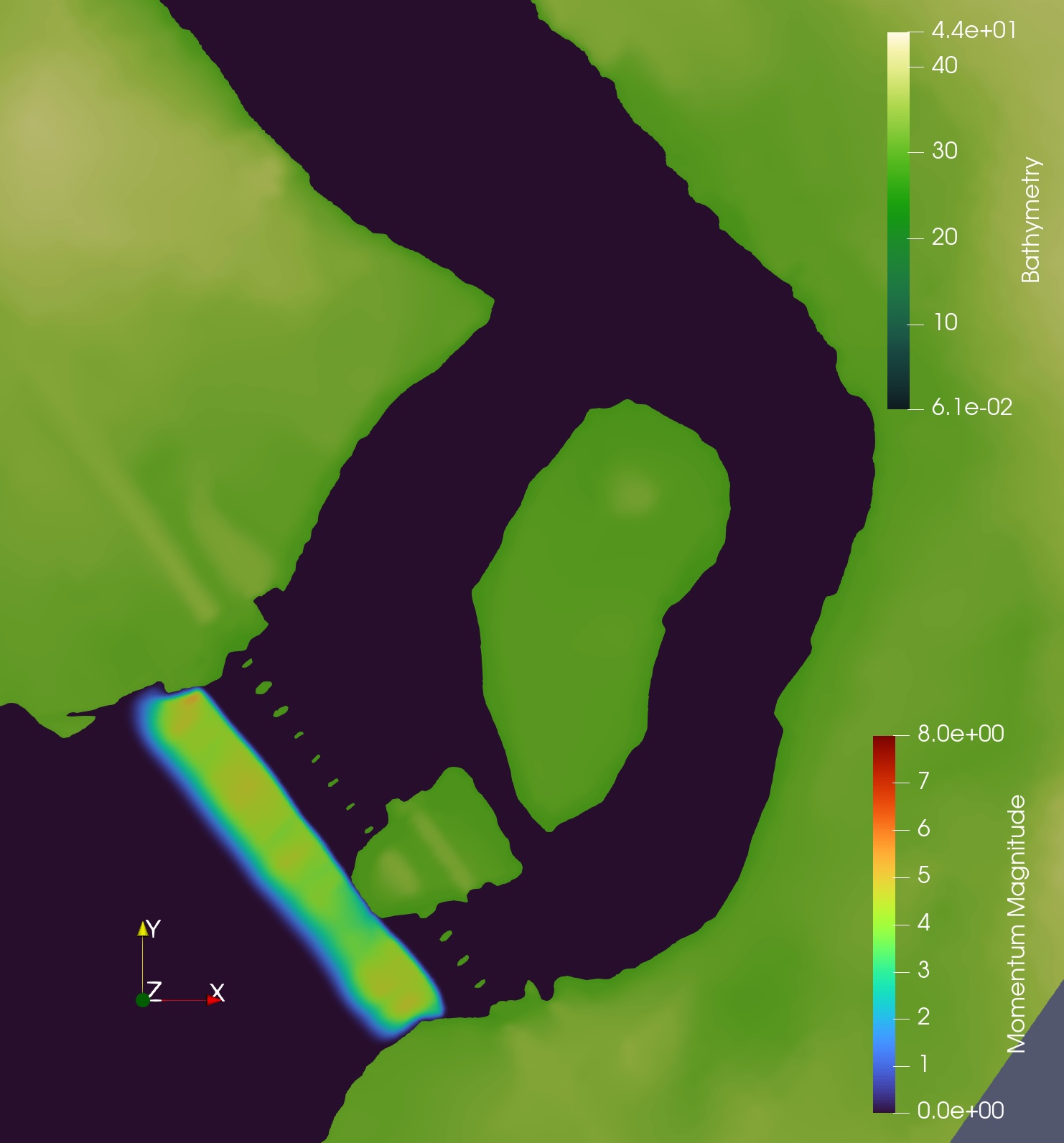}
    } 
    \subfigure[$t=35s$]{
  \includegraphics[width=.23\linewidth]{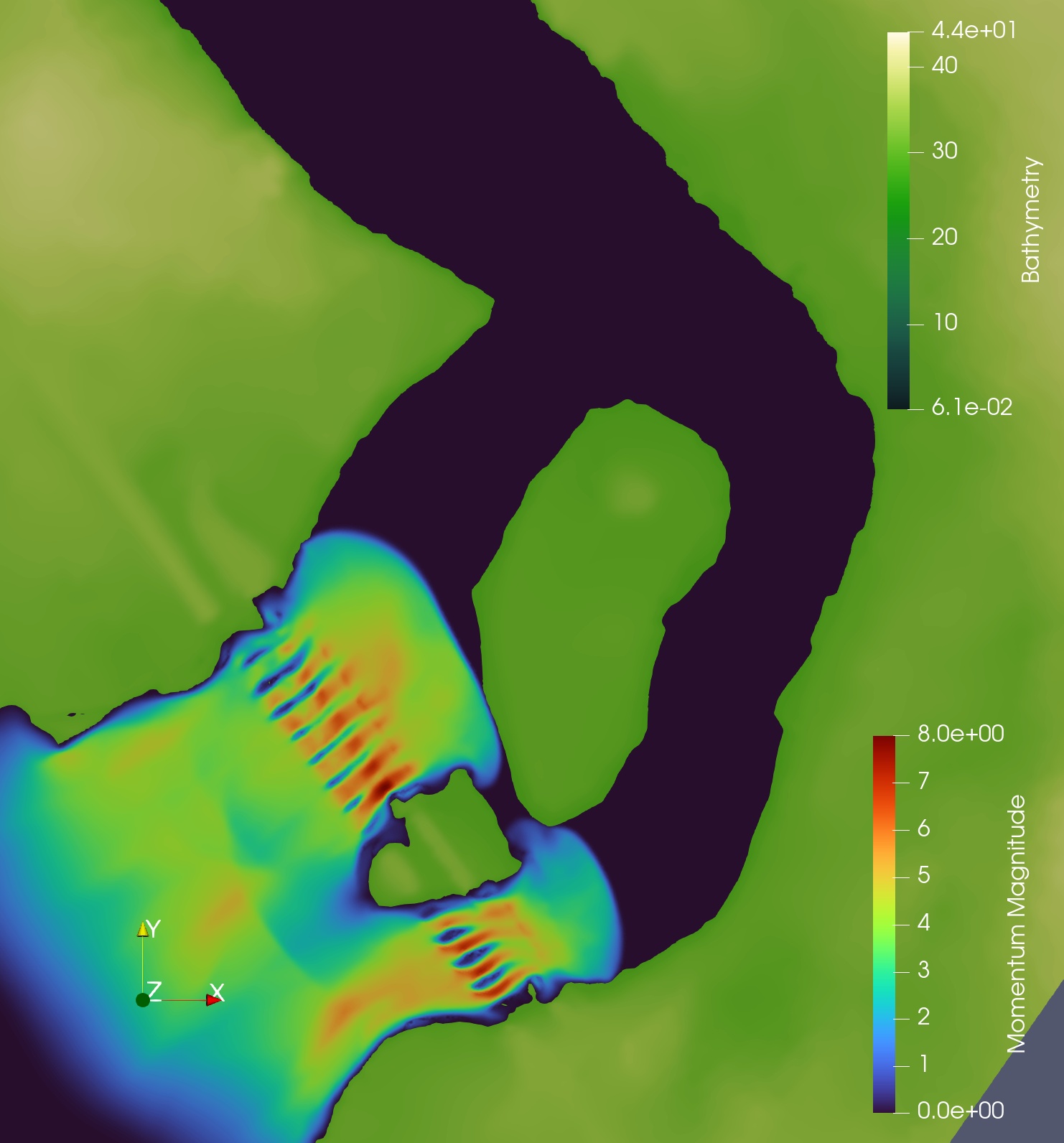}
    } 
    \subfigure[$t=70s$]{
  \includegraphics[width=.23\linewidth]{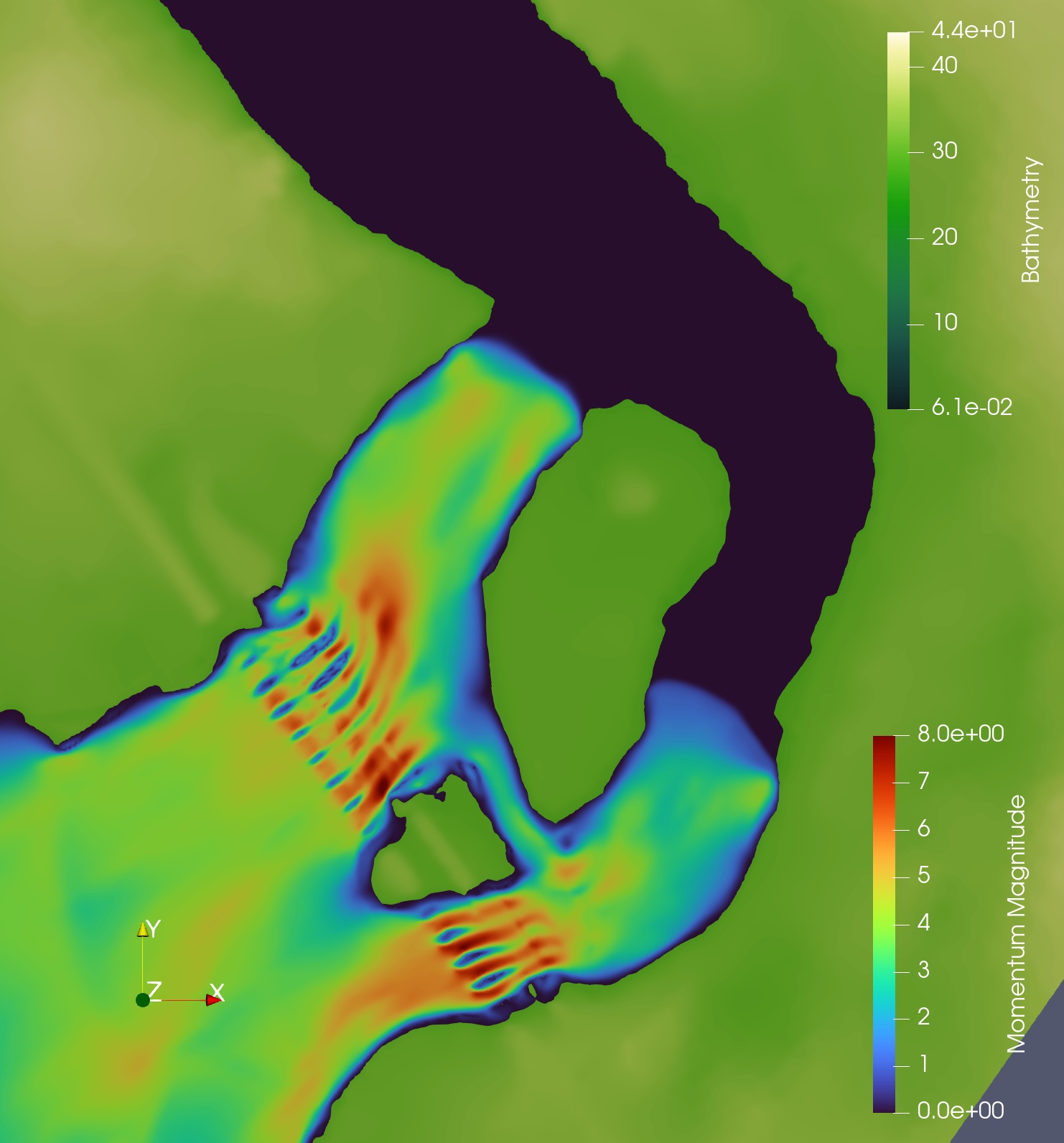}
    } 
    \subfigure[$t=105s$]{
  \includegraphics[width=.23\linewidth]{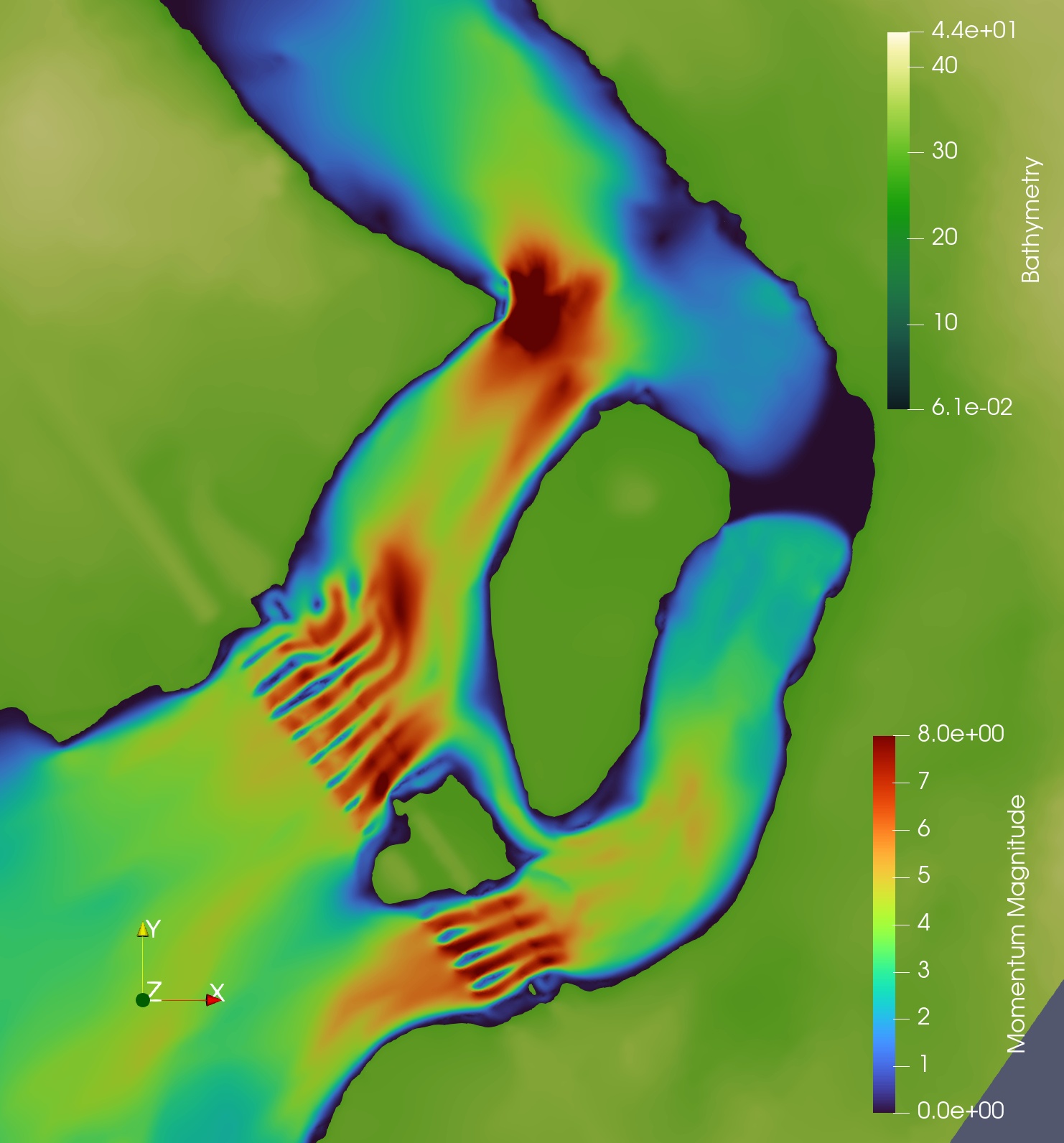}
    } 
    \subfigure[$t=140s$]{
  \includegraphics[width=.23\linewidth]{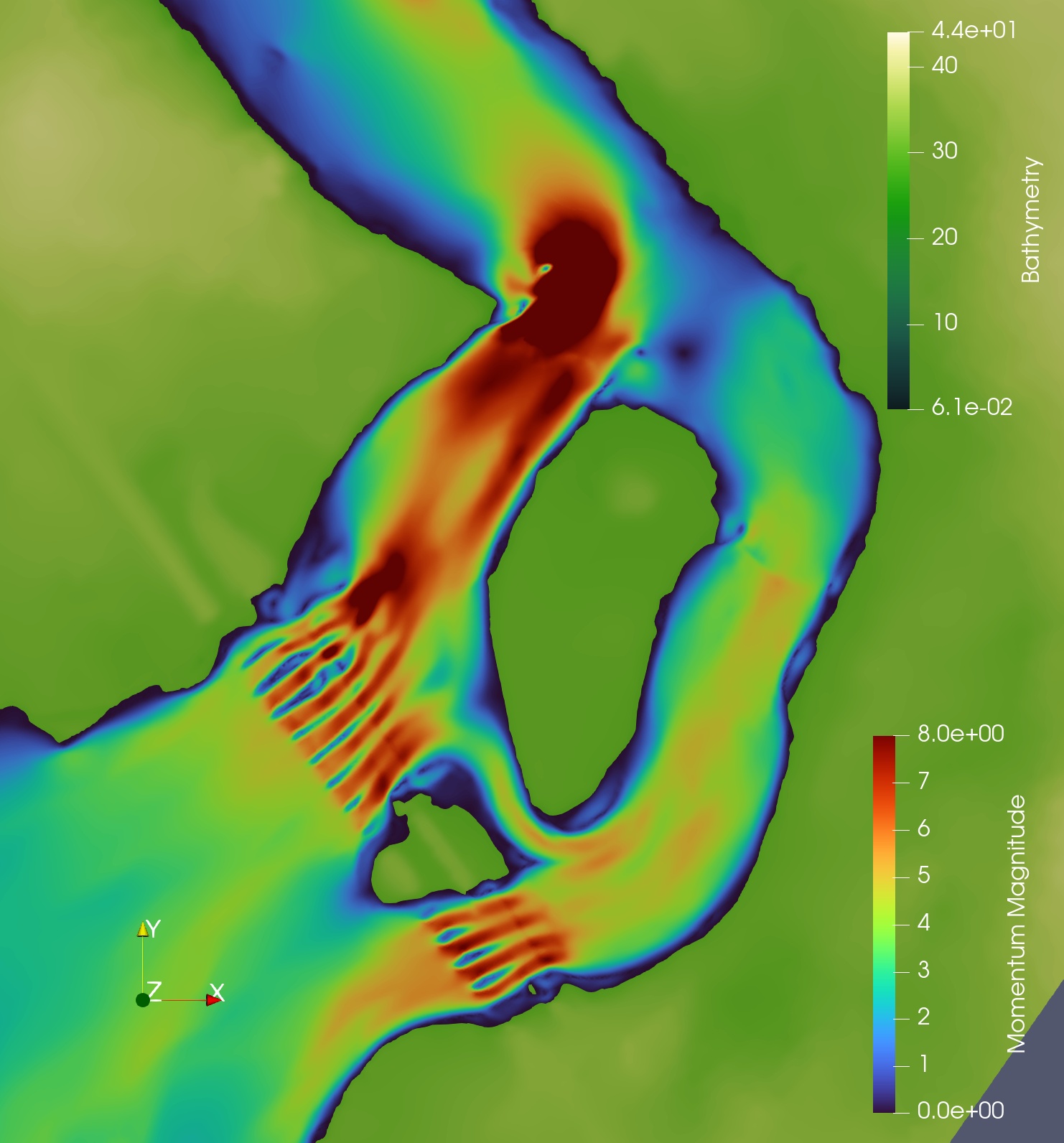}
    } 
    \subfigure[$t=175s$]{
  \includegraphics[width=.23\linewidth]{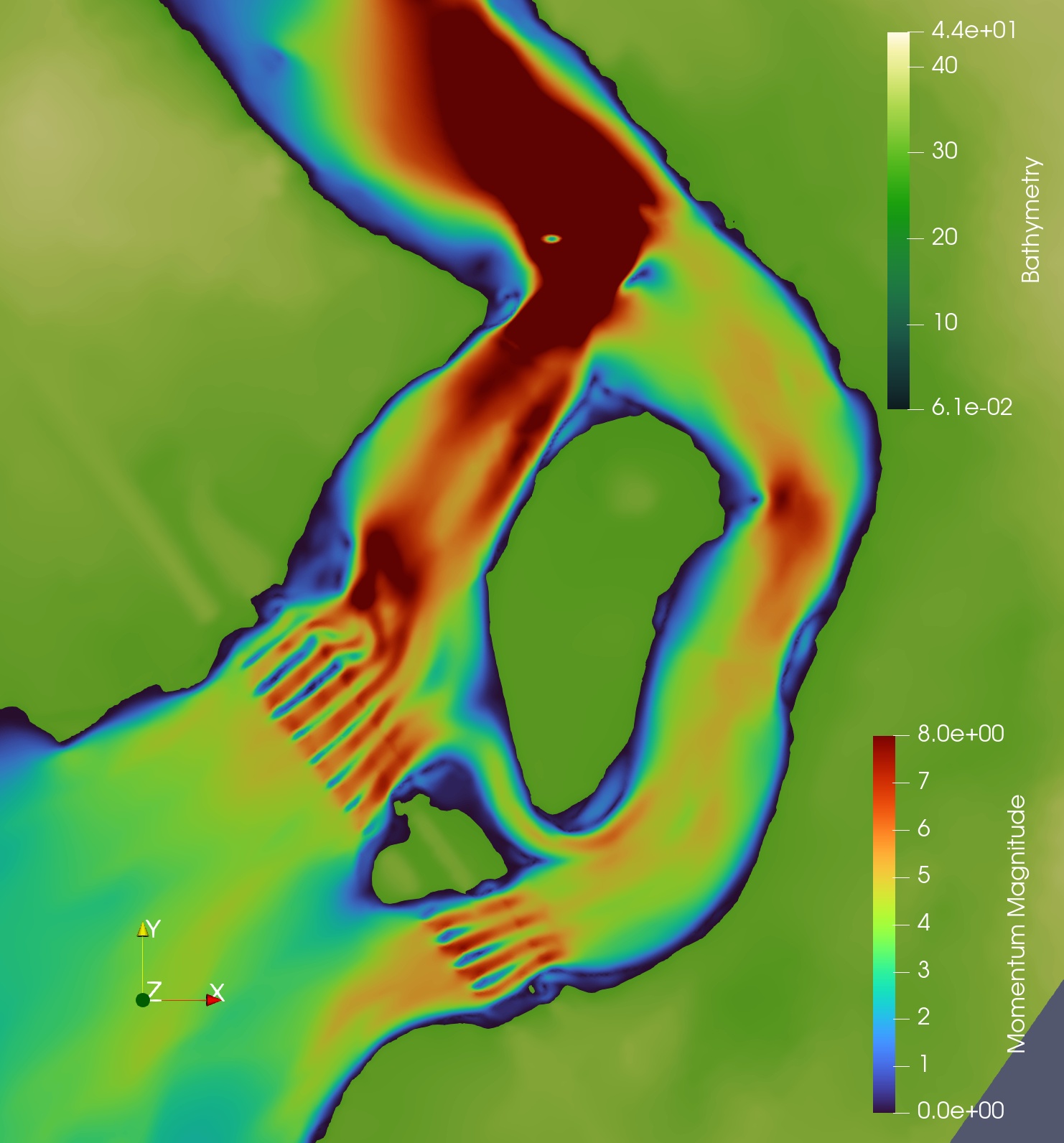}
    } 
    \subfigure[$t=200s$]{
  \includegraphics[width=.23\linewidth]{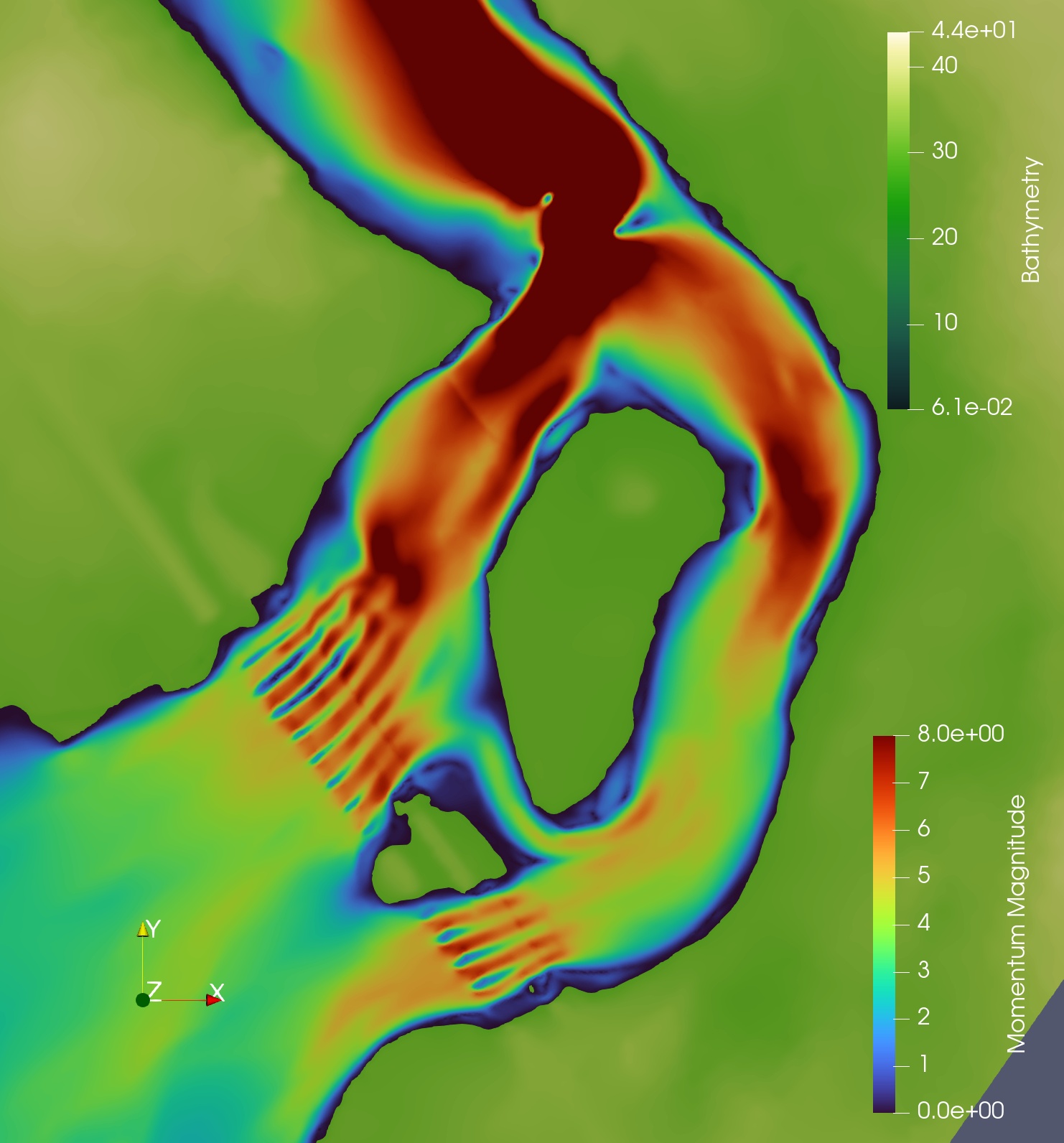}
    } 
    }
}
\end{center}
\caption{Momentum magnitude of the second order solution for the fictitious dam break on the Mille Îles River with an upstream water level of $31m$ and a downstream level of $29m$.}
\label{fig:compar_mille_dam}
\end{figure}

\begin{figure}[pos=htp]
\begin{center}
\fbox{
\parbox{0.9\linewidth} {
    \subfigure[]{
  \includegraphics[width=.48\linewidth]{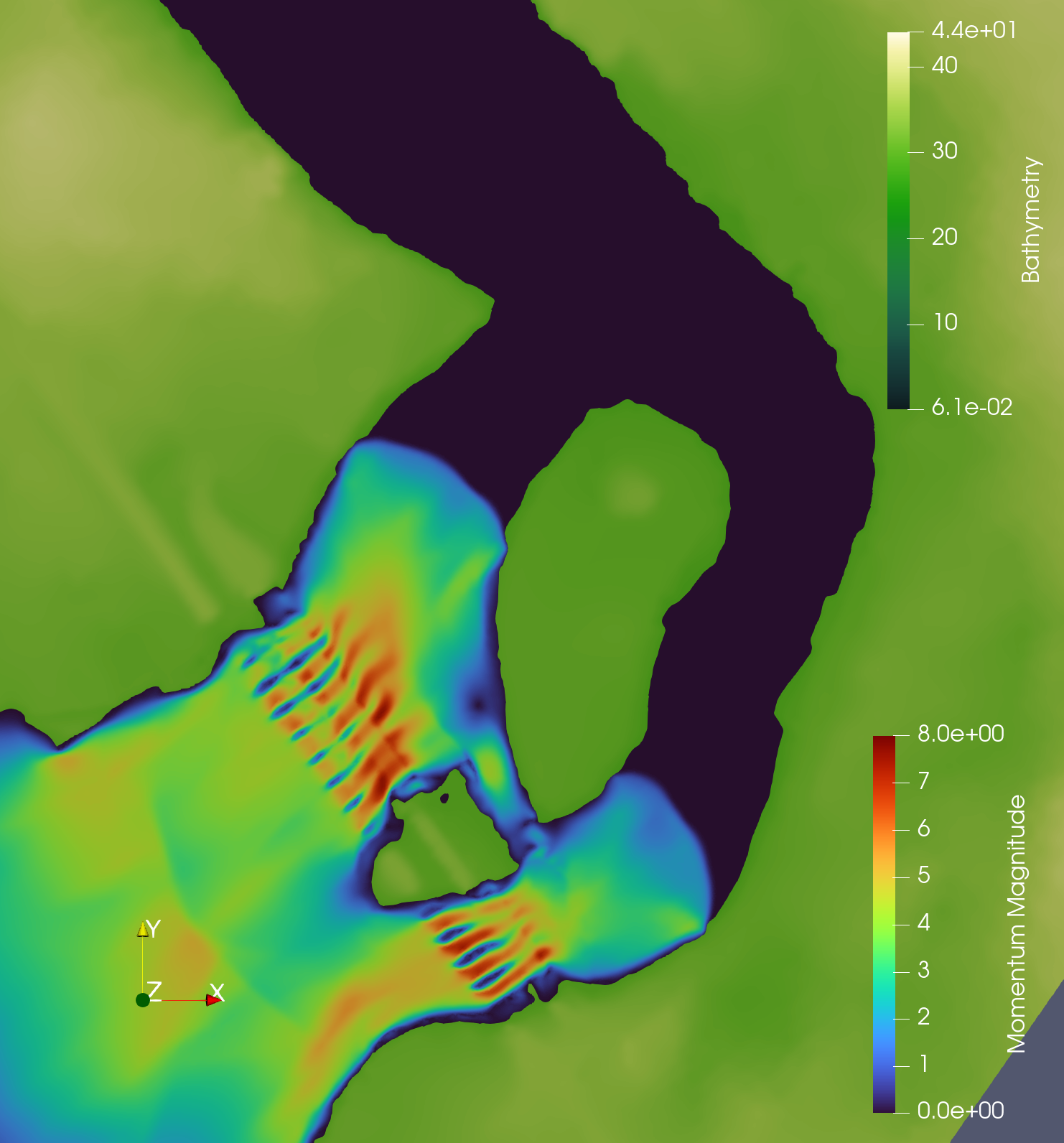}
    } 
    \subfigure[]{
  \includegraphics[width=.48\linewidth]{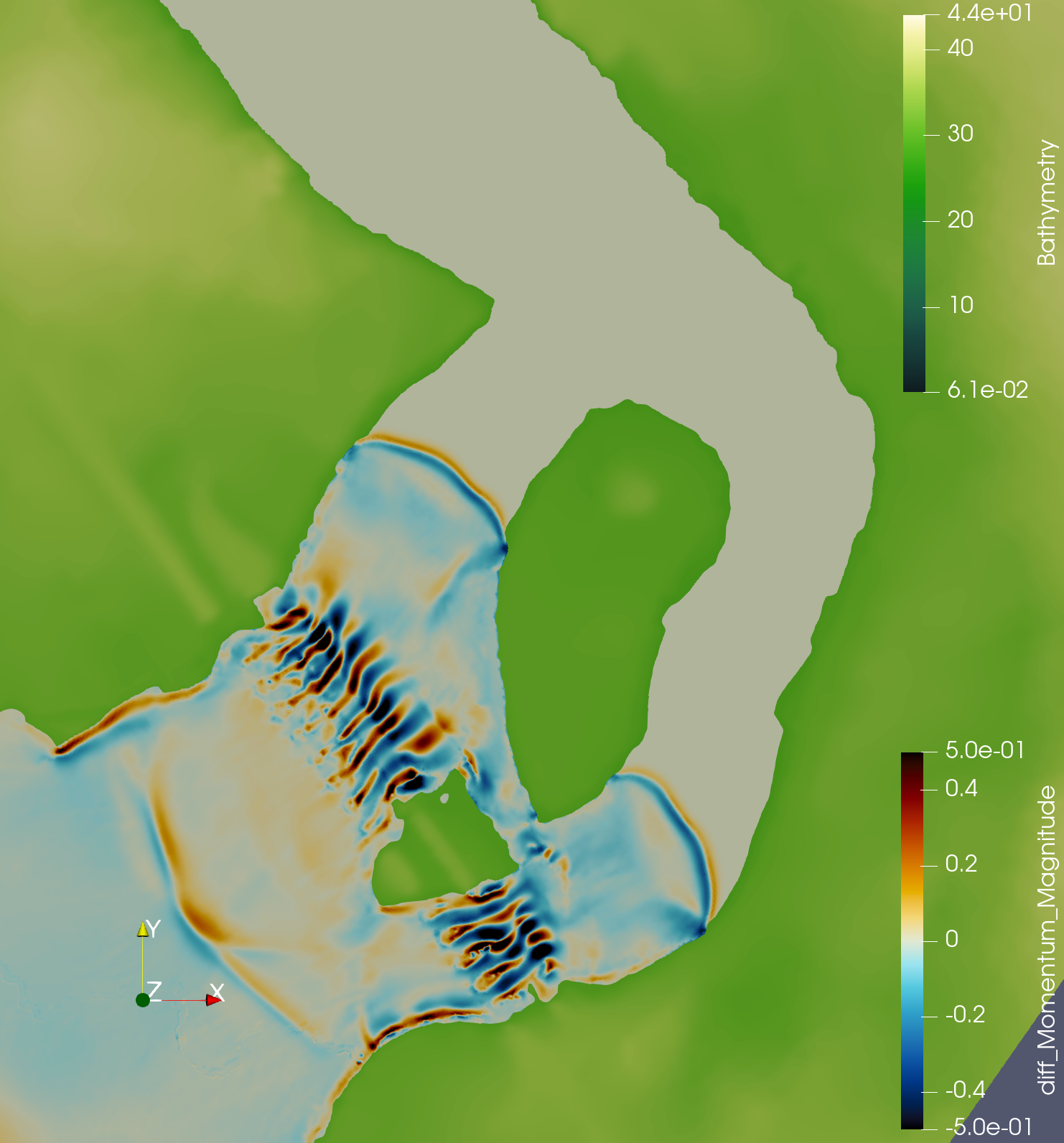}
    } 
    }
}
\end{center}
\caption{Momentum magnitude of the second order solution (a) at $t=50s$. Difference in momentum magnitude between the second- and first-order solutions at $t=50s$ (b). A video of this simulation has been attached to this paper.}
\label{fig:diff_mille_dam}
\end{figure}

Figure \ref{fig:descmilleb} shows the Mille Îles River domain colored according to the bathymetry. A close-up view of the zone that will be used for the fictitious dam break is shown in Figure \ref{fig:descmillec}. A fictitious dam break in this zone is interesting because it will show how the waves reflect off the bridge piers, the middle island, and the side of the terrain before they make a sharp left turn. The solution is initialized like the Riemann problem shown in section \ref{sec:riemprob}. Upstream of the fictitious discontinuity, the water level is set at $31m$, and downstream of the discontinuity it is set at $29m$, the initial velocity is set to zero. The results for this test case are shown in Figure \ref{fig:compar_mille_dam}. Figure \ref{fig:diff_mille_dam} shows the difference between the first- and second-order solutions at time $t=50s$. We can see a distinctive pattern at the front of the wave that is similar to what was shown for the Riemann problem on a wet bottom. In addition, Figure \ref{fig:diff_mille_dam} reveals that the flow of the first- and second-order solutions is quite different in between the piers of the bridge.\\

\begin{figure}[pos=htp]
\begin{center}
\fbox{
\parbox{0.95\linewidth} {
    \subfigure[]{
  \includegraphics[width=.31\linewidth]{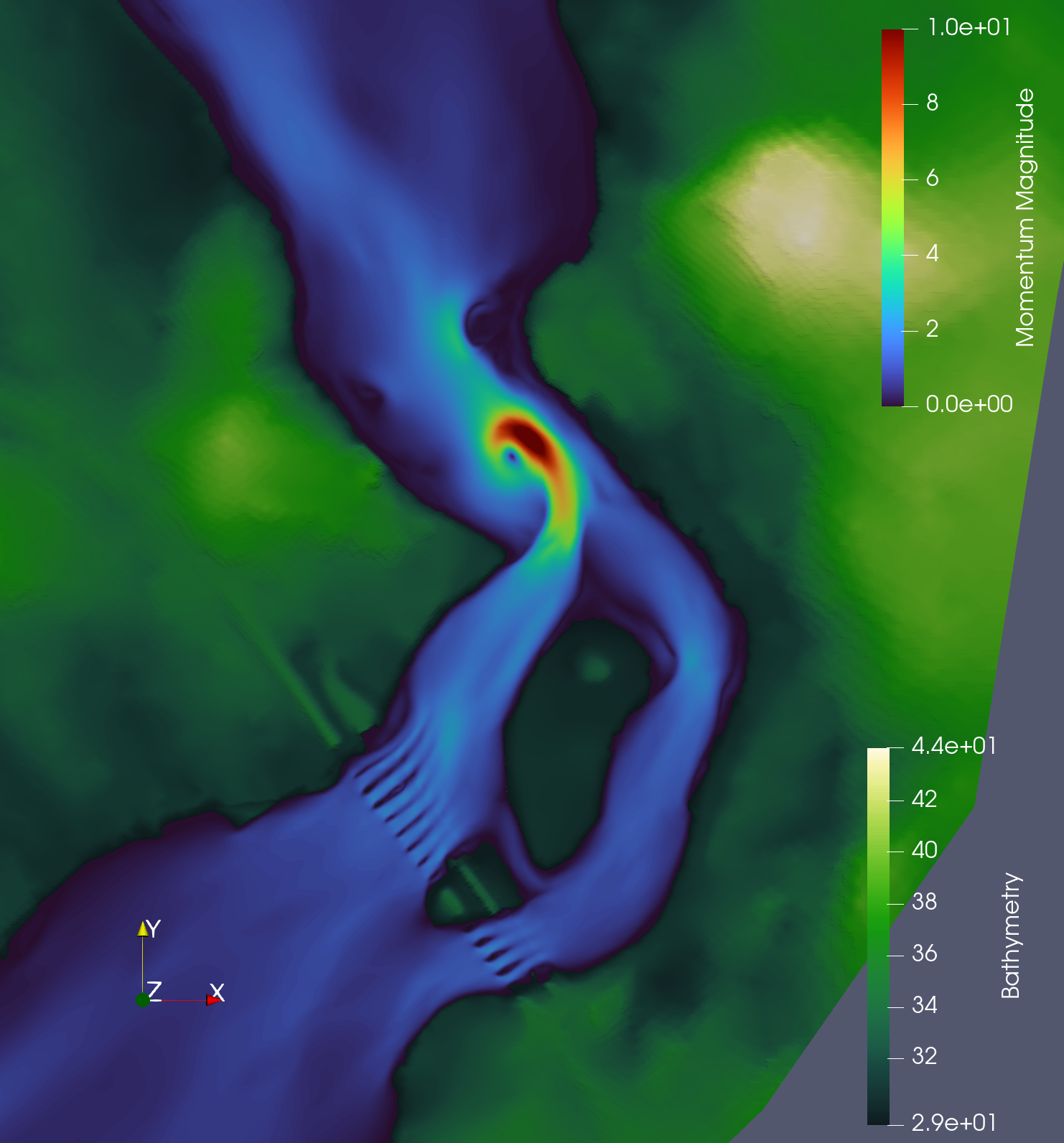}
    } 
    \subfigure[]{
  \includegraphics[width=.31\linewidth]{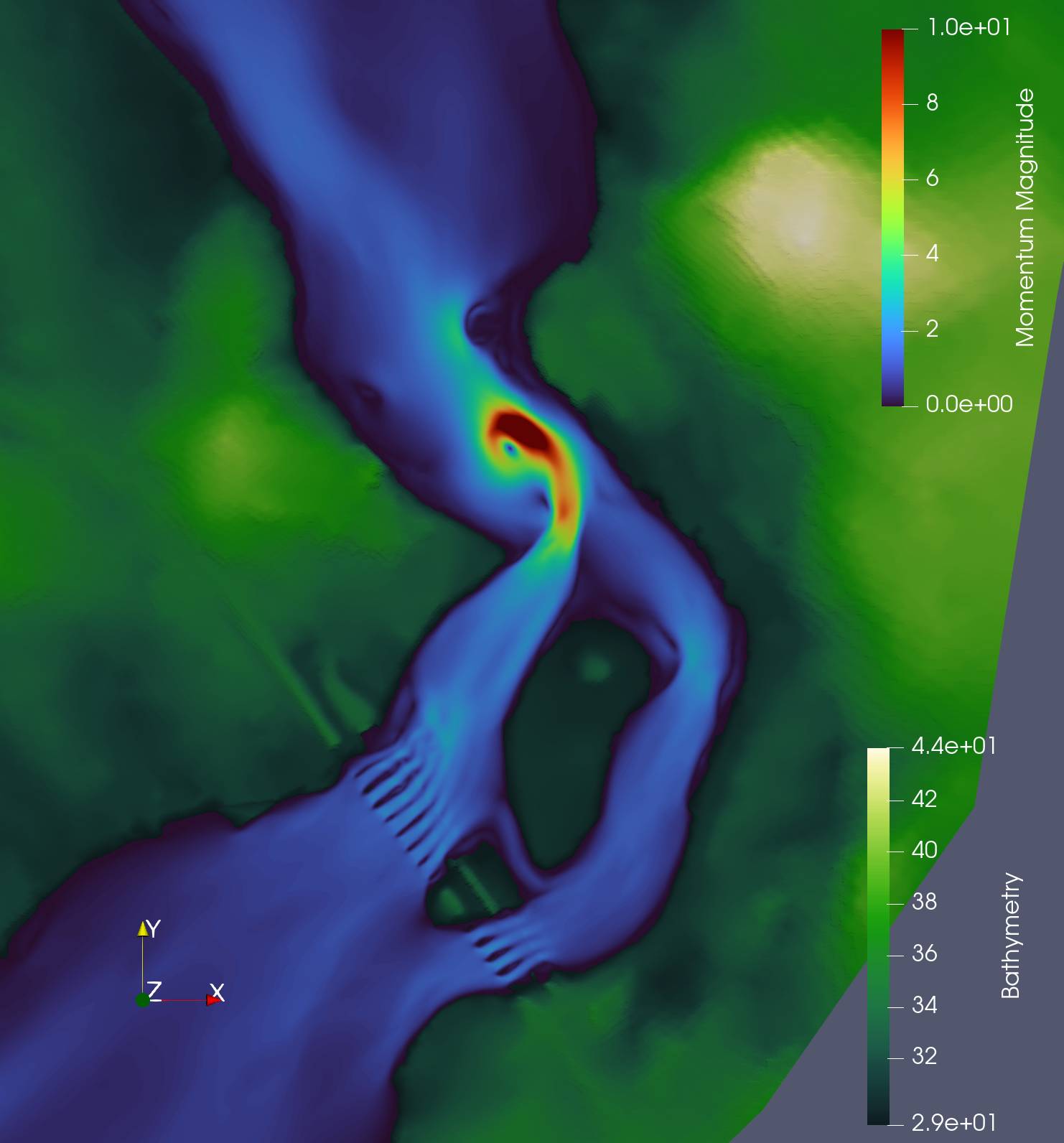}
    } 
    \subfigure[]{
  \includegraphics[width=.31\linewidth]{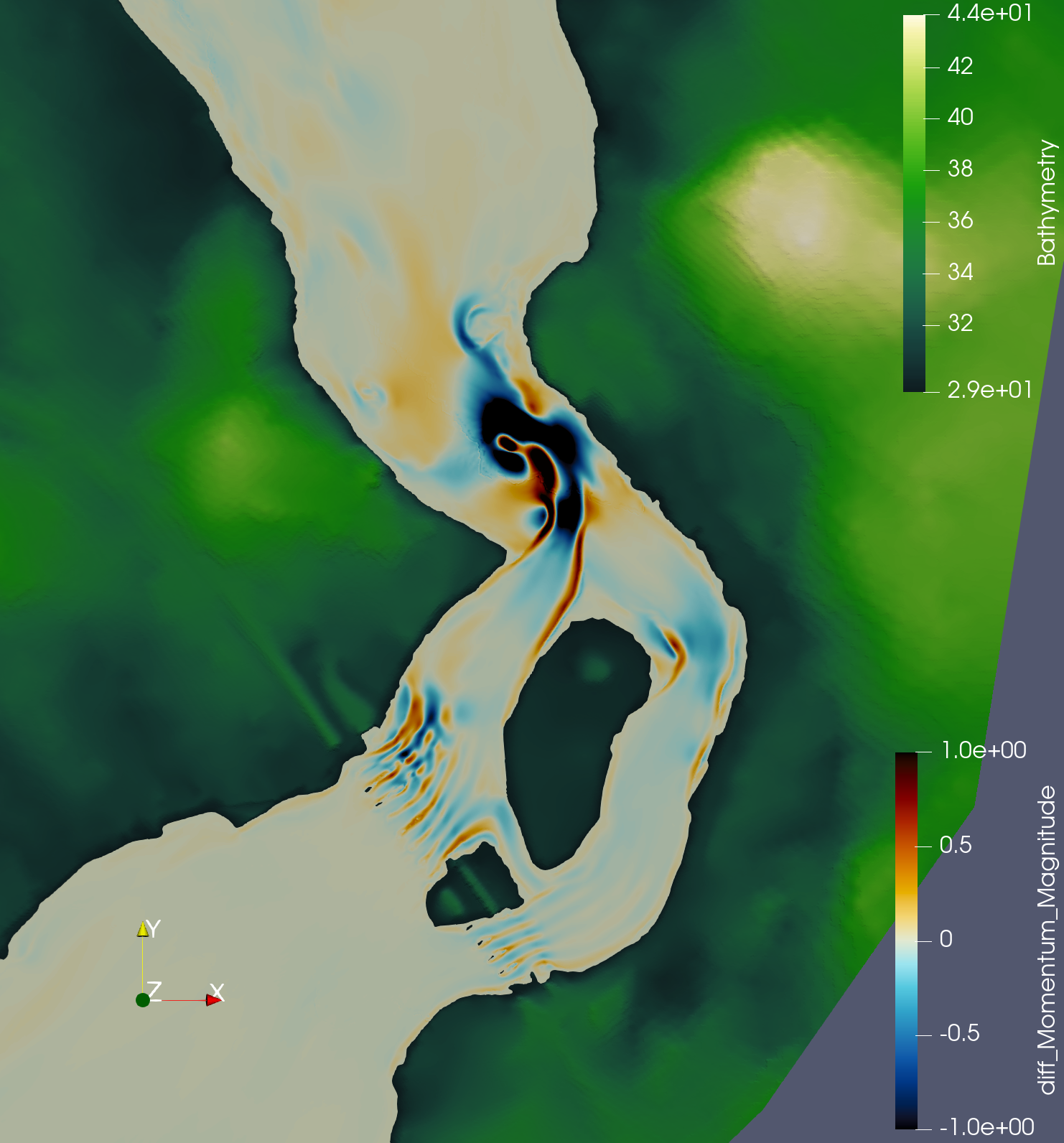}
    } 
    }
}
\end{center}
\caption{Momentum magnitude of the first (a) and second (b) order solutions near the bridge piers on the Mille Îles River domain. The difference in momentum magnitude between the two solutions (c). $t=1500s$. A video of this simulation has been attached to the paper.}
\label{fig:lastmilleup}
\end{figure}

\begin{figure}[pos=htp]
\begin{center}
\fbox{
\parbox{0.9\linewidth} {
    \subfigure[]{
  \includegraphics[width=.49\linewidth]{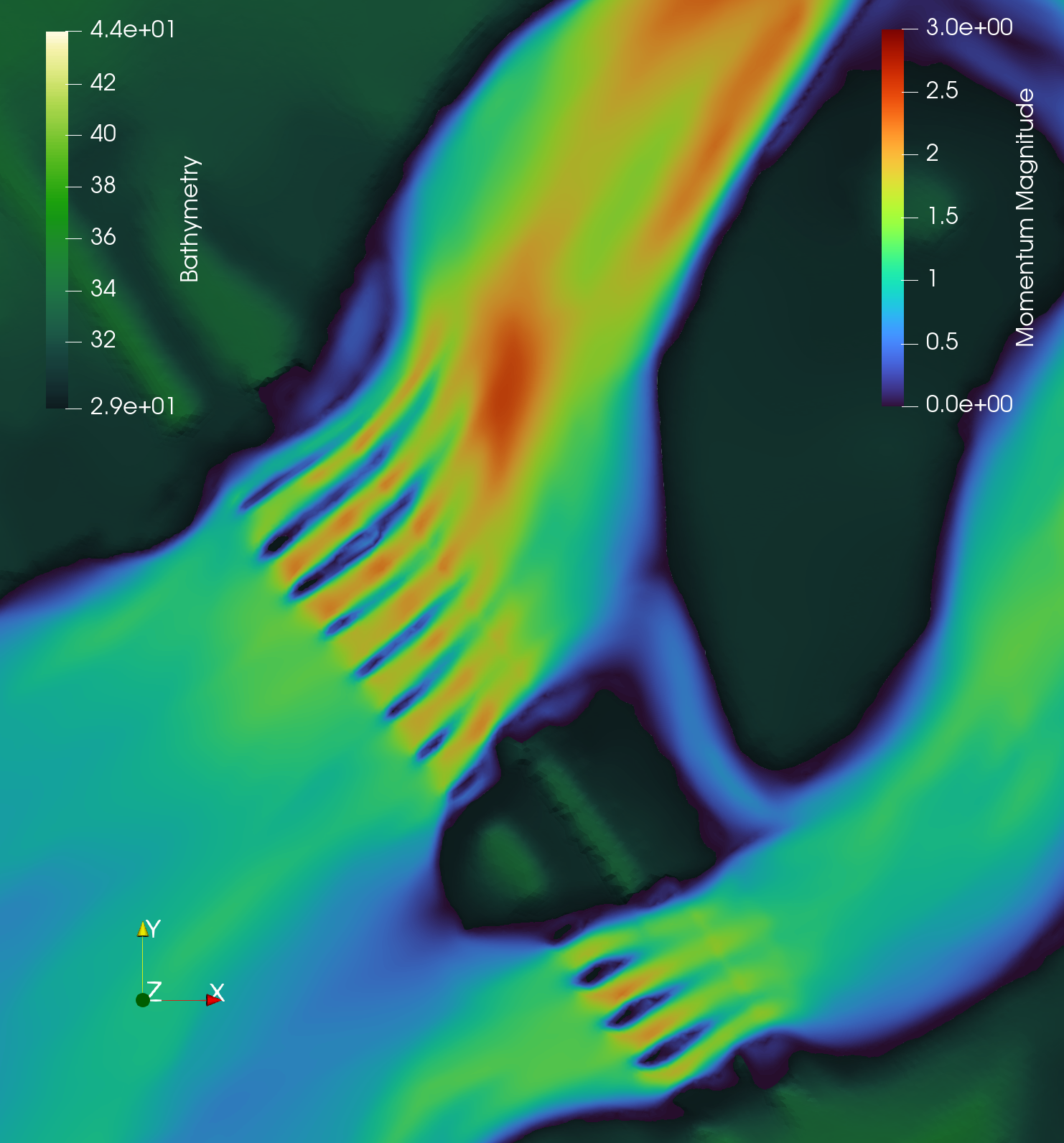}
    } 
    \subfigure[]{
  \includegraphics[width=.49\linewidth]{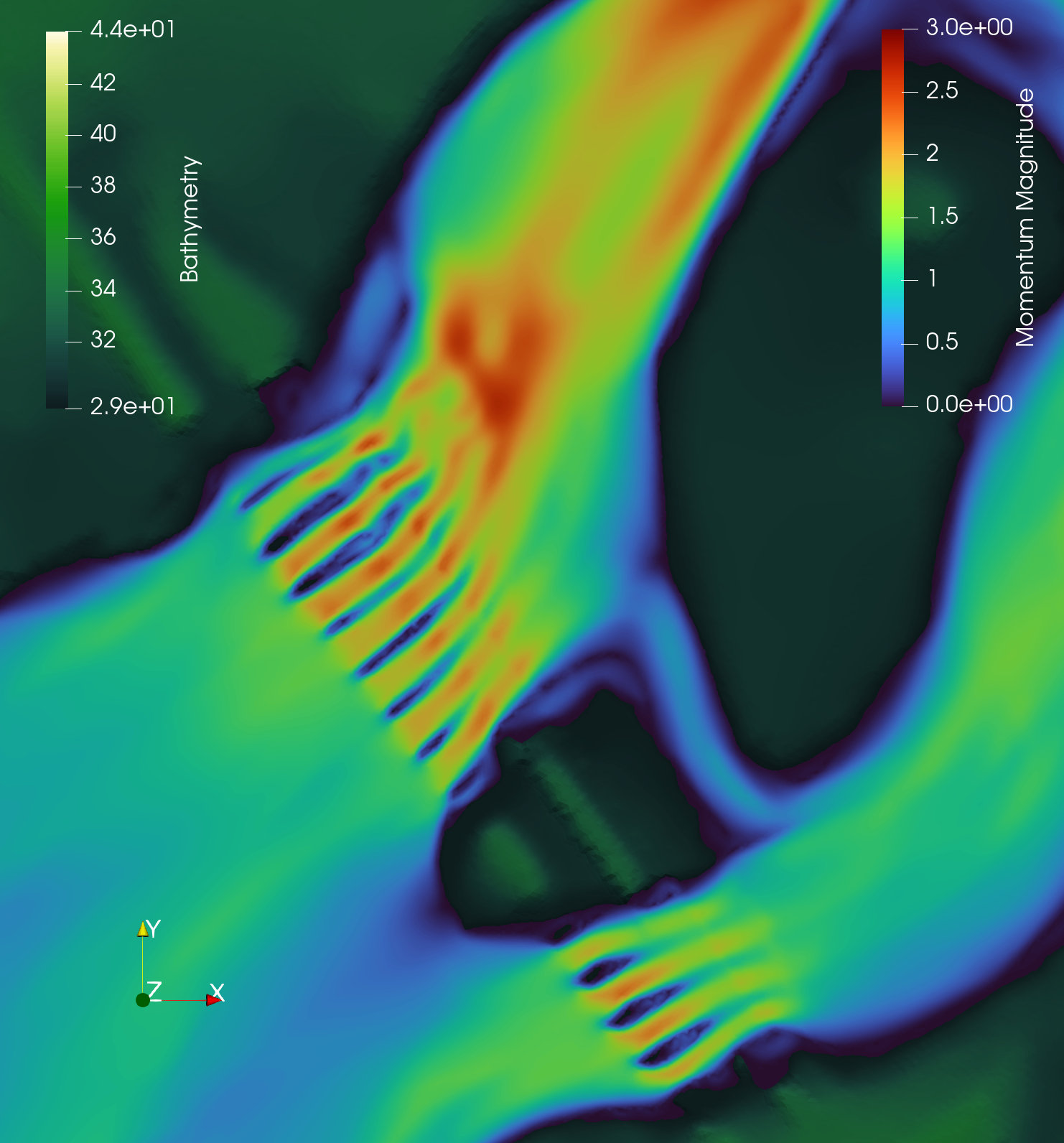}
    } 
    }
}
\end{center}
\caption{Momentum magnitude of the first (a) and second (b) order solutions near the bridge piers on the Mille Îles River domain.}
\label{fig:lastmille}
\end{figure}

We have shown the differences between the first- and second-order solutions for a fictitious dam break over the Mille Îles River but we also want to see the differences between the solutions when the river reaches a steady state. In this case, the water level is initialized at the level of the output of the domain, and a typical inflow of $800m^2.s^{-1}$ is set at the input. The solution reaches the steady state at about $t=3000s$, where we see the outflow matching the inflow. The first- and second-order solutions are similar in most places, except for the zone near the piers of the bridge, which is as expected, since it is  where the water heights and momentum gradient are the most important. A good place to show the difference between the solutions is just after the left turn taken by the river, where a vortex-like structure is created. Figure \ref{fig:lastmilleup} shows the first- and second-order solutions at time $t=1500s$ when the solutions have not yet reached the steady-state. Figure \ref{fig:lastmille} presents a close-up view of the solutions near the piers of the bridge which reveals that the turbulent motion behind the piers is better captured by the second-order solution (b) than by the first-order one (a). A video of the first- and second-order solutions has also been attached to this paper for a clearer view of the differences between the solutions.

\section{Conclusion}
In the first part of this paper, we showed how the domain decomposition and the multiple layers of ghost cells could be generated in a parallel manner. We then evaluated the performance of the ghost-layer generation process, in terms of both memory and time and for a varying number of CPU cores. Our assessment showed that the process scales near perfectly in time, and, with sufficiently large meshes, it also scales well in terms of memory. We then explained how the whole domain decomposition, including the send/receive information, is stored inside a single CGNS mesh file. We illustrated the performance of the IO operations using the simple CGNS library and the Parallel CGNS library, and explained how a combination of both is needed in order to write a usable mesh file. We showed how the use of the PCGNS library, even in a hybrid version, allows for a significant improvement in the IO time.\\

In the second part of the paper, the domain decomposition process presented in the first part was used in the context of the parallel high-order resolution of the Shallow-water equations with our in-house multi-GPU solver. The Shallow-water equations were briefly presented, along with the HLLC first-order and the MUSCL and WAF second-order methods. These methods were then utilized to show the impact and the necessity of using multiple layers of ghost cells on a classic test case of a dam break on a wet bottom. The performance of the solver was studied in detail, both for low and high overlapping versions to show how the multiple layers of ghost cells affect the computation times. It was shown that with a sufficiently high level of overlap in the solver, the computation times are not increased by the use of multiple layers of ghost cells. Finally, two real domains with complex bathymetries were studied. This evaluation showed how the second-order solution differs from the first-order one in both domains, particularly in zones with high gradients of water heights or velocities. Future work will be aimed at showing the impact of other high-order methods like ENO or WENO, as well as the impact of high-order hydrostatic reconstruction on the flow patterns of rivers on large-scale meshes. Another avenue will be to improve the the PCGNS library with a standard method of storing the send/receive information for each sub-domain and their boundary conditions.

\section*{Acknowledgements}
This research was enabled in part by funding from the National Sciences and Engineering Research Council of Canada; by bathymetry data from the Montréal Metropolitan Community (\textit{Communauté métropolitaine de Montréal}); and by computational support from Calcul Québec and Compute Canada.

\bibliography{bib}
\end{document}